\documentclass[fleqn,setspace,amsmath,amssymb]{article}
\usepackage[dvips]{graphicx}
\usepackage{setspace}
 
\newcommand{\ra}{\rangle}
\newcommand{\pa}{\partial}
\newcommand{\la}{\langle}
\newcommand{\be}{\begin{equation}}
\newcommand{\ee}{\end{equation}}
\newcommand{\bea}{\begin{eqnarray}}
\newcommand{\eea}{\end{eqnarray}}

\begin{document}
\begin{titlepage}

\begin{flushright}
\today
\end{flushright}

\vspace{1in}

\begin{center}

{\bf Core-halo quasi-stationary states in the Hamiltonian mean-field model}

\vspace{1in}

\normalsize

{Eiji Konishi\footnote{E-mail address: konishi.eiji.27c@st.kyoto-u.ac.jp}}

\normalsize
\vspace{.5in}

{\it Graduate School of Human and Environmental Studies, Kyoto University\\
 Kyoto, 606-8501, Japan}
\end{center}

\vspace{1in}

\baselineskip=24pt
\begin{abstract}
A characteristic feature of long-range interacting systems is that they become trapped in a non-equilibrium and long-lived quasi-stationary state (QSS) during the early stages of their development.
We present a comprehensive review of recent studies of the core-halo structure of QSSs, in the Hamiltonian mean-field model, which is a mean-field model of mutually coupled ferromagnetic $XY$ spins located at a point, obtained by starting from various unsteady rectangular water-bag type initial phase-space distributions.
The main result exposed in this review is that the core-halo structure can be described by the superposition of two independent Lynden-Bell distributions.
We discuss the completeness of collisionless relaxation of this double Lynden-Bell distribution by using both of Lynden-Bell entropy and double Lynden-Bell entropy for the systems at low energies per particle.
\end{abstract}

\vspace{.7in}
 
\end{titlepage}

\tableofcontents
\section{Introduction}

It is well-known that Boltzmann introduced the idea of a {\it{time-dependent distribution function}}, whose evolution obeys a kinetic equation, in order to explain the Gaussian velocity distribution of dilute equilibrium gases and show the existence of a Lyapunov function.
In the original paper, these aims were achieved by deriving the collision terms in the kinetic equation from physical considerations.\cite{Boltzmann}
After that, in thermodynamics, equilibrium was considered a proper phenomenon for collisional processes.

This situation dramatically changes when we consider non-equilibrium statistical mechanics of long-range interacting systems.
In a seminal paper, Lynden-Bell introduced {\it{collisionless (Lynden-Bell) entropy}} and then founded the ergodic theory of {\it{collisionless equilibrium}} for a coarse-grained distribution function in the context of stellar self-gravitating systems whose dynamics is violent.\footnote{In this review, we refer to the maximum Lynden-Bell entropy state as the {\it{Lynden-Bell equilibrium}}. The phrase {\it{collisionless equilibrium}} is used in a broader sense.}\cite{LB}
Interestingly, this collisionless equilibrium is induced by the potential-driven flow term only and not the collision terms in the Boltzmann equation, so it differs essentially from the collisional equilibrium.
Furthermore, as dynamical processes directly contribute to this equilibrium, it is regarded as an interface between dynamics and `thermodynamics'.
Because of these interesting properties, this collisionless equilibrium is the central subject of this review.

Lynden-Bell's theory has broadly influenced long-range physics.\cite{Text,ReviewI,ReviewII}

A characteristic feature of a long-range system is that, in the early stages of its development, it will be trapped in a non-equilibrium and long-lived {\it{quasi-stationary state}} (QSS).\cite{Text,ReviewI,ReviewII,LRR,LRT,Yamaguchi,Barre,Er}
This QSS temporally separates the collisionless regime from the collisional regime of the system and its life time diverges with the number of particles.\cite{Yamaguchi}
In particular, collisionless equilibrium states are QSSs.

In this review, we study QSSs in the {\it{Hamiltonian mean-field (HMF) model}}, which is a widely studied classical mechanical benchmark model for long-range systems.\cite{Text,ReviewI,KK,IK,Inagaki,Pichon,HMF,HMF1,HMF2}
This model deals with numerous identical particles, with unit mass, moving on a circle by the mean-field method.
The system is fully coupled and the interactions between particles depend on the cosine of their angular separation.
As a technical point, in the HMF model, if the initial state is a steady state, we need to take its dynamical stability into consideration.\cite{Yamaguchi}
In this review, we consistently consider only unsteady initial conditions, in order to study the systems that undergo their violent dynamical processes.

In the HMF model with such a setting, the question that should be addressed first is the realizability of the Lynden-Bell equilibrium state (i.e., the ergodic collisionless equilibrium state) as a QSS; in other words, the completeness of violent relaxation in the collisionless equilibrium should be examined by performing $N$-body simulations.\cite{AFBCDR,ACFR,NET1,SCDF}
The result is that, except for special cases, the QSS distributions may deviate considerably from the Lynden-Bell equilibrium one: the violent relaxation may be badly incomplete.

{{Based on this result, several years ago, Pakter and Levin began the study of the nonequilibrium core-halo structure appearing in QSSs beyond the Lynden-Bell model by following previous research.\footnote{This core-halo structure was first observed in early numerical simulations of 1D and 2D self-gravitating systems.\cite{HC,GCL,CGL,LC,Tanekusa,Mineau,Yamaguchi0,Teles2}}\cite{ReviewII,Teles2,Teles1,Teles3,PL}
In their definition, the core consists of low energy particles and decouples, on the phase space, from the halo that consists of high energy particles.
In the HMF model, the machanism for halo formation is considered to be a parametric resonance with the initial strong oscillation of the self-consistent mean-field potential.\cite{PL,BTPL}
Pakter and Levin proposed the core-halo ansatz for the one-particle energy distributions that has no fitting parameter and reproduces the simulation results well in the position and the momentum plots.
However, as mentioned above, the definition of the core-halo structure was the traditional one: namely, an {\it{attachment}} of the core and the halo on the phase space.
Concretely, the use of ``attachment'' in the ansatz refers to a completely decoupled core and halo, which are both degenerate (i.e., each has an everywhere constant phase-space density), at the maximum energy of the core.
In this review we show, in contrast, that these QSSs are actually {\it{superpositions}} of new types of core and halo that are defined by two independent Lynden-Bell equilibria.\cite{KS}
We call this equlibrium the {\it{double Lynden-Bell equilibrium}}.
Based on preceding research by the author and others\cite{PL,BTPL,KS}, we review this double Lynden-Bell scenario for QSSs with a core-halo structure arising from initial unsteady rectangular water-bag phase-space distributions with a common fine-grained level.
(Here, {\it{water-bag}} means that the phase-space distribution has a single non-zero fine-grained level $f$.
In this review, we do not discuss the multi-level version\cite{Ass,PLTop,PLmulti}.)}}

{{To clarify the significance of this scenario here, we consider the thermodynamic limit of the HMF system in the microcanonical approach: that is, we let the number of particles $N$ of the system tend to infinity while fixing the energy per particle $\hat{E}=E/N$ and the phase-space density per particle $\hat{\eta}=\eta/N$.\cite{Vlasov4,Kuzemsky}
In this limit, the collisional effects on this long-range system can be completely neglected.
Since in the formation process of the double Lynden-Bell distribution of the HMF system, the partitions of $N$, $E$ and $\eta$ into the core and the halo are fixed during violent relaxation (see Sec. 4.2.2 where the calculations performed in Sec. 2.1.2 are used), it is possible to formulate the thermodynamics corresponding to the double Lynden-Bell distribution.}}

{{Based on this standpoint, we claim that the double Lynden-Bell distribution in itself is an essentially novel type of equilibrium distribution in statistical physics.
The single Lynden-Bell distribution has the same form as the Fermi-Dirac distribution except for the former's overall fine-gained level factor.
In the double Lynden-Bell distribution at zero temperature, namely, the ground state of the system, there are two coexisting Fermi energies in a superposition: that is, the distribution refers to one kind of particle; it is not a mixture of two Lynden-Bell distributions at zero temperature for two different kinds of particles.
This review discusses the foundations of the theory of double Lynden-Bell equilibria by using the HMF model, which is the simplest model of a long-range system.}}

The organization of this review is as follows.

In the next section, after brief accounts of the Boltzmann and Vlasov equations for the collisional and collisionless regimes, respectively, we introduce the basic notions needed to describe QSSs in long-range systems: phase mixing, violent relaxation and the Lynden-Bell statistics.\cite{LB}

In Section 3, we define the HMF model and explain its Boltzmann-Gibbs collisional equilibrium structure, in particular, the caloric curve and the collisional equilibrium second order phase transition property, using the micro-canonical approach.

In Section 4, we study QSSs with a core-halo structure in the HMF model at low energies per particle.\cite{KS}
After reviewing the preceding research,\cite{ReviewII,PL,BTPL} first we follow the evolution of the system till its collisionless equilibrium to describe the formation process of the core-halo structure in the double Lynden-Bell scenario.
Second, by performing several illustrative $N$-body simulations, we corroborate the double Lynden-Bell structure of these QSSs.
Finally, we examine the deviation degree of the QSSs from the Lynden-Bell equilibrium and the completeness of the collisionless relaxation of the QSSs by using the Lynden-Bell entropy and the double Lynden-Bell entropy, respectively.
The result for the latter is that in most cases the collisionless relaxation in the double Lynden-Bell sense is incomplete.

In the final section, we summarize the results in the double Lynden-Bell scenario and give an outlook.

In the appendices, we provide supplementary calculations used in the main text.

Throughout this review, the number of particles in the simulations is assumed to be $10^4$.
The simulation time is also assumed to be $10^4$ unless otherwise noted.
We use a hat to denote per-particle normalization.

Here, as in previous work\cite{KS}, we use $\sim$ to denote a series expansion up to a finite number of terms or approximate equality between independent variables (differing from the standard meaning of equality up to a multiplicative constant of ${\cal{O}}(1)$) and use $\approx$ for other types of approximate equality.
\section{Basic Notions}
\subsection{Kinetic theory}
\subsubsection{Collisional regime}
In general macroscopic systems, the fundamental kinetic equation which governs the evolution of the time-dependent distribution functions $f$ is the Boltzmann equation.
The form of the {\it{Boltzmann equation}} is
\begin{equation}
\frac{df}{dt}=I(f,f)\;,\label{eq:Boleq}
\end{equation}
where $I$ is a quadratic functional of $f$ that represents the effect of collisions on the temporal evolution of $f$.\cite{Boltzmann}
Boltzmann's original paper considered the time-dependent kinetic energy distribution function $f(x,t)$ of gaseous molecule with kinetic energy $x$.\cite{Boltzmann}
Here, $f(x,t)dx$ is the number of molecules in a unit volume with kinetic energy in the range $(x,x+dx)$ at time $t$.
We illustrate his idea by giving the explicit form of $I(f,f)$ in this short-range case:
\begin{eqnarray}
(I(f,f))(x,t)=\int_0^\infty dx^\prime \int_0^{x+x^\prime}dy&&(f(y,t)f(x+x^\prime-y,t)\psi(y,x+x^\prime -y,x)\nonumber \\&&
-f(x,t)f(x^\prime,t)\psi(x,x^\prime,y))\;.
\end{eqnarray}
In the first term, which represents the incoming collisions, the kinetic energies of particles are within the ranges shown in the first table.
\begin{table}[htbp]
\begin{tabular}{rccc}
&&Particle $a$&Particle $b$\\
Before collision&$\cdots$&$(y,y+dy)$&$(x+x^\prime-y,x+x^\prime+dx^\prime-y)$\\
After collision &$\cdots$&$(x,x+dx)$&\\
\end{tabular}
\end{table}

\begin{flushleft}
In the second term, that represents the outgoing collisions, the kinetic energies of particles are within the ranges shown in the second table.
\end{flushleft}
 \begin{table}[htbp]
\begin{tabular}{rccc}
&&Particle $a$&Particle $b$\\
Before collision&$\cdots$&$(x,x+dx)$&$(x^\prime,x^\prime+dx^\prime)$\\
After collision &$\cdots$&$(y,y+dy)$&\\
\end{tabular}
\end{table}

The proportionality factor $\psi$ is positive-valued and depends on the three variables of the binary collisions and on the action law of gaseous molecules.

From physical considerations, it was shown that the factor $\psi$ satisfies\cite{Boltzmann}
\begin{eqnarray}
\psi(x,x^\prime,y)&=&\psi(x^\prime,x,x+x^\prime-y)\;,\label{eq:psi1}\\
\sqrt{xx^\prime}\psi(x,x^\prime,y)&=&\sqrt{y(x+x^\prime-y)}\psi(y,x+x^\prime-y,x)\label{eq:psi2}
\end{eqnarray}
for arbitrary $x$,$x^\prime$ and $y$.

The first property, Eq.(\ref{eq:psi1}), is shown in the following way.\cite{Boltzmann}
By comparing the process shown in the second table with the equivalent process shown in the third table where particles $a$ and $b$ are reversed,
\begin{table}[htbp]
\begin{tabular}{rccc}
&&Particle $a$&Particle $b$\\
Before collision&$\cdots$&$(x^\prime,x^\prime+dx^\prime)$&$(x,x+dx)$\\
After collision &$\cdots$&$(x+x^\prime-y-dy,x+x^\prime -y)$&\\
\end{tabular}
\end{table}
we obtain two expressions for the same number of collisions $dn$ within a very short time span $\tau$ in a unit volume
\begin{eqnarray}
dn&=&\tau f(x,t)dx\cdot f(x^\prime,t)dx^\prime\cdot dy\psi(x,x^\prime,y)\\
&=&\tau f(x^\prime,t)dx^\prime \cdot f(x,t)dx\cdot dy \psi(x^\prime,x,x+x^\prime-y-dy)\;.
\end{eqnarray}
Then, by dropping $dy$ from $\psi$, the first property follows.

The derivation of the second property, Eq.(\ref{eq:psi2}), is rather difficult.
We only comment that, to derive it, we assume that the force between two particles is a function of their distance and obeys the law of action and reaction.

By using these properties of $\psi$, the kinetic equation except for the flow terms can be rewritten as
\begin{eqnarray}
\biggl[\frac{\pa f(x,t)}{\pa t}\biggr]_c=\int_0^\infty dx^\prime \int_0^{x+x^\prime}dy\biggl[\frac{f(y,t)}{\sqrt{y}}\frac{f(x+x^\prime -y,t)}{\sqrt{x+x^\prime-y}}-\frac{f(x,t)}{\sqrt{x}}\frac{f(x^\prime,t)}{\sqrt{x^\prime}}\biggr]\sqrt{xx^\prime}\psi(x,x^\prime,y)\;.
\end{eqnarray}
By using this form of the kinetic equation, Boltzmann carried out calculations to show that Boltzmann's ${\cal{H}}$-function\cite{Boltzmann}
\begin{equation}
{\cal{H}}=\int_0^\infty dx \biggl[f(x,t)\biggl\{\ln \biggl[\frac{f(x,t)}{\sqrt{x}}\biggr]-1\biggr\}\biggr]
\end{equation}
is a Lyapunov function of the system, i.e., never increases over time: $d {\cal{H}}/d t\le 0$.
In this review, we do not present these calculations.
It was also shown that ${\cal{H}}$ has a negative minimum where the kinetic energy distribution function is
\begin{equation}
f_0(x,t)=C\sqrt{x} e^{-hx}\;.\label{eq:f00}
\end{equation}
Indeed, for Eq.(\ref{eq:f00}), $[\pa f_0(x,t)/\pa t]_c$ vanishes.
In terms of velocity, $f_0$ is a Gaussian distribution.
In this review, we do not explain the detailed properties of the Boltzmann equation because those lie outside of our main interest.
\subsubsection{Collisionless regime}

{{In this subsection, we consider the HMF system whose phase-space variables are a particle's position $\theta$ on the circle and its canonical conjugate momentum $p$.}}

{{In the collisionless regime, the interparticle interaction of a long-range system, represented by the motion of the particles in the self-consistent mean-field, is described by the Vlasov equation\cite{Vlasov5,Vlasov6,Vlasov0,BH} that just drops the collision term $I(f,f)$ from the Boltzmann equation Eq.(\ref{eq:Boleq}) and can be derived from the Bogoliubov-Born-Green-Kirkwood-Yvon hierarchy by using a perturbative expansion.}}

{{In long-range systems, it is known that the collision term in the Boltzmann equation is of order $1/N$, and thus the collisional evolution is slow for a system with numerous particles.\cite{Vlasov1,Vlasov2,Vlasov3}
Due to this long-range nature, the complete vanishing of the collision term is achieved in the thermodynamic limit $N\to \infty$ where $\hat{E}$ and $\hat{\eta}$ are fixed in the micro-canonical approach.
In the long-range context, while the collisionless effects are collective, the collision effects are due to granularity, that is, they are a {{finite $N$ correction}}.\cite{ReviewI}}}

The {\it{Vlasov equation}} of the phase-space distribution function $f(\theta,p,t)$ is
\begin{equation}
\frac{df}{dt}=\frac{\pa f}{\pa t}+p\frac{\pa f}{\pa \theta}+F(\theta)\frac{\pa f}{\pa p}=0\;,\ \ F(\theta)=-\frac{\pa \Phi(\theta)}{\pa\theta}\;,
\end{equation}
where $\Phi(\theta)$ is the self-consistent mean-field potential energy function which is the average over the distribution function.\cite{Vlasov5,Vlasov6,Vlasov0,BH}
So, this is a non-linear equation of the distribution function.
Of course, this equation is for the fine-grained distribution function and does not hold for the coarse-grained distribution function.

We now look into the basic consequences of the Vlasov equation.
First, we show that the continuity equation follows from the Vlasov equation.
To see this, we integrate the Vlasov equation with respect to momentum:
\begin{equation}
\int_{-\infty}^\infty \frac{\pa f}{\pa t}dp+\int_{-\infty}^\infty p\frac{\pa f}{\pa \theta}dp+\int_{-\infty}^\infty F(\theta)\frac{\pa f}{\pa p}dp=0\;.
\end{equation}
Here,
\begin{eqnarray}
({\rm{1st\ Term}})&=&\frac{\pa }{\pa t}\int_{-\infty}^\infty f dp \\
&=&\frac{\pa \varrho}{\pa t}\;,\ \ \varrho=\int_{-\infty}^\infty f dp \;,\\
({\rm{2nd\ Term}})&=&\frac{\pa }{\pa \theta}\int_{-\infty}^\infty (p f)dp \\
&=&\frac{\pa(\varrho v)}{\pa \theta}\;,\ \ v=\frac{1}{\varrho}\int_{-\infty}^\infty (pf) dp \;,\\
({\rm{3rd\ Term}})&=&\int_{-\infty}^\infty \frac{\pa }{\pa p}(fF)dp \\
&=&(fF)|_{-\infty}^\infty \\
&=&0\;.
\end{eqnarray}
This leads to the differential continuity equation
\begin{equation}
\frac{\pa \varrho}{\pa t}+\frac{\pa}{\pa \theta}(\varrho v)=0\;,
\end{equation}
 and the total mass $N$ is conserved:
 \begin{equation}
 \frac{dN}{dt}=0\;.
 \end{equation}

Next, we show that for a compact container space such as the circle in the HMF model, energy conservation follows from the Vlasov equation.
To see this, we take the second order moment of the Vlasov equation with respect to momentum
\begin{eqnarray}
\frac{d}{dt}\int_0^{2\pi}d\theta\int_{-\infty}^\infty dp (p^2f)+\int_0^{2\pi}d\theta\int_{-\infty}^\infty dp \frac{\pa}{\pa \theta}(p^3f)+\int_0^{2\pi}d\theta \int_{-\infty}^\infty dp \biggl(p^2\frac{\pa}{\pa p}(Ff)\biggr)=0\;.
\end{eqnarray}
Here,
\begin{eqnarray}
({\rm{1st\ Term}})&=&2\frac{dK}{dt}\;,\\
({\rm{2nd\ Term}})&=&0\;,\\
({\rm{3rd\ Term}})&=&-2\int_0^{2\pi}d\theta\int_{-\infty}^\infty dp (pFf)\\
&=&-2\int_0^{2\pi} F(\varrho v)d\theta\\
&=&2\int_0^{2\pi} \frac{\pa \Phi}{\pa \theta}(\varrho v)d\theta \\
&=&-2\int_0^{2\pi} \Phi\frac{\pa (\varrho v)}{\pa \theta}d\theta \\
&=&2\int_0^{2\pi} \Phi\frac{\pa \varrho}{\pa t}d\theta\;.
\end{eqnarray}
From the above equations we obtain
\begin{equation}
\frac{dK}{dt}+\int_0^{2\pi} \Phi\frac{\pa \varrho}{\pa t}d\theta=0\;.
\end{equation}
Since, in general, the distribution function used in the definition of $\Phi$ is the same as that of the system\footnote{This argument fails for the core and the halo in double Lynden-Bell distributions.}, the second term is $\frac{d}{dt}\int (\Phi/2)\varrho d\theta$. Thus, the total energy $E$ is conserved:
\begin{equation}
\frac{dE}{dt}=0\;.
\end{equation}

Apart from the Vlasov equation, the relation
\begin{eqnarray}
\frac{d\varepsilon}{dt}&=&\frac{\pa \varepsilon}{\pa p}\frac{dp}{dt}+\frac{\pa \varepsilon}{\pa \Phi}\frac{d\Phi}{dt}\;,\ \ \varepsilon=\frac{p^2}{2}+\Phi(\theta,t)\\
&=&-p\frac{\pa \Phi}{\pa \theta}+\frac{d\Phi}{dt}\\
&=&-p\frac{\pa \Phi}{\pa \theta}+\frac{\pa \Phi}{\pa t}+\frac{\pa \Phi}{\pa \theta}\frac{d\theta}{dt}\\
&=&\frac{\pa \Phi}{\pa t}
\end{eqnarray}
holds.
Namely, in the collisionless regime, the one-particle energy $\varepsilon$ can vary its value via only the time-dependent potential.
As particles gain or lose energy, the range of the energy distribution spreads and strong spatial and temporal oscillations of the potential facilitate microscopic phase mixing.

Here, we explain {\it{phase mixing}}.
As an example, we consider an ensemble of $N$ identical harmonic oscillators with unit spring constant.
On the phase space, each harmonic oscillator draws a circular orbit, but for different orbits, there are different angular frequencies $\omega$.
Then, for two oscillators $i$ and $j$, the difference between the angular variables $\vartheta(t)$ on their circular orbits is
\begin{equation}
\vartheta_i(t)-\vartheta_j(t)=2\pi (\omega_i-\omega_j)t\;.
\end{equation}
Namely, this difference increases linearly with time.
For the whole $N$-body system, the winding number in the totality of circular orbits increases monotonically, and on the phase space a corresponding vortex emerges.
This process continues till $f$ relaxes to being a function of $\varepsilon$ only.
Of course, in a single circular orbit, there is no mixing.
This is the prototype of the phase-mixing phenomenon.\footnote{To be exact, we note that if we take the limit $N\to\infty$, the phase mixing would never end in the fine-grained sense; but in the coarse-grained sense we regard it as ending.} Due to this phase mixing, the potential of the long-range system may oscillate strongly.
This process is called {\it{violent relaxation}}\cite{LB}.

If the initial distribution is a spatially inhomogeneous water-bag one, the HMF system undergoes phase mixing and violent relaxation.
In Fig. 1, we show the phase-mixing process in such a case taken from the HMF system. (Using symbols that will be introduced later, this system is the case of $M_0=0.53$, $\hat{\eta}=0.15$ and $\hat{E}=0.4984$.)
At each time the contours of the inscribed and circumscribed Vlasov stationary water-bag states, that is, the dense energy water-bag distributions, for the distribution are indicated by red and green curves.

\begin{figure}[htbp]
\includegraphics[width=0.33\hsize,bb=0 0 370 390]{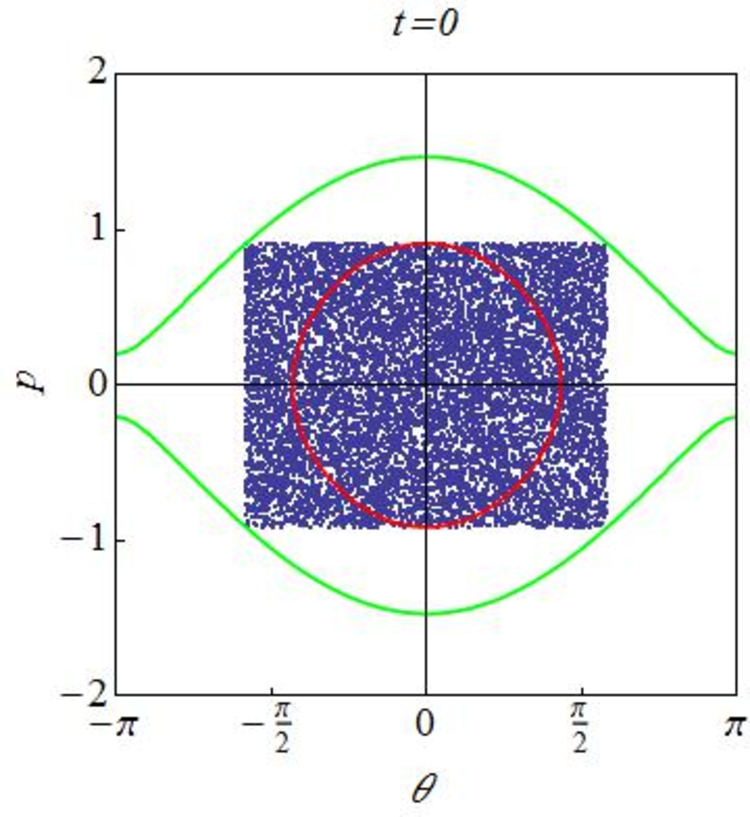}\includegraphics[width=0.33\hsize,bb=0 0 370 390]{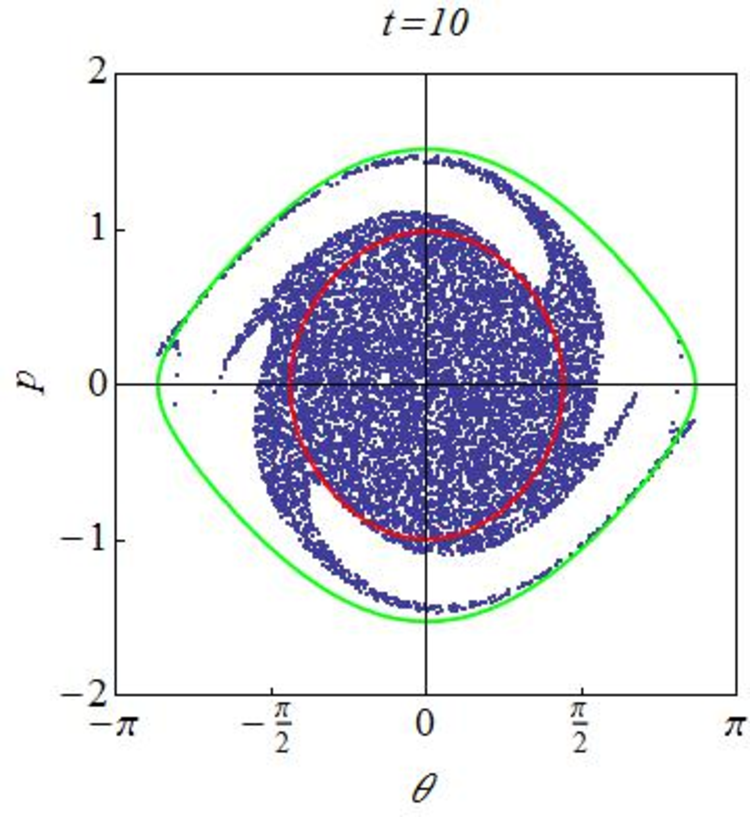}\includegraphics[width=0.33\hsize,bb=0 0 370 390]{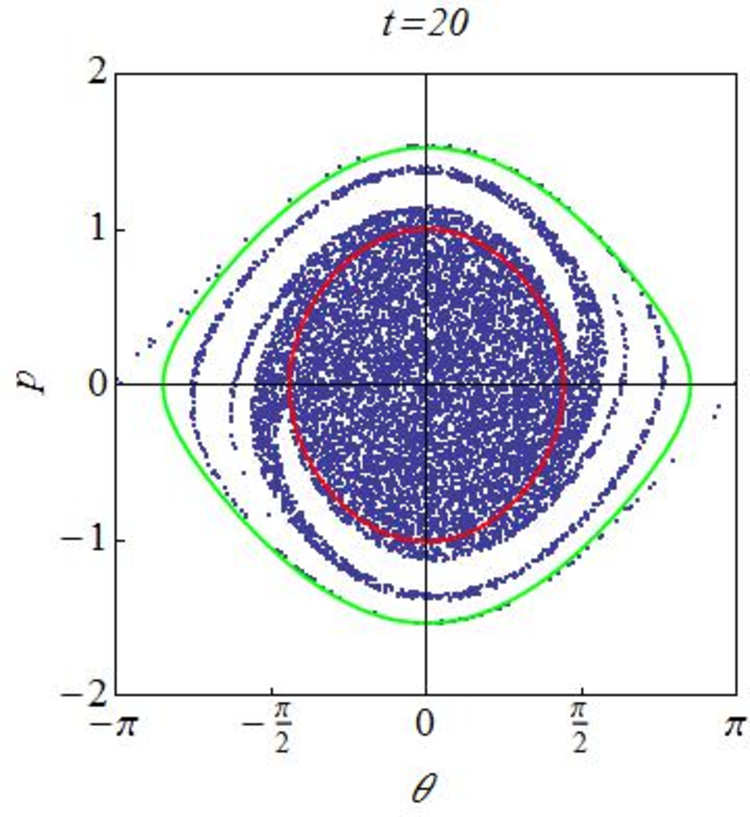}\\
\includegraphics[width=0.33\hsize,bb=0 0 370 390]{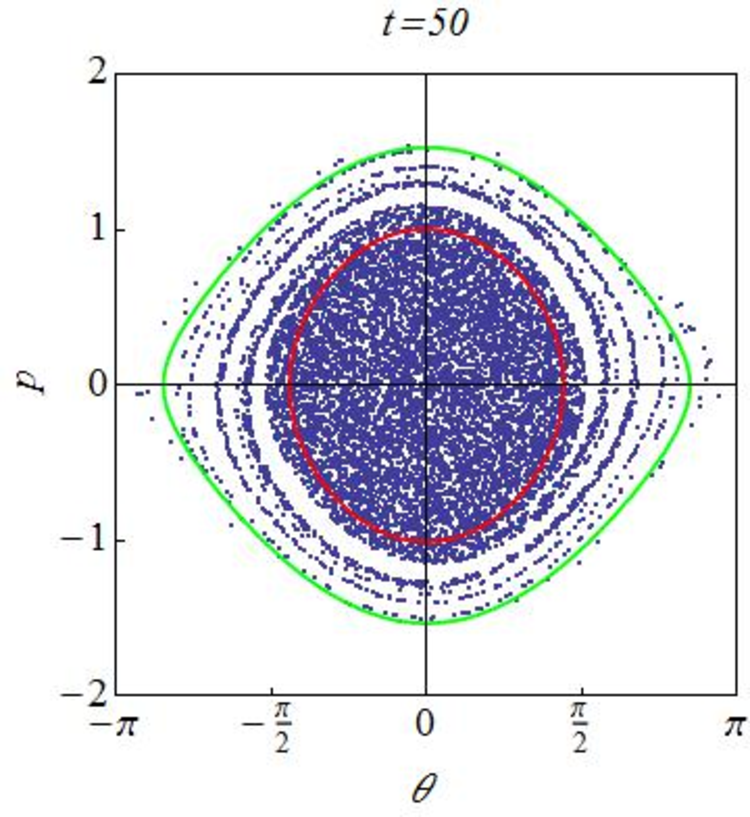}\includegraphics[width=0.33\hsize,bb=0 0 370 390]{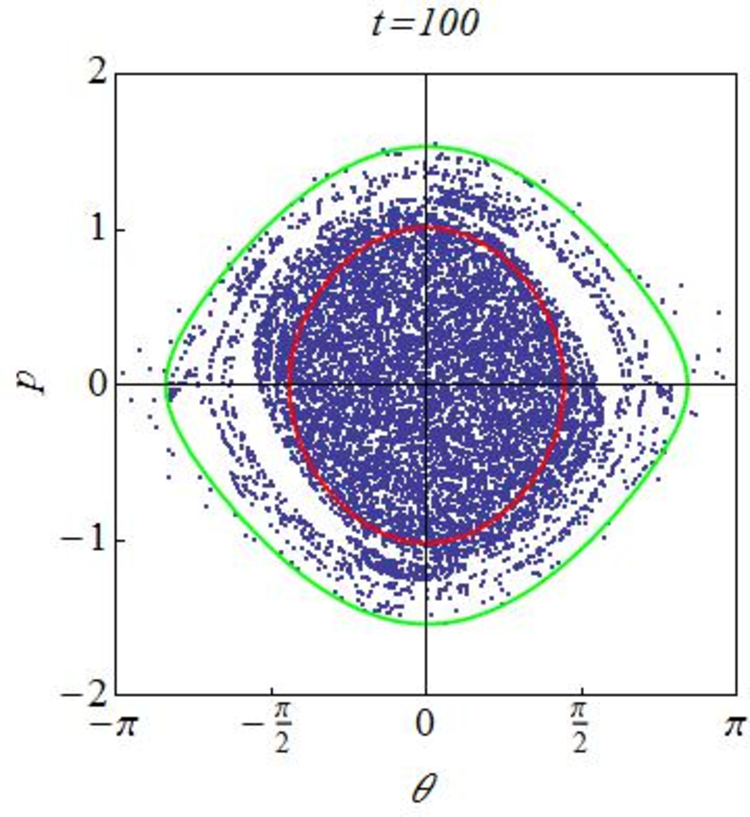}\includegraphics[width=0.33\hsize,bb=0 0 370 390]{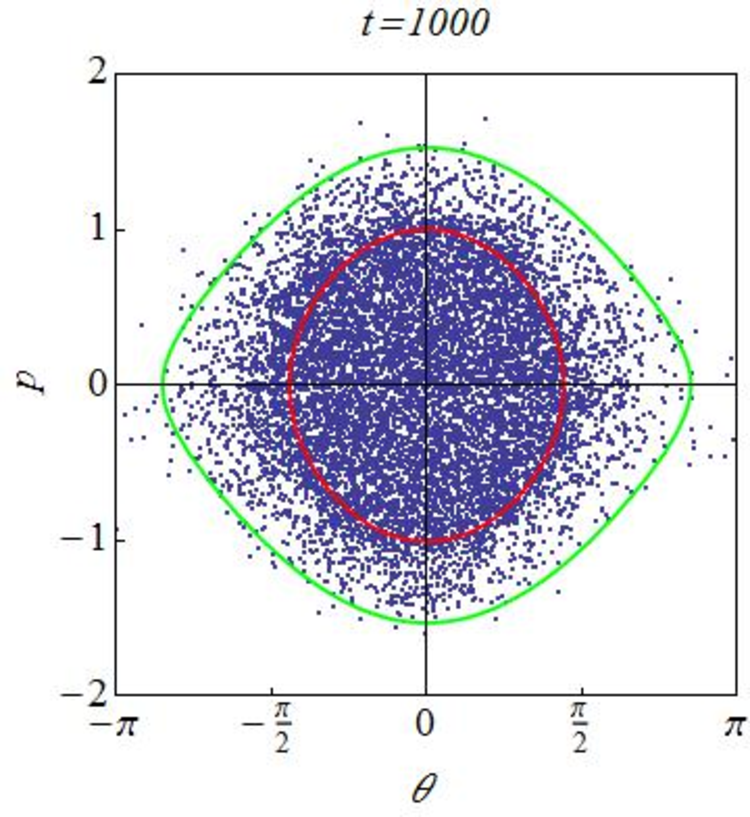}\\
\includegraphics[width=0.33\hsize,bb=0 0 370 390]{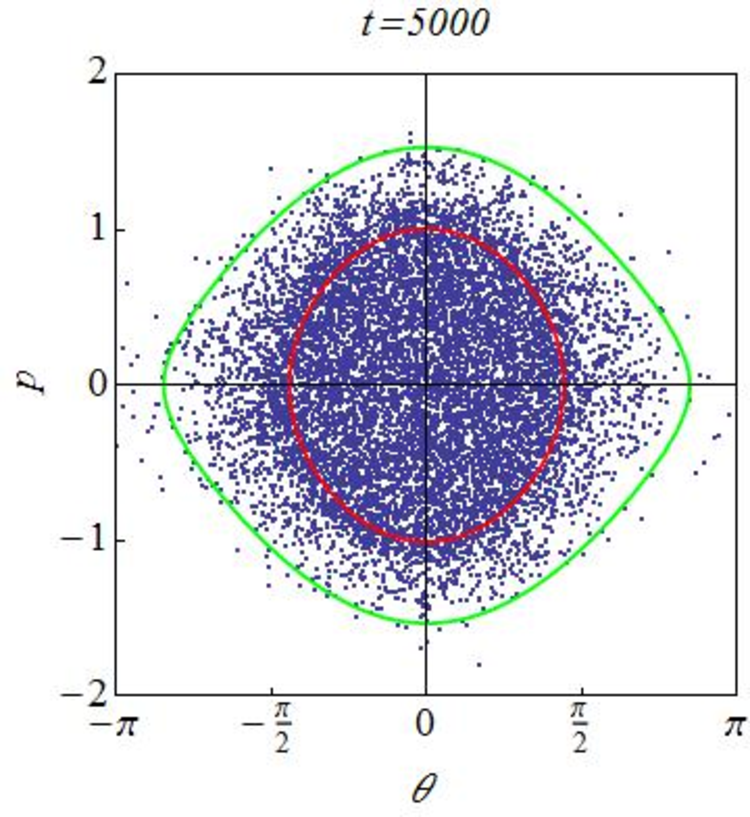}\includegraphics[width=0.33\hsize,bb=0 0 370 390]{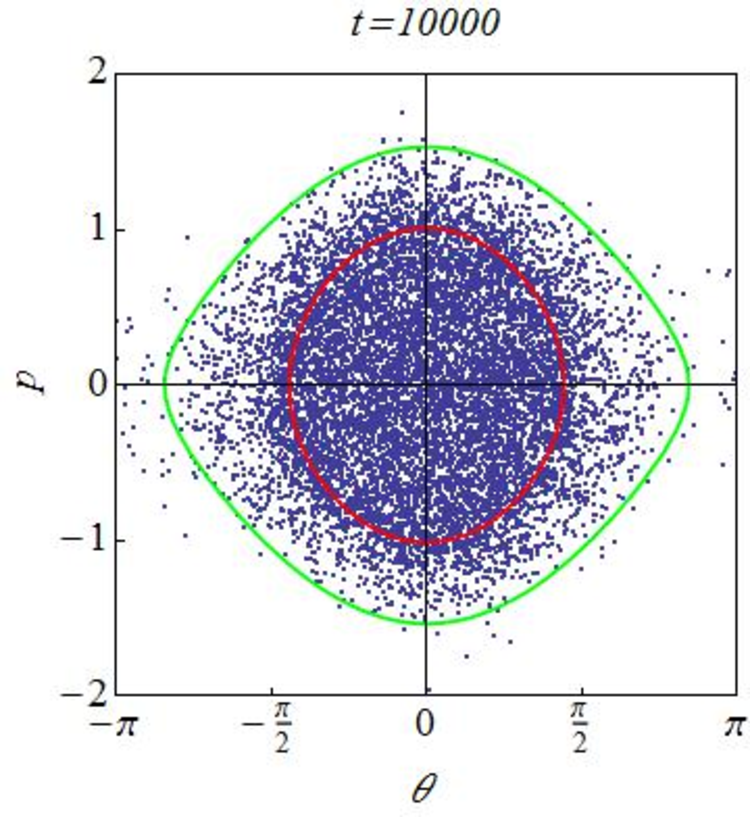}\includegraphics[width=0.33\hsize,bb=0 0 370 390]{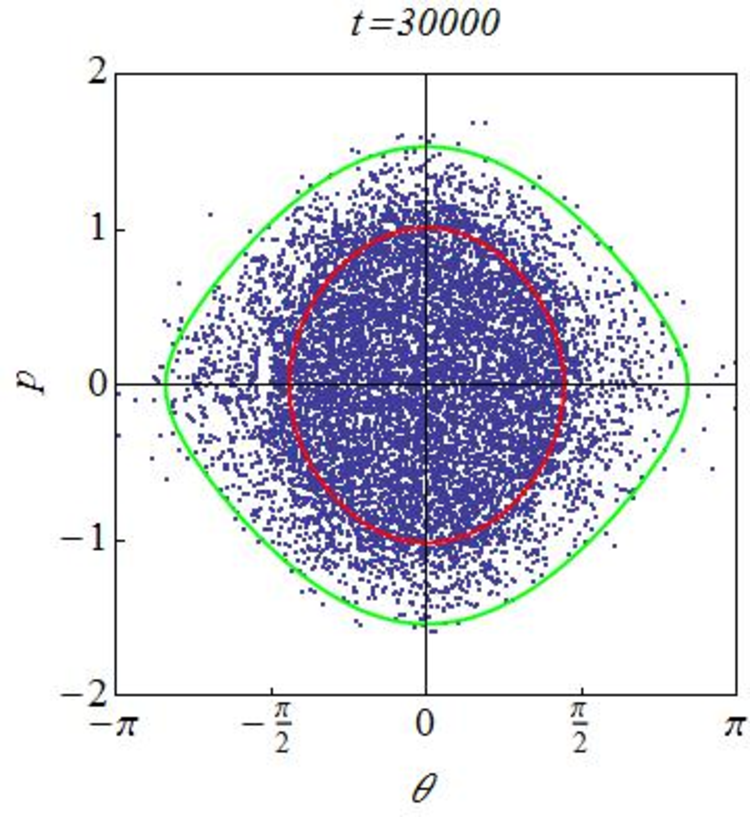}
\caption{An example of phase mixing in the HMF model.
The figures show the evolution of the system on the phase space.
The green and red one-particle energy contour curves change according to the change of the magnetization (i.e., the self-consistent mean-field potential).
}
\end{figure}

As shown in Fig. 1, the phase-space region sandwiched between the inscribed and circumscribed Vlasov stationary water-bag states forms the halo of the system.

\subsection{Lynden-Bell statistics}
As well as collisional equilibrium statistical mechanics, there is an established theory of an `equilibrium' statistical mechanics of collisionless QSSs based on the violent relaxation process.
This is the {\it{Lynden-Bell statistics}}.\cite{LB,LT}
By considering the statistics of QSSs, it predicts the most probable QSS by entropy maximization.
Due to the Vlasov equation $df/dt=0$, the fine-grained distribution function $f$ is an integral of motion and the Vlasov fluid elements move on the phase space as an incompressible fluid.
By phase mixing, $f$ relaxes to a function of the one-particle energy function $\varepsilon(\theta,p)$ that is the collisionless equilibrium state.
In the following, we assume that the system is a water-bag, that is, it has only two levels, $f$ and $0$.
So, on the phase space, we decompose the fine-grained distribution into unsuperposed Vlasov elements with the same level.
We assume that the totality of Vlasov elements satisfies the following three conditions.
\begin{enumerate}
\item The system conserves the total mass $N$.
\item The system conserves the total energy $E$.
\item The system conserves the phase-space density $\eta$ of the Vlasov elements due to the incompressibility. We denote the area of every Vlasov element by $\omega$.
\end{enumerate}

The Lynden-Bell statistics counts the configurations of Vlasov elements on the phase space.
To do this, we divide the phase space into identical micro-cells.
The area of micro-cells is assumed to be that of the Vlasov elements, that is, $\omega$.
When a Vlasov element occupies a micro-cell, the density of this micro-cell is $\eta$; otherwise it is $0$. This is the exclusion principle as a consequence of the Vlasov incompressibility.
The Lynden-Bell statistics treats the macroscopic structure in the same way as the usual statistics does.
We focus on the $P$ identical macro-cells on the phase space that are assemblies of $\nu$ micro-cells and average (coarse-grain) the fine-grained distribution on them.
In the $i(=1,2,\ldots,P)$ th macro-cell, we use $n_i(\le \nu)$ to denote the number of Vlasov elements which occupy this macro-cell.

\begin{figure}[htbp]
\begin{center}
\includegraphics[width=0.5\hsize,bb=0 0 260 200]{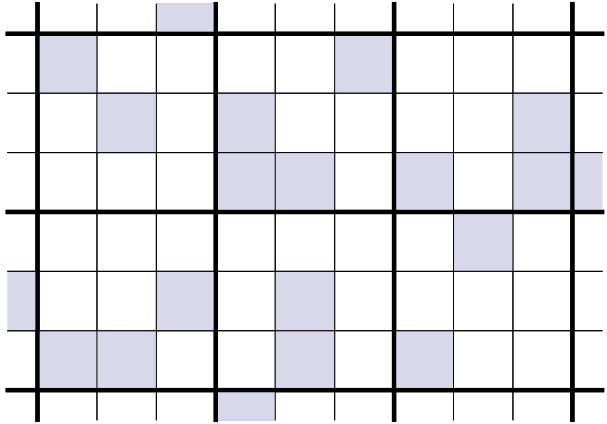}
\caption{An example of the phase space division ($\nu=9$). Thick and thin boxes represent macro-cells and micro-cells, respectively. The shaded micro-cells are occupied by Vlasov elements.}
\end{center}
\end{figure}

In the following, we assume ergodicity.

Now, we calculate the entropy in the Lynden-Bell statistics.
First of all, we calculate the number of states for configurations of Vlasov elements on the phase space.
To do this, we consider the partition number of counting the assignments of $N$ Vlasov elements to the $P$ macro-cells by $n_i(i=1,2,\ldots,P)$ number of Vlasov elements.
By assuming distinguishability between Vlasov elements, and by regarding each macro-cell as a single unit to be counted, the $n_i!$ arrangements of Vlasov elements inside the $i$-th macro-cell ($i=1,2,\ldots,P$) need to be ignored.
Thus, the total partition number of this counting is
\begin{equation}
\frac{N!}{\prod_{i=1}^P n_i!}\;.\label{eq:LB1}
\end{equation}

However, the configuration of the Vlasov elements inside each macro-cell is not determined.
So, we need to consider the partition number for each configuration of the $n_i$ Vlasov elements inside the $i(=1,2,\ldots,P)$ th macro-cell as that of the corresponding micro-cells inside each macro-cell.
Namely, we need to multiply Eq.(\ref{eq:LB1}) by the partition number for $n_i$ elements in $\nu$ sites for all $i$.
Here, regarding the configuration of $\nu$ micro-cells in a macro-cell for $i=1,2,\ldots,P$, the partition number for the vacant part of macro-cell, where none of the $n_i$ Vlasov elements is present, that is, $(\nu-n_i)$ micro-cells, needs to be ignored.
So, the relevant result is
\begin{equation}
\ _\nu P_{n_i}=\frac{\nu !}{(\nu-n_i)!}\;.
\end{equation}

Thus, the total partition number $W$ for assigning $\{n_i\}$ Vlasov elements to the macro-cells is
\begin{equation}
W=\frac{N!}{\prod_{i=1}^Pn_i!}\prod_{i=1}^P\frac{\nu!}{(\nu-n_i)!}\;.
\end{equation}
We recall that the system is macroscopic.
Using Stirling's formula, $\ln W$ is given approximately by
\begin{eqnarray}
\ln W\approx N(\ln N-1)-\sum_{i=1}^P[n_i(\ln n_i-1)+(\nu-n_i)(\ln(\nu-n_i)-1)-\nu(\ln\nu-1)]\;.
\end{eqnarray}

From now on, we use $f$ to denote the {\it{coarse-grained}} distribution function.
The distribution function $f$ that is coarse-grained by units of macro-cells with area $\nu\omega$ is defined by
\begin{equation}
f(\theta_i,p_i)=f_i=\frac{\eta n_i \omega}{\nu\omega}=\frac{\eta n_i}{\nu}\;.\label{eq:cg}
\end{equation}
By using this quantity, we express $W$ as
\begin{eqnarray}
\ln W&\approx&N(\ln N-1)-\sum_{i=1}^P\biggl(\frac{\nu}{\eta}\biggl(f_i\biggl(\ln \biggl(\frac{\nu f_i}{\eta}\biggr)-1\biggr)+(\eta -f_i)\biggl(\ln\biggl(\frac{\nu}{\eta}(\eta-f_i)\biggr)-1\biggr)\biggr)\nonumber \\&& -\nu(\ln\nu-1)\biggr)\;.
\end{eqnarray}

The variation of $-(\eta/\nu)\ln W$ with respect to $f$ is
\begin{eqnarray}
&&\sum_{i=1}^P\delta \biggl(f_i\biggl(\ln \biggl(\frac{\nu f_i}{\eta}\biggr)-1\biggr)+(\eta -f_i)\biggl(\ln\biggl(\frac{\nu}{\eta}(\eta-f_i)\biggr)-1\biggr)\biggr)\\
&&=\sum_{i=1}^P\biggl(\delta f_i\biggl(\ln\biggl(\frac{\nu f_i}{\eta}\biggr)-1\biggr)+\delta f_if_i\frac{1}{f_i}-\delta f_i\biggl(\ln \frac{\nu}{\eta}(\eta-f_i)-1\biggr)-(\eta-f_i)\delta f_i\frac{1}{\eta-f_i}\biggr)\\
&&=\sum_{i=1}^P\biggl(\delta f_i\ln\frac{\nu f_i}{\eta}-\delta f_i\ln\biggl(\frac{\nu}{\eta}(\eta-f_i)\biggr)\biggr)\\
&&=\sum_{i=1}^P\delta f_i\ln\frac{f_i}{\eta-f_i}\;.
\end{eqnarray}
Since we coarse-grain the distributions by the unit of macro-cell ($\nu\omega$), the continuum limit procedure ($\nu\to 0$) is
\begin{equation}
\sum_{i=1}^P(\cdots)_i\to\frac{1}{\nu\omega}\int_0^{2\pi}d\theta\int_{-\infty}^\infty dp (\cdots)(\theta,p)\;.
\end{equation} 
By takng this limit and using Lagrange multiplier methods, under the two constraints (i.e., conservation of total mass $N$ and total energy $E$)
\begin{eqnarray}
N&=&\int_0^{2\pi} d\theta\int_{-\infty}^{\infty} dp f(\theta,p)\;,\\
{E}&=&\int_0^{2\pi} d\theta \int_{-\infty}^{\infty} dp \biggl(\frac{p^2}{2}+\frac{1}{2}\Phi(\theta)\biggr)f(\theta,p)\;,
\end{eqnarray}
we obtain an equation for the maximization of $\ln W$:
\begin{equation}
\int_0^{2\pi} d\theta\int_{-\infty}^{\infty} dp \frac{1}{\omega}\frac{\delta f}{\eta}\biggl(\ln\frac{f}{\eta-f}+\alpha +\beta \varepsilon\biggr)=0\;.
\end{equation}
Since the choice of $\delta f$ is arbitrary, the bracketed part of the integrand needs to be zero.

Consequently, the most probable QSS distribution is
\begin{eqnarray}
f&=&\eta\frac{\exp(-\alpha-\beta \varepsilon)}{1+\exp(-\alpha -\beta \varepsilon)}\\
&=&\frac{\eta}{\exp(\beta(\varepsilon-\mu))+1}\;,\ \ \mu=-\frac{\alpha}{\beta}\;.\label{eq:Lynden-Bell}
\end{eqnarray}
This is called the {\it{Lynden-Bell distribution}}.

The resemblance of the Lynden-Bell distribution to the Fermi-Dirac distribution is due to the exclusion principle for the Vlasov elements on each micro-cell, arising from their incompressibility.
The reduced entropy in the continuum limit is
\begin{equation}
S=-\int_0^{2\pi} d\theta \int_{-\infty}^\infty dp\biggl(\frac{f}{\eta}\ln \frac{f}{\eta}+\biggl(1-\frac{f}{\eta}\biggr)\ln\biggl(1-\frac{f}{\eta}\biggr)\biggr)\;.
\end{equation}
In the following, we refer this entropy as the {\it{Lynden-Bell entropy}}.

The Lynden-Bell distribution has the four parameters: the phase-space density $\eta$, multipliers $\beta$ and $\mu$ and the stationary value of the magnetization (in the HMF case) in the one-particle energy.\cite{AFBCDR,ACFR}
In this review, we refer the {\it{Lynden-Bell equilibrium}} by the solution in the form Eq.(\ref{eq:Lynden-Bell}) of the three conservation laws for mass, energy and fine-grained phase-space density and the self-consistency condition for the magnetization.

The main statement in this review is that the real QSS that undergoes violent relaxation is, in general, {\bf{not}} the Lynden-Bell equilibrium but a superposition of two independent Lynden-Bell distributions.
We corroborate this scenario in Section 4.
\section{The Hamiltonian Mean-Field Model}
\subsection{Basic properties}
\subsubsection{Definition}
The Hamiltonian mean-field (HMF) model is a widely studied classical mechanical toy model of long-range systems.\cite{Text,ReviewI,HMF,HMF1,HMF2}
The $N$-body HMF model considers $N$ identical fully coupled interacting particles with unit mass on a circle.
Their dynamics is governed by the Hamiltonian
\begin{equation}
H=\sum_{i=1}^N\frac{p_i^2}{2}+\frac{\epsilon}{2N}\sum_{i,j=1}^N(1-\cos(\theta_i-\theta_j))\;,\label{eq:Ham}
\end{equation}
where the angle $\theta_i$ is the orientation of the $i$ th particle and $p_i$ is its canonical conjugate momentum.\cite{HMF}

From this Hamiltonian, we obtain the coupled canonical equations
\begin{equation}
\dot{\theta}_i=p_i\;,\ \ \dot{p}_i=-\frac{\epsilon}{2N}\Biggl(\sum_{j=1}^N\sin(\theta_i-\theta_j)-\sum_{j=1}^N\sin(\theta_j-\theta_i)\Biggr)\;.
\end{equation}
These can be unified as
\begin{equation}
\ddot{\theta}_i=-\frac{\epsilon}{N}\sum_{j=1}^N\sin(\theta_i-\theta_j)\;.\label{eq:EOM}
\end{equation}

To clarify the physical meaning of the HMF model, we compare it with the familiar Heisenberg $XY$ model\cite{Text,Heisenberg} that considers two-dimensional spins distributed over a square lattice, which has the Hamiltonian
\begin{equation}
H^\prime=-\epsilon\sum_{\la i,j\ra}\vec{s}^i\cdot \vec{s}^j\;,\ \ \vec{s}^i=(s^i_x,s^i_y)\;,
\end{equation}
where the variables $s^i_x$ and $s^i_y$ are $\cos \theta_i$ and $\sin\theta_i$, respectively and $\la i,j\ra$ denotes a pair of adjoining sites.
In the Heisenberg $XY$ model, for $\epsilon>0$, when all spins have the same direction, the energy of the system is at its minimum. Thus the ground state of the system is ferromagnetic.
For $\epsilon<0$, the ground state is anti-ferromagnetic: the directions of adjacent spins are opposite.

In the HMF model, by setting
\begin{eqnarray}
{\vec{m}}^i&=&(m_x^i,m_y^i)\\
&=&(\cos\theta_i,\sin\theta_i)\;,
\end{eqnarray}
the interaction Hamiltonian with the exception of the constant term is
\begin{equation}
H_{{\rm{int}}}=-\frac{\epsilon}{2N}\sum_{i,j=1}^N\vec{m}^i\cdot \vec{m}^j\;.
\end{equation}

The formal resemblance between these two models is clear.
However, a significant difference between them is that, in the HMF model, the particle interactions are not only between particles with adjacent indices, as in the Heisenberg $XY$ model, but between any pair of particles.

Summarizing the above arguments, we characterize the interaction of the HMF particles in the following two ways.
\begin{enumerate}
\item It is a long-range interaction depending on only distance on the circle and all particles are fully coupled.
For consistency, it is periodic for each variable $\theta_i$ on the circle.
\item It resembles the Heisenberg $XY$ spin exchange interaction.
Namely, it is proportional to the inner product between two {\it{spins}} ${\vec{m}}^i$ and ${\vec{m}}^j$.
\end{enumerate}

In the following, we set $\epsilon=1$.

We now return to the equation of motion, Eq.(\ref{eq:EOM}).
By introducing
\begin{equation}
M_x=\frac{1}{N}\sum_{j=1}^N\cos\theta_j\;,\ \ M_y=\frac{1}{N}\sum_{j=1}^N\sin\theta_j\;,
\end{equation}
and
\begin{equation}
M=\sqrt{M_x^2+M_y^2}\;,\ \ \tan\phi=\frac{M_y}{M_x}=\frac{\sum_{j=1}^N\sin\theta_j}{\sum_{j=1}^N\cos\theta_j}\;,
\end{equation}
Eq.(\ref{eq:EOM}) becomes
\begin{equation}
\ddot{\theta}_i=- M\sin(\theta_i-\phi)\;,
\end{equation}
where the modulus $M$ represents the magnetization by analogy with the Heisenberg $XY$ model because in both models there is no external magnetic field.

Here, we used
\begin{eqnarray}
M\sin(\theta_i-\phi)&=&\sqrt{M_x^2+M_y^2}\sin(\theta_i-\phi)\\
&=&\frac{M_x\sin(\theta_i-\phi)}{\cos \phi}\\
&=&M_x\sin\theta_i-M_x\tan\phi\cos\theta_i\\
&=&\frac{1}{N}\Biggl(\sum_{j=1}^N\cos \theta_j\sin \theta_i-\sum_{j=1}^N\sin\theta_j\cos \theta_i\Biggr)\\
&=&\frac{1}{N}\sum_{j=1}^N\sin(\theta_i-\theta_j)\;.
\end{eqnarray}

The HMF model significantly has both equilibrium\cite{HMF} and non-equilibrium\cite{ACFR,NET1,PL,CDFR,NET2,NET3} phase transition properties with the Boltzmann-Gibbs magnetization $M_\ast$ and the QSS magnetization $M_s$ of the spins ${\vec{m}}$, respectively, as the order parameters.
A remarkable difference is that the latter property depends not only on $\hat{E}$ but also on the initial magnetization $M_0$,\cite{ACFR,NET1,NET2} while the former property depends on only $\hat{E}$.
In this review, we will discuss the former property in Section 3.2 but do not discuss the latter property.

\subsubsection{Mean-field methods}
For $N\gg 1$, we can ignore the granularity in the distribution, and the system is described by the time-dependent one-particle distribution function $f(\theta,p,t)$.\footnote{The description manner in the following part of Section 3 follows that of Taruya's unpublished article.\cite{Taruya}}
This description uses mean-field methods.
For convenience we repeat the definitions of the following quantities (in the following, we omit the parameter $t$):
\begin{eqnarray}
{\rm{Number\ of\ particles}}:N&=&\int_0^{2\pi}d\theta \int_{-\infty}^\infty dpf(\theta,p)\;,\label{eq:2T}\\
{\rm{Total\ energy}}:E&=&\int_0^{2\pi}d\theta \int_{-\infty}^\infty dp\biggl(\frac{p^2}{2}+\frac{1}{2}\Phi(\theta)\biggr)f(\theta,p)\;,\label{eq:3}\\
{\rm{Magnetization}}:\vec{M}&=&\frac{1}{N}\int_0^{2\pi}d\theta \int_{-\infty}^\infty dp(\cos \theta,\sin \theta)f(\theta,p)\;.\label{eq:4}
\end{eqnarray}

Here, $\Phi(\theta)$ represents the potential energy function
\begin{equation}
\Phi(\theta)=\frac{1}{N}\int_0^{2\pi}d\theta^\prime \int_{-\infty}^\infty dp^\prime \{1-\cos (\theta-\theta^\prime)\}f(\theta^\prime,p^\prime)\;,\label{eq:5T}
\end{equation}
and the force acting on the $i$ th particle is
\begin{equation}
F_i=-\frac{\pa \Phi}{\pa \theta_i}\;.
\end{equation}

The potential energy function $\Phi(\theta)$ satisfies 
\begin{eqnarray}
\frac{d^2\Phi(\theta)}{d\theta^2}&=&\frac{1}{N}\int_0^{2\pi} d\theta^\prime\int_{-\infty}^\infty dp^\prime \cos (\theta-\theta^\prime)f(\theta^\prime, p^\prime)\\
&=&\frac{d^2(1-\Phi(\theta))}{d \theta^2}\\
&=&1-\Phi(\theta)\;.\label{eq:6}
\end{eqnarray}
The general solution of Eq.(\ref{eq:6}) is
\begin{equation}
\Phi(\theta)=1+M\cos (\theta +\gamma)\;,\label{eq:Phi}
\end{equation}
where $M$ and $\gamma$ are arbitrary constants. Owing to the translational invariance of Eq.(\ref{eq:Ham}) with respect to $\theta$, we can fix the phase constant to $\gamma=\pi$.
$M$ characterizes the clustering of the distribution (see Fig. 3).
The distribution for $M=0$ is uniform.
For $M\neq 0$, there is clustering towards a particular direction, which breaks the translational symmetry.

\begin{figure}[htbp]
\begin{center}
\includegraphics[width=0.4\hsize,bb=0 0 260 265]{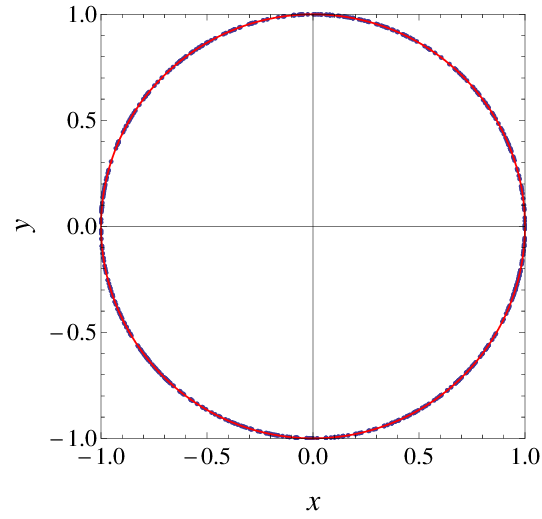}\ \ \ \ \includegraphics[width=0.4\hsize,bb=0 0 260 265]{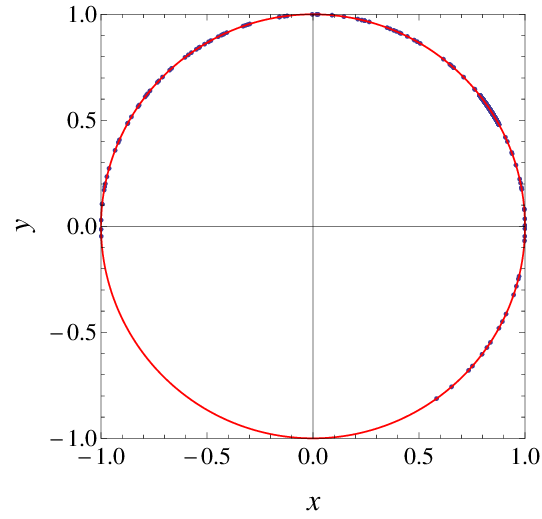}
\caption{(left) Non-clustering ($M=0.05$) and (right) clustering states ($M=0.85$) of the HMF system. ($N=500$)}
\end{center}
\end{figure}

\subsection{Boltzmann-Gibbs equilibrium}
In the micro-canonical approach, the Boltzmann-Gibbs entropy is written by using the one-particle distribution.
 In the mean-field method, as in Eq.(\ref{eq:Phi}), we replace the many-body effects by a self-consistent mean-field and describe it using two phase-space variables, $\theta$ and $p$. The Boltzmann-Gibbs entropy is
\begin{eqnarray}
S_{BG}&=&-N\int_0^{2\pi}d\theta \int_{-\infty}^\infty dp\biggl\{\frac{f(\theta,p)}{N}\biggr\}\ln\biggl\{\frac{f(\theta,p)}{N}\biggr\}\\
&=&-N\int_0^{2\pi}d\theta \int_{-\infty}^\infty dph (\ln h)\ \ (f(\theta,p)=Nh(\theta,p))\;.\label{eq:11}
\end{eqnarray}
By using the Lagrange multiplier method, the extremum equation for Eq.(\ref{eq:11}) for fixed $N$ and $E$ is
\begin{eqnarray}
\delta \biggl(S_{BG}-\alpha \biggl(\int_0^{2\pi} d\theta \int_{-\infty}^\infty dph(\theta,p)-1\biggr)-\beta \biggl(\int_0^{2\pi} d\theta \int_{-\infty}^\infty dp\biggl(\frac{p^2}{2}+\frac{1}{2}\Phi\biggr)f-E\biggr)\biggr)=0\;.\label{eq:12}
\end{eqnarray}
That is,
\begin{equation}
\int_0^{2\pi} d\theta \int_{-\infty}^\infty dp\biggl(-N(\ln h+1)-\alpha -\beta N\biggl(\frac{p^2}{2}+\Phi\biggr)\biggr)\delta h=0\;.
\end{equation}
Here, since $\Phi$ contains $f$ in its integral, the variation of $\int d\theta \int dp\frac{1}{2}\Phi f$ with respect to $f$ is $\int d\theta \int dp\Phi \delta f$.
This equation is independent of the choice of $\delta h$.
Thus, we have
\begin{equation}
-(\ln h +1)-\frac{\alpha}{N}-\beta \biggl(\frac{p^2}{2}+\Phi\biggr)=0\;,
\end{equation}
which can be rewritten as
\begin{equation}
\ln h=-1-\frac{\alpha}{N}-\beta\biggl(\frac{p^2}{2}+\Phi\biggr)\;.
\end{equation}
From this equation, the extremum solution of Eq.(\ref{eq:12}) is
\begin{eqnarray}
f&=&Nh\\
&=&Ne^{-1-\frac{\alpha}{N}}e^{-\beta (\frac{p^2}{2}+\Phi)}\\
&=&Ae^{-\beta (\frac{p^2}{2}+\Phi)}\;.\label{eq:13}
\end{eqnarray}

The numerical constant $A$ in Eq.(\ref{eq:13}) is related to the constant $M$ in $\Phi$.
By using Eq.(\ref{eq:Phi}), we substitute Eq.(\ref{eq:13}) into Eq.(\ref{eq:2T}).
Then, by introducing the $n$-th first-kind deformed Bessel function
\begin{equation}
I_n(x)=\frac{1}{2\pi}\int_0^{2\pi} e^{x\cos \theta}\cos(n\theta)d\theta\;,
\end{equation} the equation
\begin{eqnarray}
N&=&\int_0^{2\pi} d\theta \int_{-\infty}^\infty dp f(\theta,p)\\
&=&A\int_0^{2\pi} d\theta e^{-\beta \Phi}\int_{-\infty}^\infty dp e^{-\beta\frac{p^2}{2}}\\
&=&\sqrt{\frac{2\pi}{\beta}}A\int_0^{2\pi} d\theta e^{-\beta (1-M\cos \theta)}\label{eq:14}
\end{eqnarray}
can be rewritten as
\begin{equation}
\frac{A}{N}=\sqrt{\frac{\beta}{(2\pi)^3}}\frac{1}{e^{-\beta}I_0(\beta M)}\;.
\end{equation}
Moreover, by substituting Eq.(\ref{eq:13}) into Eq.(\ref{eq:5T}), since
\begin{eqnarray}
\Phi&=&1-M\cos\theta\\
&=&\frac{1}{N}\int_0^{2\pi} d\theta^\prime \int_{-\infty}^\infty dp^\prime \{1-\cos(\theta-\theta^\prime)\}f(\theta^\prime,p^\prime)\\
&=&1-\frac{1}{N}\int_0^{2\pi} d\theta^\prime \int_{-\infty}^\infty dp^\prime (\cos\theta \cos\theta^\prime +\sin\theta \sin \theta^\prime)f(\theta^\prime, p^\prime)\;,
\end{eqnarray}
we obtain
\begin{eqnarray}
M&=&\sqrt{\frac{2\pi}{\beta}}\frac{A}{N}\int_0^{2\pi} d\theta^\prime \cos \theta^\prime e^{-\beta (1-M\cos \theta^\prime)}\ \ ({\rm{coeff\ of}}\ \cos\theta)\;,\label{eq:15}\\
0&=&\sqrt{\frac{2\pi}{\beta}}\frac{A}{N}\int_0^{2\pi} d\theta^\prime \sin \theta^\prime e^{-\beta (1-M\cos\theta^\prime)}\ \ ({\rm{coeff\ of}}\ \sin\theta)\;.\label{eq:16}
\end{eqnarray}
From the first equation (\ref{eq:15}), we obtain the non-trivial relation
\begin{eqnarray}
M&=&\frac{\int_0^{2\pi}d\theta \cos\theta e^{-\beta (1-M\cos\theta)}}{2\pi e^{-\beta }I_0(\beta M)}\\
&=&\frac{I_1(\beta M)}{I_0(\beta M)}\;.\label{eq:16}
\end{eqnarray}
At low temperature, a second-order phase transition occurs.\cite{HMF}
Its critical temperature is the solution of the next equation:
\begin{eqnarray}
\biggl(\frac{d}{d \beta_c M}\frac{I_1(\beta_c M)}{I_0(\beta_c M)}\biggr)\biggl|_{M=0}&=&\frac{1}{2}\label{eq:half}\\
&=&\frac{d M}{d (\beta_c M)}\\
&=&\frac{1}{\beta_c}\\
&=&T_c\;.
\end{eqnarray}
Here, Eq.(\ref{eq:half}) can be verified by expanding $I_0(z)$ and $I_1(z)$ in Taylor series of $z$, and extracting the first two terms.
Near to the critical temperature, the behavior of $M$ with respect to $T-T_c$ can be found by arranging ${d}(I_1/I_0)/{d(\beta M)}=\frac{1}{\beta}$.
From
\begin{equation}
I_\nu(z)=\biggl(\frac{z}{2}\biggr)^\nu\sum_{n=0}^\infty \frac{(z/2)^{2n}}{n!\Gamma(\nu+n+1)}\;,\end{equation}
we obtain
\begin{eqnarray}
I_0(\beta M)&=&1+\biggl(\frac{\beta M}{2}\biggr)^2+{\cal{O}}(M^4)\;,\\
I_1(\beta M)&=&\frac{\beta M}{2}+\biggl(\frac{\beta M}{2}\biggr)^3\frac{1}{2}+{\cal{O}}(M^5)\;.\end{eqnarray}
Thus,
\begin{eqnarray}
\biggl(\frac{d}{d\beta M}\frac{I_1(\beta M)}{I_0(\beta M)}\biggr)\biggl|_{M=0}&=&\Biggl(\frac{d}{d\beta M}\biggl(\frac{\frac{\beta M}{2}+\frac{(\beta M)^3}{16}}{1+(\frac{\beta M}{2})^2}\biggr)\Biggr)\Biggl|_{M=0}\\
&=&\Biggl(\frac{(\frac{1}{2}+\frac{3(\beta M)^2}{16})(1+\frac{(\beta M)^2}{4})-(\frac{\beta M}{2}+\frac{(\beta M)^3}{16})\frac{\beta M}{2}}{(1+(\frac{\beta M}{2})^2)^2}\Biggr)\Biggl|_{M=0}\\
&=&\Biggl(\frac{\frac{1}{2}+\frac{3(\beta M)^4}{64}+\frac{5(\beta M)^2}{16}-\frac{(\beta M)^2}{4}-\frac{2(\beta M)^4}{64}}{(1+(\frac{\beta M}{2})^2)^2}\Biggr)\Biggl|_{M=0}\\
&=&\Biggl(\frac{\frac{1}{2}+\frac{1}{16}(\beta M)^2+\frac{1}{64}(\beta M)^4}{(1+\frac{(\beta M)^2}{4})^2}\Biggr)\Biggl|_{M=0}\\
&=&\frac{1}{2}
\end{eqnarray}
holds. So, it follows that
\begin{equation}
\frac{1}{\beta_c}=\frac{1}{2}\Leftrightarrow T_c=\frac{1}{2}\;.
\end{equation}
As 
\begin{equation}
\frac{\frac{M}{2T}+\frac{M^3}{16T^3}}{1+\frac{M^2}{4T^2}}=\frac{M}{2T}-\frac{M^3}{16T^3}+{\cal{O}}(M^4)\;,
\end{equation}
the behavior of the magnetization around the critical temperature is given by the solution of
\begin{equation}
\frac{M}{2T}-\frac{M^3}{16T^3}=M\;,
\end{equation}
which is
\begin{equation}
M=4T\sqrt{T_c-T}\;.\label{eq:BBol}
\end{equation}

The self-consistency condition Eq.(\ref{eq:16}) is the most important equation determining the equilibrium configuration. 
Eq.(\ref{eq:16}) has solutions $M=0$ and $M\neq0$.
In the following, we denote the non-zero solution by $M=M_\ast(\beta)$, which we call the {\it{magnetization}} by analogy with the Heisenberg $XY$ model.

We derive the relation between total energy and the magnetization.
By substituting Eq.(\ref{eq:13}) into the definition Eq.(\ref{eq:3}), the total energy is
\begin{equation}
E=\frac{1}{2}\int_0^{2\pi} d\theta \int_{-\infty}^\infty dp \{p^2+(1-M_\ast\cos \theta)\}Ae^{-\beta(\frac{p^2}{2}+1-M_\ast \cos \theta )}\;.\label{eq:Ecal}
\end{equation} 
We calculate each term in Eq.(\ref{eq:Ecal}).
The first term is
\begin{eqnarray}
\frac{1}{2} \int_0^{2\pi} d\theta \int_{-\infty}^\infty dp p^2 Ae^{-\beta (\frac{p^2}{2}+1-M_\ast \cos\theta)}&=&\frac{A}{2}\sqrt{\frac{2\pi}{\beta}}\beta^{-1}\int_0^{2\pi}d\theta  e^{-\beta (1-M_\ast\cos\theta)}\\
&&\biggl(-\frac{d }{d\beta}\sqrt{\frac{2\pi}{\beta}}=\frac{1}{2\beta}\sqrt{\frac{2\pi}{\beta}}\biggr)\\
&=&\frac{N}{2\beta}\;.
\end{eqnarray}
The second term is
\begin{eqnarray}
\frac{1}{2} \int_0^{2\pi} d\theta \int_{-\infty}^\infty dp Ae^{-\beta (\frac{p^2}{2}+1-M_\ast \cos\theta)}&=&\frac{A}{2}\sqrt{\frac{2\pi}{\beta}}\int_0^{2\pi}d\theta e^{-\beta (1-M_\ast\cos\theta)}\\
&=&\frac{N}{2}\;.
\end{eqnarray}
The third term is 
\begin{eqnarray}
-\frac{1}{2} \int_0^{2\pi} d\theta \int_{-\infty}^\infty dp M_\ast \cos \theta Ae^{-\beta (\frac{p^2}{2}+1-M_\ast \cos\theta)}&=&-\frac{A}{2}\sqrt{\frac{2\pi}{\beta}}M_\ast\int_0^{2\pi}d\theta \cos \theta  e^{-\beta (1-M_\ast\cos\theta)}\\
&=&-\frac{NM_\ast^2}{2}\\ &&\biggl(M_\ast=\frac{I_1(\beta M_\ast)}{I_0(\beta M_\ast)}\biggr)\;.
\end{eqnarray}
As a result, we obtain
\begin{equation}
E=\frac{N}{2\beta}\{1+\beta (1-M_\ast^2)\}\;.\label{eq:17}
\end{equation}
In the calculation, we used Eq.(\ref{eq:15}).
As an important point, in the simulation, we fix $E$ and $N$.
Here, the non-zero $M_\ast$ is determined by giving the temperature.

\begin{figure}[htbp]
\begin{center}
\includegraphics[width=0.4\hsize,bb=0 0 260 279]{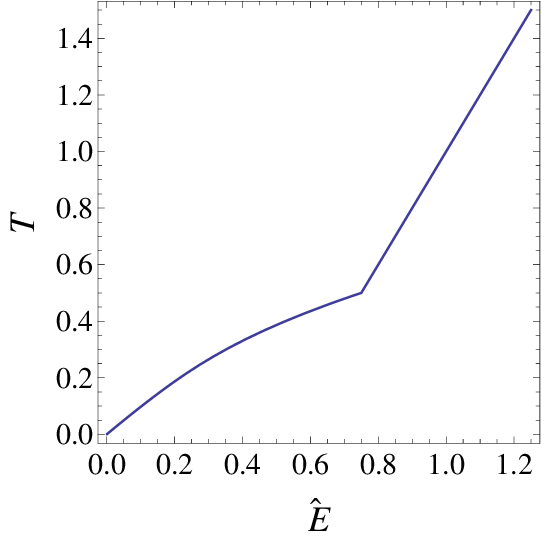}\ \ \ \ \includegraphics[width=0.4\hsize,bb=0 0 260 279]{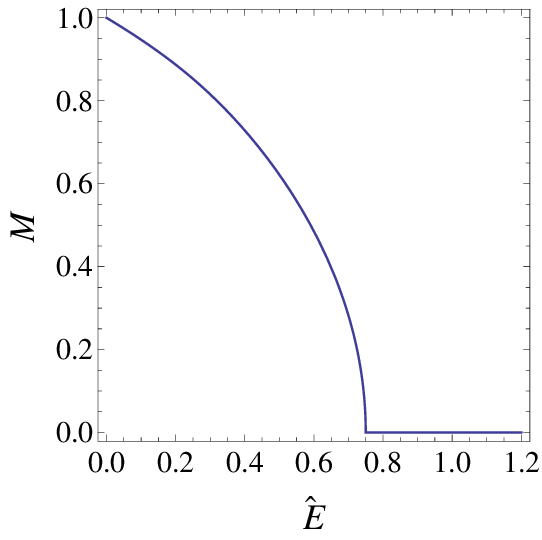}
\caption{The caloric curve and the magnetization curve of the HMF model in the Boltzmann-Gibbs statistics.}
\end{center}
\end{figure}

Due to the relation in Eq.(\ref{eq:15}), the magnetization given by Eq.(\ref{eq:4}) is 
\begin{eqnarray}
\vec{M}&=&\frac{A}{N}\sqrt{\frac{2\pi}{\beta}}\int_0^{2\pi}d\theta (\cos\theta,\sin \theta)e^{-\beta (1-M_\ast \cos \theta)}\\&=&(M_\ast,0)\;.\label{eq:18}
\end{eqnarray}
Here, the second component vanishes.
This can be seen by changing the angular integral to $\int_{-\pi}^\pi d\theta(\cdots)$, and the fact that the integrand is an odd function with respect to $\theta$.

\section{QSSs with Core-Halo Structure}
In this section, we review the studies of QSSs that have the core-halo structure on the phase space at low energies per particle.
The core-halo structure appears ubiquitously in long-range systems.\cite{ReviewII}
For a long while, in the general context, the distributions of the core and the halo had been considered as an {\it{attachment}}.
Namely, in the central core region in the phase space, there is no halo particle, and vice-versa.
However, according to this traditional standpoint, the resultant distribution from the simulation has a distorted form (see Fig. 12) that cannot be explained.
In this section, we expose the author's proposal\cite{KS} of a view of the core and halo distributions as a {\it{superposition}}, and corroborate its collisionless equilibrium state, that is, the double Lynden-Bell state, with illustrative results from $N$-body simulations.
\subsection{Review of research preceding the double Lynden-Bell scenario}
\subsubsection{Pakter and Levin's ansatz}
The study of the core-halo structure of the QSSs in the HMF model was begun by Pakter and Levin.\cite{PL}
They explained the origin of the core-halo structure in the following way.
In the violent relaxation process, when the magnetization of the system is macroscopically damped, the particles that are parametrically resonant with this oscillation gain energy and form the high-energy halo.
As a result, due to energy conservation, the remaining particle move to the low-energy region and due to the Vlasov incompressibility (i.e., the conservation of the fine-grained phase-space density $\eta$), as in the Fermi-degeneration phenomenon, they form the low-energy dense core.

\begin{figure}[htbp]
\includegraphics[width=0.33\hsize,bb=0 0 260 179]{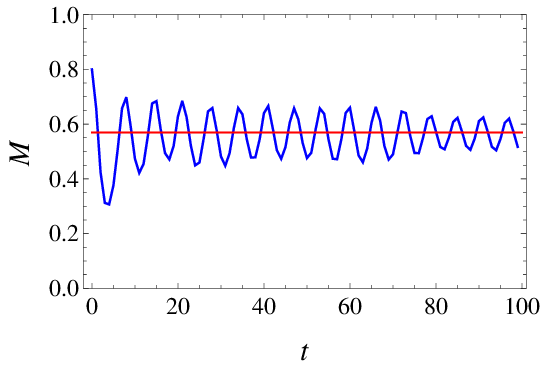}\includegraphics[width=0.33\hsize,bb=0 0 260 179]{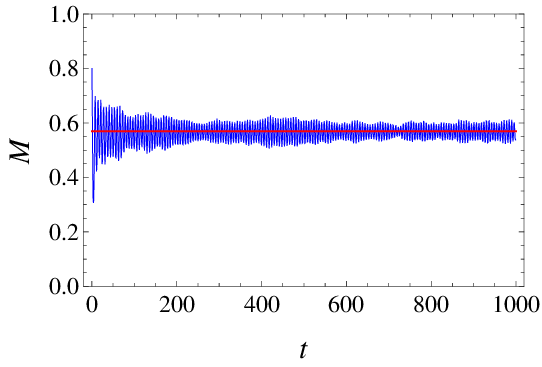}
\includegraphics[width=0.33\hsize,bb=0 0 260 179]{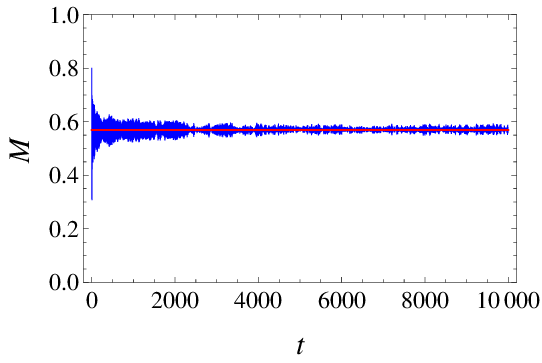}
\caption{Oscillation of the magnetization for $M(t=0)=0.8$, $\hat{E}=0.5419$, $\hat{\eta}=0.15$ (blue curves).
The red line represents the magnetization at the Boltzmann-Gibbs equilibrium.}
\end{figure}

To represent the coarse-grained QSS distribution function, they made an ansatz for the core-halo distribution
\begin{eqnarray}
\hat{f}_s(\theta,p)&=&\hat{\eta}(\Theta(\varepsilon_F-\varepsilon)+\chi\Theta(\varepsilon_h-\varepsilon)\Theta(\varepsilon-\varepsilon_F))\;,\label{eq:fsPL}\\\varepsilon(\theta,p,M_s)&=&\frac{p^2}{2}+1-M_s\cos\theta\;.
\end{eqnarray}
According to the traditional idea, the core-halo distribution was written in attachment form not as a superposition.

In Eq.(\ref{eq:fsPL}), there are four parameters.
\begin{enumerate}
\item $\chi$ is the ratio between the halo and core (diluted) phase-space densities. \item $M_s$ is the stationary value of the magnetization.
\item $\varepsilon_h$ is the maximum one-particle energy of the halo.
\item $\varepsilon_F$ is the core's `Fermi energy'.
\end{enumerate}

In the next few paragraphs, we give a clue to determine $\varepsilon_h$ by using the test particle model.
First, the equation of motion of the HMF system for an individual particle is
\begin{equation}
\ddot{\theta}=-M(t)\sin\theta\;.\label{eq:eq1}
\end{equation}
Here, as pointed out in \cite{PL}, for short elapsed time (first one or two periods of the oscillation of the magnetization), the statistical correlations between momentum $p$ and phase variable $\theta$ can be approximately ignored.
Moreover, by neglecting the average of high frequency quantities such as $\la \cos 2\theta\ra$, under the initial conditions $M(0)=M_0$ and $\dot{M}(0)=0$ an approximate single equation of motion for $M(t)$ over a short elapsed time is obtained. Then, we solve this numerically.

We derive this equation of motion following Pakter and Levin.
\begin{eqnarray}
\ddot{M}&=&\left\la \frac{d^2}{dt^2}\cos\theta\right\ra\\
&=&\la -\sin \theta \ddot{\theta}\ra+\la -\cos \theta \dot{\theta}^2\ra\\
&=&\la \sin^2 \theta\ra M-\la p^2\cos\theta\ra\\
&\approx&\frac{1}{2}M-\la p^2\ra\la \cos\theta\ra\\
&=&\frac{1}{2}M-(2{\hat{E}}-1+M^2)M\\
&=&-M\biggl(2{\hat{E}}+M^2-\frac{3}{2}\biggr)\;.\label{eq:eq2}
\end{eqnarray}

This equation can be interpreted as the equation of motion of a `particle' with coordinate $M$ and conserved energy under the double-well potential
\begin{equation}
V(M)=\frac{1}{4}M^4+\biggl({\hat{E}}-\frac{3}{4}\biggr)M^2\;.
\end{equation}
Here, the `velocity' $\dot{M}$ is $0$ at $t=0$. Thus, this motion is temporally periodic for a particle with conserved energy $V(M_0)$.
This is different from the actual motion of the magnetization that is damped by transferring the energy of the density wave to each resonant particle.

\begin{figure}[htbp]
\begin{center}
\includegraphics[width=0.5\hsize,bb=0 0 260 179]{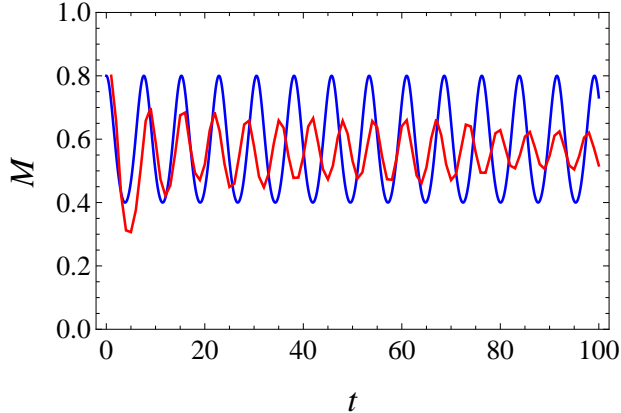}
\caption{The real and reduced oscillations of the magnetization for $M_0=0.8$, $\hat{\eta}=0.15$, $\hat{E}=0.5419$ (red and blue curves, respectively).}
\end{center}
\end{figure}

Now, we consider the way to determine the four parameters in Eq.(\ref{eq:fsPL}): $\varepsilon_h,\varepsilon_F,\chi$ and $M_s$.
As just mentioned, by the macroscopic damping of the magnetization, the energy of the density wave is transferred to the resonant particles.
Pakter and Levin considered that these resonant particles form the halo.
In actuality, the oscillation of the magnetization is significantly damped within the initial one or two periods.
Thus, within this time scale, the width of the halo is determined.
As the scheme, first, we determine the two period reduced oscillation of the magnetization described by Eq.(\ref{eq:eq2}).
Next, we run the probe test particles, which have no effect on the mean-field potential, under the equations of motion Eqs.(\ref{eq:eq1}) and (\ref{eq:eq2}) starting from a random initial distribution within the water-bag distribution on the phase space.
Based on the temporal evolution of the reduced magnetization, when $\ddot{\theta}(t)$ (Eq.(\ref{eq:eq1})) and the initial conditions $\theta(0),p(0)$ are determined, $\theta(t)$ and $p(t)=\dot{\theta}(t)$ will be determined.
The maximum energy of the probe test particles determines $\varepsilon_h$.

Once $\varepsilon_h$ is determined, the remaining parameters, $\varepsilon_F,\chi$ and $M_s$, are the solutions of the following three constraints (the conservation laws of mass and energy and the self-consistency condition on the magnetization):
\begin{eqnarray}
\int_0^{2\pi}d\theta \int_{-\infty}^\infty dp \hat{f}_s(\theta,p)&=&1\;,\\
\int_0^{2\pi} d\theta \int_{-\infty}^\infty dp \hat{f}_s(\theta,p)\biggl(\frac{p^2}{2}+\frac{1-M_s\cos \theta}{2}\biggr)&=&{\hat{E}}\;,\\
\int_0^{2\pi} d\theta \int_{-\infty}^\infty dp \hat{f}_s(\theta,p)\cos \theta&=&M_s\;.
\end{eqnarray}
Here, the energy per particle ${\hat{E}}$ is fixed.

An explication of these equations will be given in Section 4.2.4.
\subsubsection{Generalized virial condition (GVC)}
In general, it has been reported that ergodicity, which the Lynden-Bell equilibrium requires, may be broken\cite{ReviewI,Er,Er1,Er2}.
In the present case, this ergodicity breaking is induced by the parametric resonance of particles with the initial oscillation of the magnetization\cite{BTPL}.

This is because, after the macroscopic oscillation of the magnetization ends, the dynamics becomes regular and so the time averaging becomes regular.
However, due to the emergence of the Lynden-Bell statistically highly improbable\cite{NET2} halo, the statistical averaging becomes irregular.
Thus ergodicity, which asserts the equivalence between time averaging and statistical averaging, is broken.
The only mechanics which can move the halo region is the parametric resonance for the macroscopic oscillation of the magnetization.
Thus, after this ends, there is no mechanics which relaxes the halo and other regions.
That is, in global aspect, due to the existence of the halo, phase mixing between the core and the halo becomes insufficient.

Then, Benetti et al. considered that, for an initial magnetization $M_0$ for which the initial oscillation has as small an amplitude as possible (for the given $\hat{E}$), a Lynden-Bell equilibrium may arise.\cite{BTPL}

In general long-range systems (e.g., self-gravitating systems and non-neutral plasma systems), if the virial condition does not hold, the balance between the kinetic energy $K$ and potential energy $V$ is lost and the mean-field potential oscillates and a resonance emerges.
On the other hand, if the virial condition holds, there is no resonance.
When the system starts from a non-steady state, the system undergoes density oscillation.
After the relaxation, a QSS is achieved and then, the virial condition will be satisfied.\cite{ReviewII}

Benetti et al. tried to apply this role of the virial condition to the HMF model.\cite{BTPL}

In their context, the virial condition is used to discuss the deviation from the steady state.
So, the time averaging in virial is not long-term but is only over the time interval for which the steadiness is defined.
The virial condition can be written using macroscopic quantities only when the potential $V$ is a homogeneous function (we set $n=\deg V$).
However, the self-consistent mean-field potential of the HMF model is a cosine function and the virial condition cannot be applied. In the HMF model, we can formulate the condition, corresponding to $R=1$ for the {\it{virial number}} $R=2K/(nV)$, on the initial distribution only. That is, in the HMF model, there is no index corresponding to the virial number $R$.\cite{ReviewII,BTPL}

In the following, for the HMF model, we derive a condition, corresponding to the virial condition, as the initial condition ($\theta_0,p_0,M_0,\ldots$) for which the initial oscillation of the magnetization disappears.
In actuality, this condition is the one for which the envelope of the distribution is steady in the initial elapsed time.

We consider the initial water-bag distribution
\begin{equation}
\hat{f}_0(\theta,p)=\frac{1}{2\theta_0}\Theta(\theta_0-|\theta|)\frac{1}{2p_0}\Theta(p_0-|p|)\;.\label{eq:Water-Bag}
\end{equation}
To formulate our {\it{generalized virial condition}} (GVC)\cite{BTPL}, we define the temporal envelope of the distribution by
\begin{equation}
\theta_e(t)=\sqrt{3\la \theta^2\ra}\;,\label{eq:5}
\end{equation}
which takes values in $[0,\pi]$.
In Eq.(\ref{eq:5}), the factor $\sqrt{3}$ comes from
\begin{eqnarray}
\frac{1}{2\theta_0}\int_{-\theta_0}^{\theta_0}\theta^2&=&\frac{1}{2\theta_0}\frac{2}{3}\theta_0^3\\
&=&\frac{1}{3}\theta_0^2
\end{eqnarray}
 and by imposing $\theta_e(0)=\theta_0$.

To determine the temporal evolution of this envelope, we need approximations.
Concretely, we make two assumptions:
\begin{enumerate}
\item The distribution of $\theta$ is assumed to be within $[-\theta_e(t),\theta_e(t)]$. For instance, the position integral is restricted to this interval.
Thus, the magnetization $M$ becomes
\begin{eqnarray}
M(t)&=&\frac{1}{2\theta_e(t)}\int_{-\theta_e(t)}^{\theta_e(t)}d\theta \cos \theta \\
&=&\frac{\sin\theta_e(t)}{\theta_e(t)}\;.\label{eq:7}
\end{eqnarray}
\item We neglect the statistical correlation between position and momentum: $f(\theta,p)=f_\theta(\theta)f_p(p)$. Thus,
\begin{equation}
\la\theta p\ra=\la \theta\ra\la p\ra=0\;.
\end{equation}
Here, we used the dynamical inversion symmetry of the distribution through the origin: $\la \theta\ra=\la p\ra=0$.
\end{enumerate}

Based on these approximations, due to
\begin{eqnarray}
\dot{\theta_e}=\frac{3\la \theta p\ra}{\theta_e}\;,
\end{eqnarray}
the second temporal derivative of Eq.(\ref{eq:5}) is
\begin{eqnarray}
\ddot{\theta_e}&=&3\biggl(-\frac{1}{\theta_e^2}\biggr)\dot{\theta_e}\la \theta p\ra+\frac{3}{\theta_e}\dot{(\la \theta p\ra)}\\
&=&-\frac{9}{\theta_e^3}\la \theta p\ra^2+\frac{3}{\theta_e}(\la p^2\ra+\la \theta \ddot{\theta}\ra)\;.
\end{eqnarray}
Here, by using assumption 1 and the HMF equation of motion
\begin{eqnarray}
\la \theta\ddot{\theta}\ra&=&\frac{-M}{2\theta_e}\int_{-\theta_e}^{\theta_e}\theta \sin \theta d\theta\\
&=&\frac{M}{2\theta_e}(\theta \cos \theta -\sin \theta)|_{-\theta_e}^{\theta_e}\\
&=&\frac{M}{2\theta_e}(2\theta_e\cos \theta_e-2\sin\theta_e)\\
&=&M\cos \theta_e-M\frac{\sin\theta_e}{\theta_e}\\
&=&M\cos\theta_e-M^2
\end{eqnarray}
holds. From this, the assumption 2 and energy conservation
\begin{equation}
\la p^2\ra=2{\hat{E}}+M^2(t)-1\;,
\end{equation}
it follows that
\begin{eqnarray}
\ddot{\theta}_e&=&\frac{3}{\theta_e}(\la p^2\ra+\la \theta \ddot{\theta}\ra)\\
&=&\frac{3}{\theta_e(t)}(2{\hat{E}}+M(t)\cos\theta_e(t)-1)\;.\label{eq:GVC1}
\end{eqnarray}

  This equation is the equation of motion of a particle with coordinate $\theta_e$, initial velocity $\dot{\theta}_e(0)$ and conserved energy under the potential.
Thus, its motion is temporally periodic.\footnote{Strictly speaking, we need to note the following points.
The potential of $\theta_e$ in this equation is not a periodic function within $[0,\pi]_{\theta_e}$.
If there is a large oscillation so that $\theta_e$ exceeds $\pi$, we need to make the potential periodic artificially.
However, for the purpose of generalizing the virial condition, we don't need to consider such cases.}

Here, ideally, if the oscillation amplitude of $\theta_e$ becomes $0$, the behavior of $\theta_e$ and $M$ is predicted to be steady (temporally constant) due to Eq.(\ref{eq:7}).
In actuality, we give the weaker, approximate condition
\begin{equation}
\dot{\theta_e}(0)=\ddot{\theta_e}(0)=0\label{eq:GVC2}
\end{equation}
 and call it the GVC.
 Due to Eqs.(\ref{eq:7}), (\ref{eq:GVC1}) and (\ref{eq:GVC2}), the GVC becomes
\begin{equation}
(2{\hat{E}}-1)\theta_0+\sin\theta_0\cos\theta_0=0\;.
\end{equation}

By solving this equation for given $E$, we obtain ${\theta}_0=\theta_0({\hat{E}})$.
Moreover, by combining with Eq.(\ref{eq:7}), we obtain $M_0=M_0({\hat{E}})$.
Thus, we have found how to choose the initial magnetization $M_0$ which satisfies the GVC for given $\hat{E}$.

As already mentioned, the motion of envelope ($\theta_e(t),M_e(t)$) is periodic as long as $\theta_e$ is confined within the potential.
But, the actual $M(t)$ is damped due to energy transfer by the parametric resonance.

\subsection{Double Lynden-Bell scenario}
In this subsection, we discuss the double Lynden-Bell scenario.
The full contents of Sections 4.2.2 and 4.2.3 (except for Figs. 7, 8 and 10), and Figs. $9\sim 15$ (except for Fig. 10) are quoted from previous work\cite{KS}.
\subsubsection{GVC and the residual energy}
In the following, we consider the rectangular water-bag distribution in Eq.(\ref{eq:Water-Bag}).

Due to
\begin{eqnarray}
\frac{1}{4\theta_0p_0}\int_{-p_0}^{p_0}dp\int_{-\theta_0}^{\theta_0} d\theta \cos \theta &=&\frac{\sin \theta_0}{\theta_0}\\
&=&M_0
\end{eqnarray}
and
\begin{eqnarray}
\frac{1}{4\theta_0 p_0}\int_{-p_0}^{p_0}dp\int_{-\theta_0}^{\theta_0}d\theta \frac{p^2}{2}&=&\frac{p^2_0}{6}\\
&=&\hat{E}-\frac{1}{2}(1-M_0^2)\;,
\end{eqnarray}
$\theta_0,p_0$ and $\hat{\eta}$ are determined by $\hat{E}$ and $M_0$ via the relations
\begin{eqnarray}
\frac{\sin \theta_0}{\theta_0}&=&M_0\;,\label{eq:Rel1}\\
p_0&=&\sqrt{6\hat{E}-3(1-M_0^2)}\;,\label{eq:Rel2}\\
\hat{\eta}&=&\frac{1}{4\theta_0p_0}\;.\label{eq:Rel3}
\end{eqnarray}

The GVC on Eq.(\ref{eq:Water-Bag}) is
\begin{equation}
{\rm{GVC}}:(2{\hat{E}}-1)\theta_0+\sin\theta_0\cos \theta_0=0\;.
\end{equation}
By using relations (\ref{eq:Rel1}), (\ref{eq:Rel2}) and (\ref{eq:Rel3}), the GVC becomes:
\begin{eqnarray}
0&=&\biggl({\frac{p_0^2}{3}}-{M_0^2}\biggr)\theta_0+\sin\theta_0\cos \theta_0\\
&=&\biggl({\frac{1}{48\hat{\eta}^2\theta_0^2}}-{\frac{\sin\theta_0^2}{\theta_0^2}}\biggr)\theta_0+\sin\theta_0\cos \theta_0\;.
\end{eqnarray}
Then, multiplying by $\theta_0$ on both sides, we obtain
\begin{equation}
\frac{1}{48\hat{\eta}^2}-\sin^2\theta_0+\sin\theta_0\cos \theta_0\theta_0=0\;.\end{equation}

Next, the minimization condition on the energy per particle with respect to $M_0$ for fixed $\hat{\eta}$ is
\begin{equation}
\frac{\pa {\hat{E}}}{\pa M_0}|_{\hat{\eta}:{\rm{fix}}}=0\;,\ \ {\hat{E}}={\frac{1}{96\hat{\eta}^2\theta_0^2}}+\frac{1-M_0^2}{2}\;.
\end{equation}
Now
\begin{eqnarray}
\frac{\pa {\hat{E}}}{\pa M_0}|_{\hat{\eta}:{\rm{fix}}}&=&-M_0+\frac{\pa }{\pa M_0}\biggl(\frac{1}{96\hat{\eta}^2\theta_0^2}\biggr)_{\hat{\eta}:{\rm{fix}}}\\
&=&-{M_0}+\biggl(\frac{\pa M_0}{\pa \theta_0}\biggr)^{-1}{\frac{\pa }{\pa \theta_0}\biggl(\frac{1}{96\hat{\eta}^2\theta_0^2}\biggr)_{\hat{\eta}:{\rm{fix}}}}\\
&=&-\frac{\sin\theta_0}{\theta_0}+\biggl(\frac{-\sin \theta_0+\cos \theta_0\theta_0}{\theta_0^2}\biggr)^{-1}\biggl(\frac{-1}{48\hat{\eta}^2\theta_0^3}\biggr)\\
&=&-\frac{\sin\theta_0}{\theta_0}+\frac{1}{48\hat{\eta}^2(\sin \theta_0-\cos \theta_0\theta_0)\theta_0}
\\
&=&0\;,
\end{eqnarray}
which reduces to
\begin{equation}
\frac{1}{48\hat{\eta}^2}-\sin^2\theta_0+\sin \theta_0\cos \theta_0\theta_0=0\;.
\end{equation}
This matches the GVC.
This expression of the GVC will be used in the context of the minimization of the residual total energy of the system against the energy of the Vlasov stationary water-bag state: see Section 4.2.3.
\subsubsection{Typical temporal evolution}
In this subsection, we convert Pakter and Levin's scenario of core-halo formation in the attachment form to the double Lynden-Bell scenario in the superposition form.

From this standpoint, the formation process of the core-halo structure of an HMF system consists of four steps.

First, as the result of a trigger, which will create a chemical potential gap between the core and the halo, by the parametric resonance of the system with the initial strong oscillation of the magnetization, the core-halo structure starts forming\cite{PL}.

Second, after a {{dynamical}} process facilitated by particle and energy exchanges between the core and the halo, the distribution relaxes to a {{steady}} superposition of two components: that is, the core and the halo.
Due to its long-range nature, the potential $\Phi(\theta)$ is common to the core and halo distributions.
Here, we denote the fine-grained core and halo distributions by $f_c$ and $f_h$, respectively.

The dynamical relaxation between the core and the halo
\begin{equation}
df_a/dt\to 0\;,\ a=c,h\label{eq:dr}
\end{equation}
plays the role of the Vlasov fluid property of incompressibility for each component $a=c,h$.
This relaxation converges the total mass $N_a$ and the diluted phase-space density $\eta_a$ for each $f_a$ under the condition $\eta=\eta_c+\eta_h$, where $\eta$ denotes the fine-grained phase-space density of the system.

Thirdly, the magnetization stabilizes, and the system enters the QSS regime.

Finally, phase-mixing converges.
That is, $f$ and $f_a$ closely approximate functions of $\varepsilon$ only.
Then, due to Eq.(\ref{eq:dr}), $\pa f_a(\varepsilon) /\pa t\to 0$ holds.
Consequently, the total energy $E_a$ of each $f_a$ converges.
At this time, the core-halo formation is complete.

\begin{figure}[htbp]
\includegraphics[width=0.33\hsize,bb=0 0 370 390]{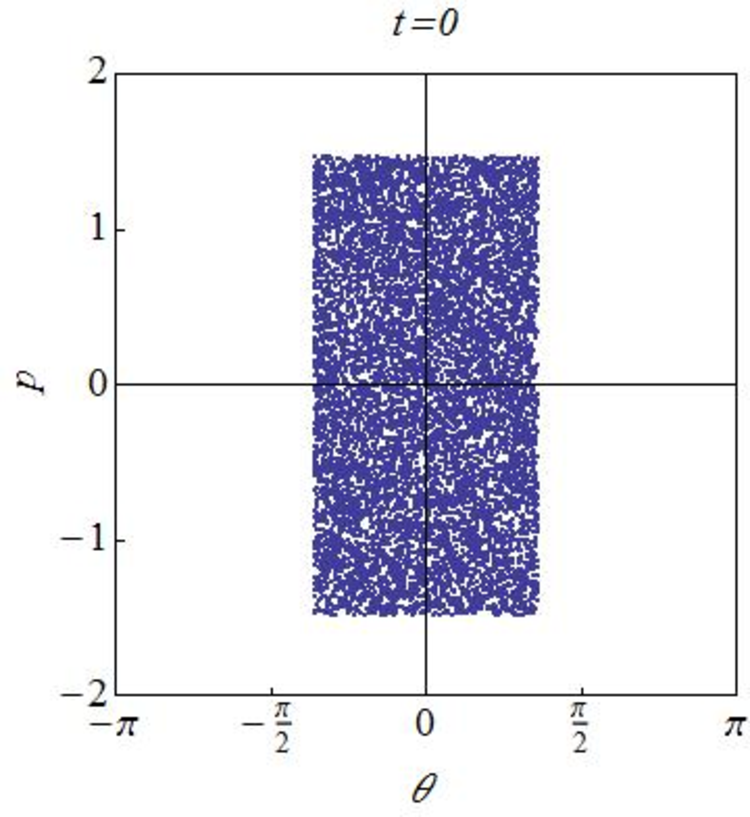}\includegraphics[width=0.33\hsize,bb=0 0 370 390]{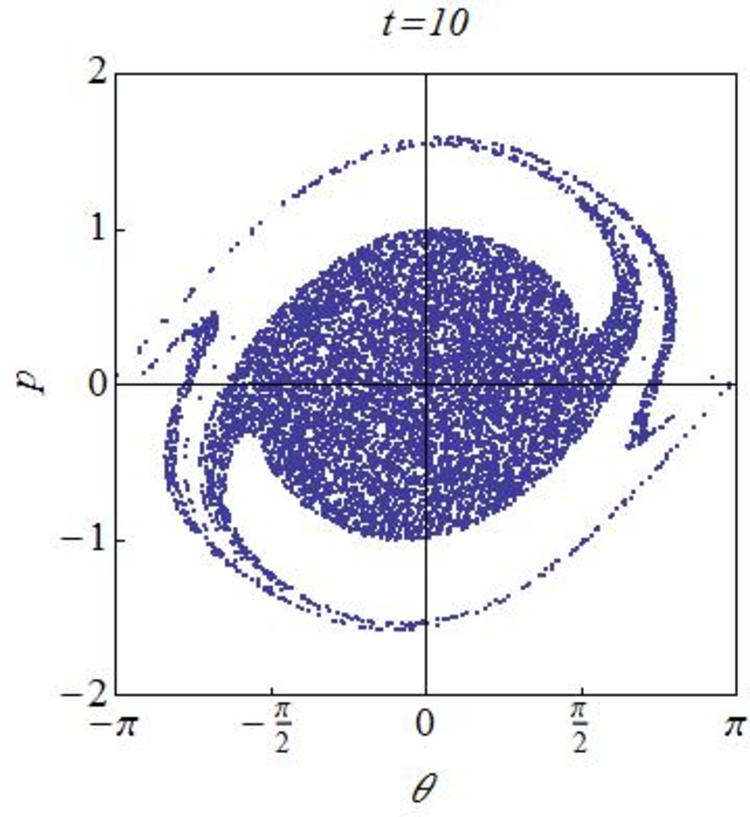}\includegraphics[width=0.33\hsize,bb=0 0 370 390]{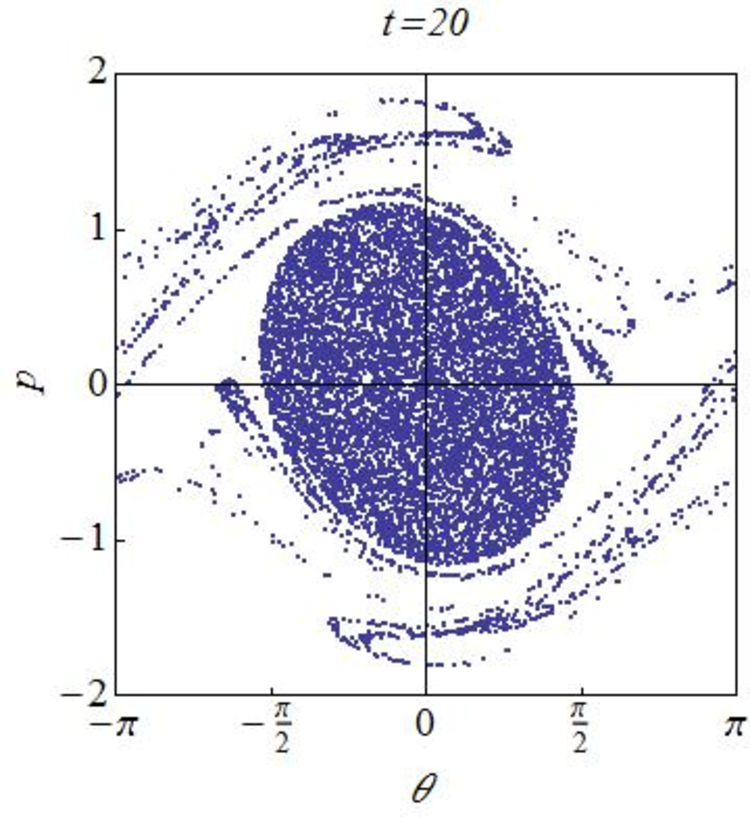}\\
\includegraphics[width=0.33\hsize,bb=0 0 260 179]{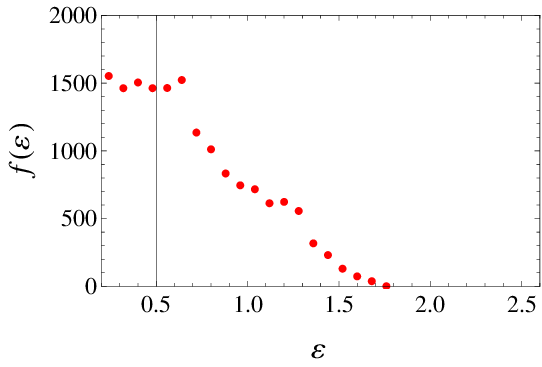}\includegraphics[width=0.33\hsize,bb=0 0 260 179]{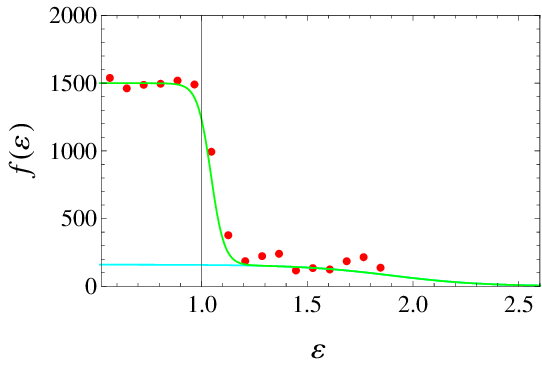}\includegraphics[width=0.33\hsize,bb=0 0 260 179]{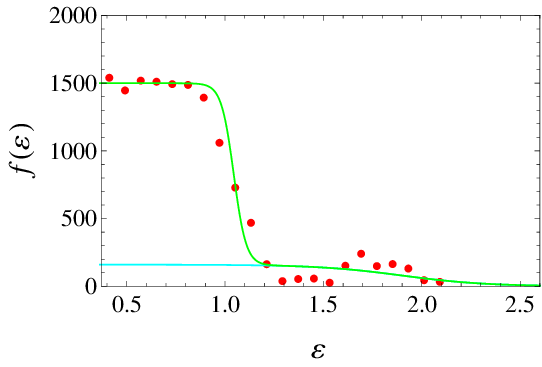}\\
\includegraphics[width=0.33\hsize,bb=0 0 370 390]{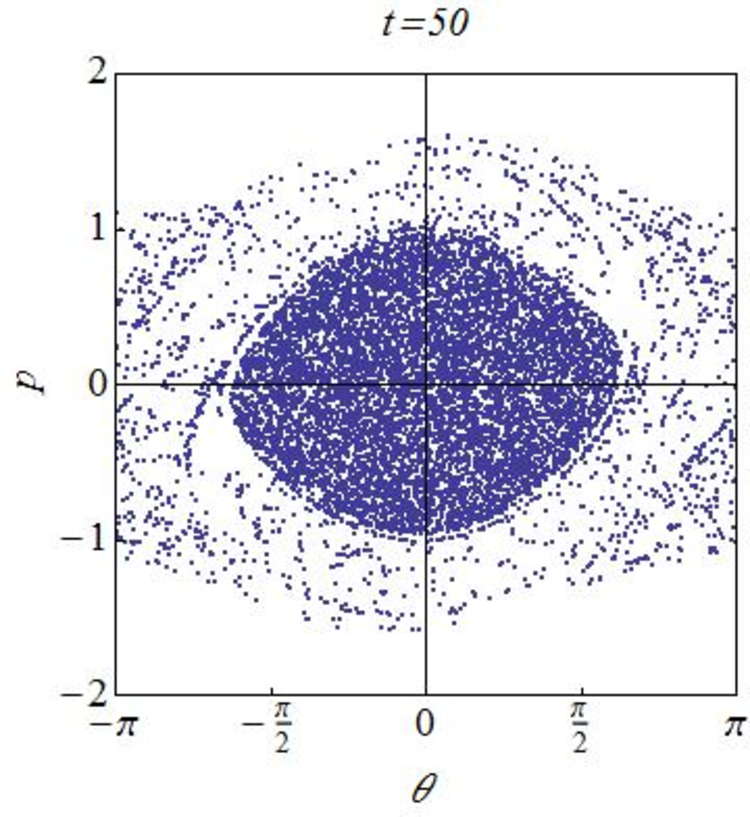}\includegraphics[width=0.33\hsize,bb=0 0 370 390]{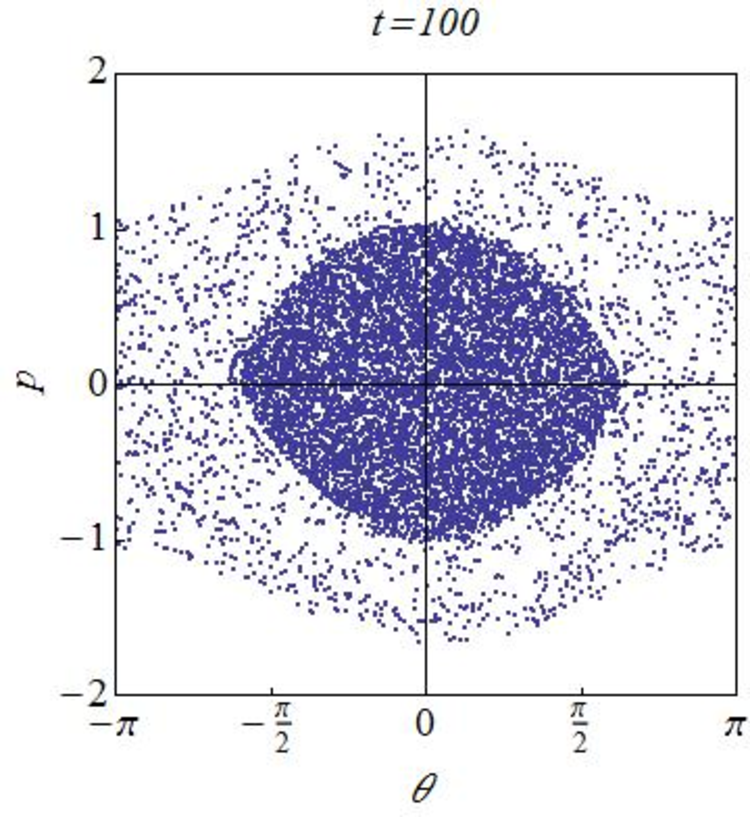}\includegraphics[width=0.33\hsize,bb=0 0 370 390]{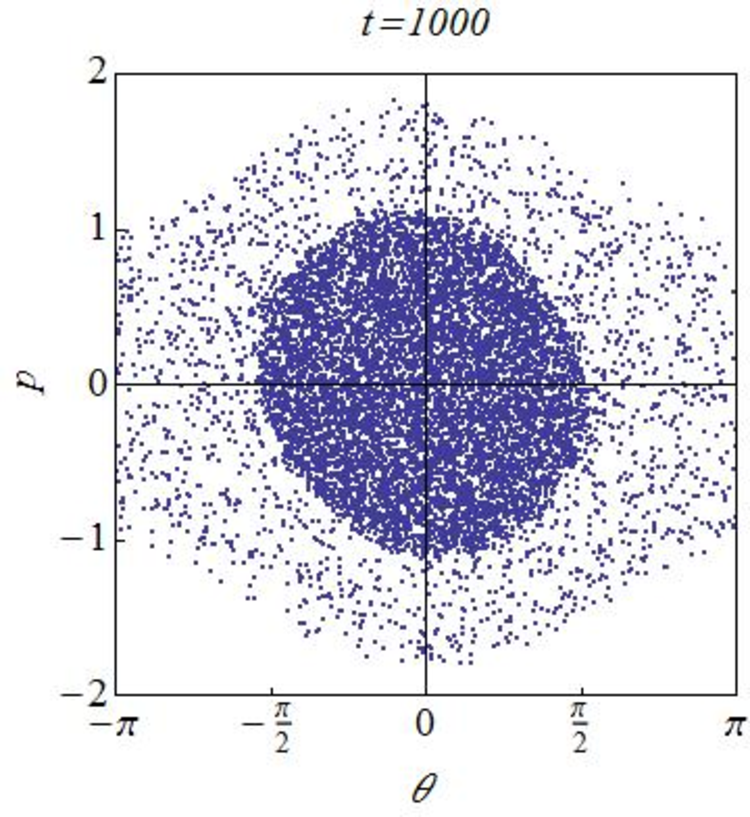}\\
\includegraphics[width=0.33\hsize,bb=0 0 260 179]{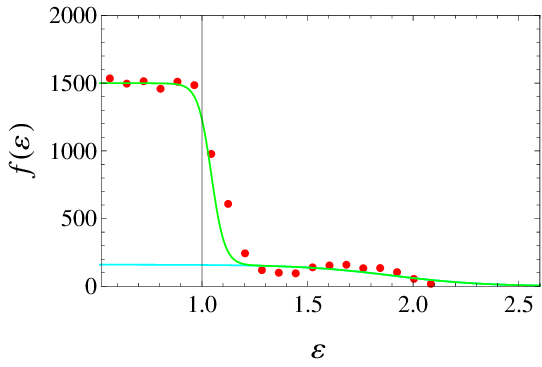}\includegraphics[width=0.33\hsize,bb=0 0 260 179]{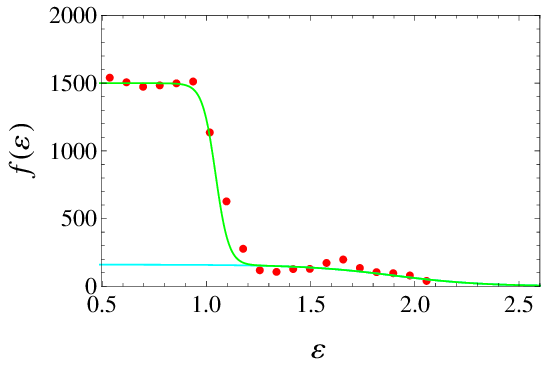}\includegraphics[width=0.33\hsize,bb=0 0 260 179]{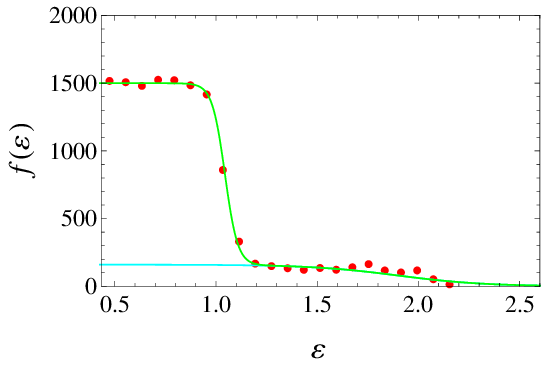}
\caption{Phase mixing ($M_0=0.8$, $\hat{\eta}=0.15$, $\hat{E}=0.5419$).
The upper figures show the evolution of the system on the phase space.
The lower figures show the corresponding $f(\varepsilon)$.
The green and cyan curves represent the full and halo parts of the double Lynden-Bell fitting of the single run simulation results (red dots) at $t=10^4$.
The fitting method will be explained in Fig. 13.}
\end{figure}

\begin{figure}[htbp]
\includegraphics[width=0.33\hsize,bb=0 0 370 390]{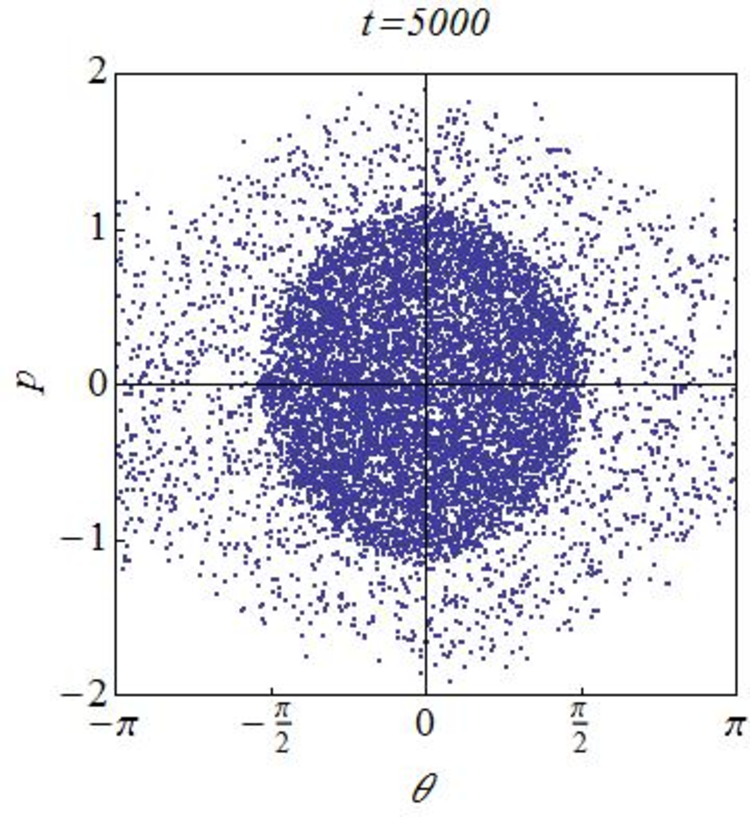}\includegraphics[width=0.33\hsize,bb=0 0 370 390]{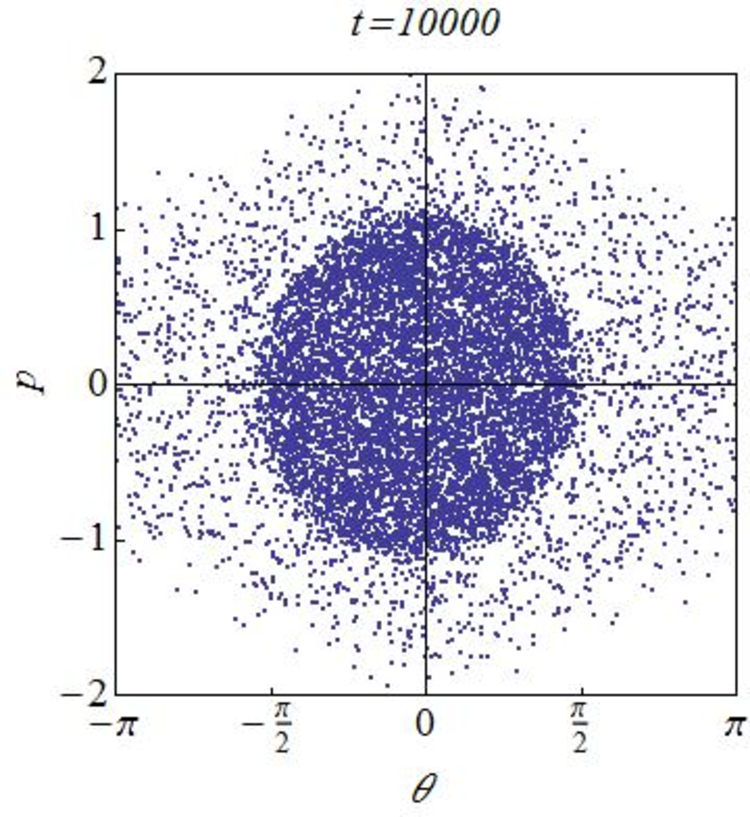}\includegraphics[width=0.33\hsize,bb=0 0 370 390]{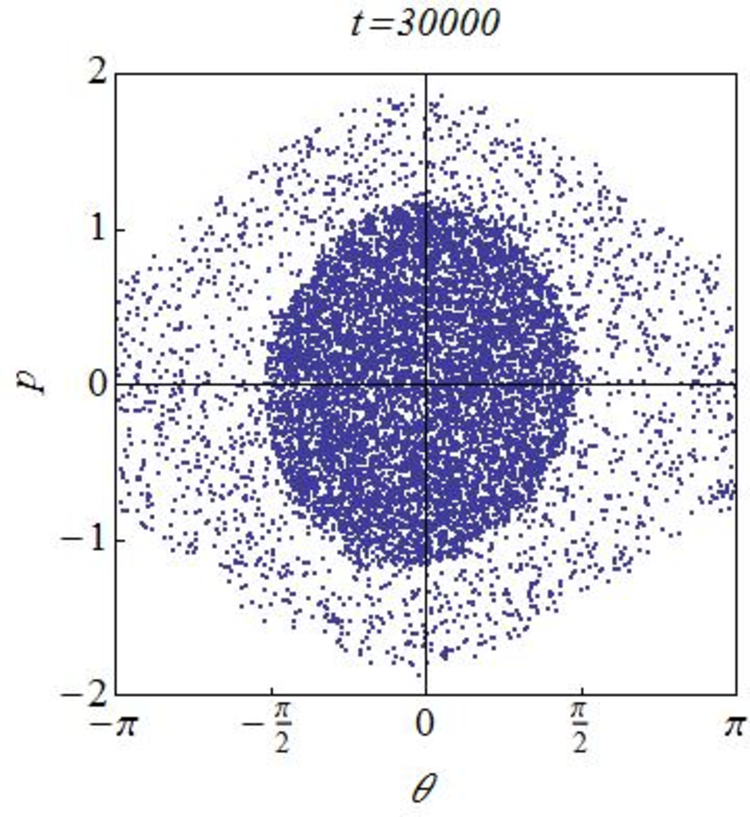}\\
\includegraphics[width=0.33\hsize,bb=0 0 260 179]{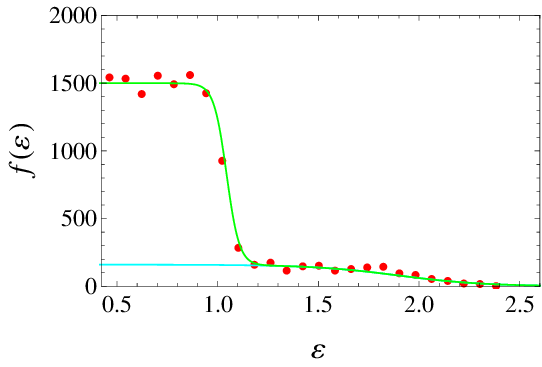}\includegraphics[width=0.33\hsize,bb=0 0 260 179]{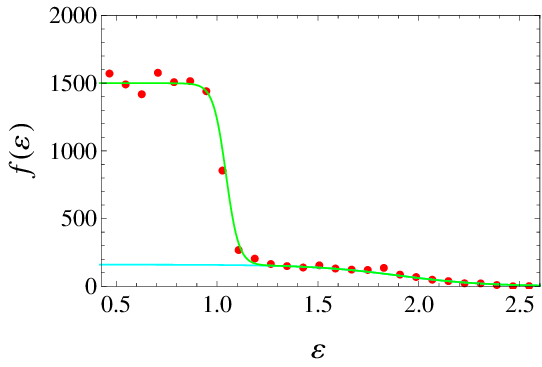}\includegraphics[width=0.33\hsize,bb=0 0 260 179]{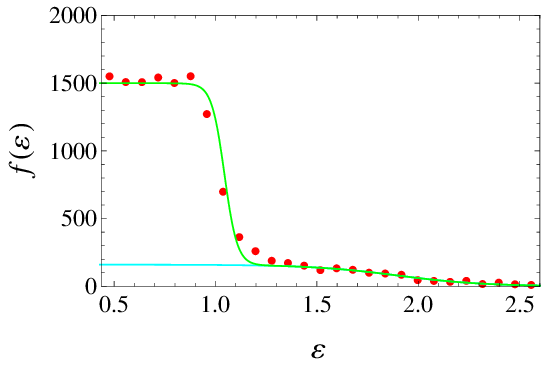}
\caption{Phase mixing (continued).}
\end{figure}

Based on this process, we derive the core-halo QSS distribution and its corresponding entropy, by following the discussion of collisionless ergodic relaxation by Lynden-Bell.\cite{LB}

The phase space is divided into macro-cells, that is, assemblies of micro-cells, and incompressible Vlasov elements occupy micro-cells.
From now on, while $\eta$ denotes the fine-grained phase-space density, we consistently denote the {\it{coarse-grained}} (macro-cell level) core and halo distributions by
\begin{eqnarray}
{f}_c(\theta_i,p_i)&=&{f}_{c,i}=\frac{\eta m_i \omega}{\nu\omega}=\frac{\eta_c m_i}{\nu_c}\;,\ \ \eta_c=\frac{\eta}{\nu}\nu_c\;,\label{eq:f1}\\
{f}_h(\theta_i,p_i)&=&{f}_{h,i}=\frac{\eta n_i \omega}{\nu\omega}=\frac{\eta_h n_i}{\nu_h}\;,\ \ \eta_h=\frac{\eta}{\nu}\nu_h\;,\label{eq:f2}
\end{eqnarray}
where $i$ labels macro-cells ($i=1,2,\ldots,P$), $m_i$ and $n_i$ are the numbers of Vlasov elements occupying the $i$-th macro-cell, $\nu$ is the number of micro-cells in each macro-cell, and $\omega$ is the area of each micro-cell.

In the process described above, the following partitions are fixed:
\begin{equation}
N=N_c+N_h\;,\ E=E_c+E_h\;,\ \nu=\nu_c+\nu_h\;.\label{eq:constraint}
\end{equation}
The third partition in Eq.(\ref{eq:constraint}) is kept for the ratios in the continuum limit $\nu\to 0$.

The total partition number of the configurations of Vlasov elements in the phase space is
\begin{eqnarray}
W&=&W_{mix} W_{lb}^{(c)} W^{(h)}_{lb}\;,\label{eq:W}\\
W_{mix}&=&\frac{N!}{N_c!N_h!}\prod_{i=1}^P\frac{\nu!}{\nu_c!\nu_h!}\;,\\
W_{lb}^{(c)}&=&\frac{N_c!}{\prod_{i=1}^P m_i!}\prod_{i=1}^P\frac{\nu_c!}{(\nu_c-m_i)!}\;,\\
W_{lb}^{(h)}&=&\frac{N_h!}{\prod_{i=1}^P n_i!}\prod_{i=1}^P\frac{\nu_h!}{(\nu_h-n_i)!}\;,
\end{eqnarray}
where $W_{mix}$ is the partition number of mixing the core and the halo and $W_{lb}^{(a)}$ are the Lynden-Bell partition numbers for the core and the halo.
Using Eqs.(\ref{eq:f1}) and (\ref{eq:f2}), the total partition number $W$ can be expressed as a functional of coarse-grained distributions ${f}_{c,i}$ and ${f}_{h,i}$.

The procedure for the maximization of the entropy in terms of these distributions is\footnote{We omit the details of the calculations in the following few sentences, since these are parallel to those in the single Lynden-Bell case.}
\begin{equation}
\delta_{{f}_{c,i}} \ln W=0\;,\ \ \delta_{{f}_{h,i}} \ln W=0
\end{equation}
under the constraints in Eq.(\ref{eq:constraint}).
We introduce two kinds of Lagrange multiplier, $\alpha_a$ and $\beta_a$, where $a=c,h$, for fixed particle number $N_a$ and energy $E_a$, respectively.

Under the continuum limit ($\nu\to 0$), the total entropy reduces to
\begin{equation}
S=S^{(c)}+S^{(h)}\;,\label{eq:S}
\end{equation}
where each $S^{(a)}$ is the Lynden-Bell entropy\cite{LB}
\begin{equation}
S^{(a)}=-\int d\theta dp \biggl(\frac{f_a}{\eta_a}\ln \frac{f_a}{\eta_a}+\biggl(1-\frac{f_a}{\eta_a}\biggr)\ln\biggl(1-\frac{f_a}{\eta_a}\biggr)\biggr)\label{eq:LB}
\end{equation}
 for $a=c,h$.

The maximization solution of Eq.(\ref{eq:S}) is the double Lynden-Bell distribution
\begin{equation}
f(\theta,p)=\sum_{a=c,h}\frac{\eta_a}{\exp(\beta_a(\varepsilon(\theta,p,M_s)-\mu_a))+1}\;,\label{eq:dfs}
\end{equation}
where $\mu_a=-\alpha_a/\beta_a$ is the chemical potential of the core or the halo. At this point, the double Lynden-Bell scenario is complete.

Here, readers may think that since the partition number $W$ is the product in Eq.(\ref{eq:W}), the distribution function Eq.(\ref{eq:dfs}) would also be a product. 
However, the fine grains of the distribution functions $f_c$ and $f_h$ do not share any micro-cells.
Thus, $f$ is a superposition, that is, $f=f_c+f_h$.

\subsubsection{$N$-body simulations}

 In the $N$-body simulation described in the following subsections, the initial phase-space distribution function $\hat{f}_0(\theta,p)$ is the uniform water-bag type distribution over the rectangle $[-\theta_0,\theta_0]\times [-p_0,p_0]$, namely
\begin{equation}
\hat{f}_0(\theta,p)=\hat{\eta}\Theta(\theta_0-|\theta|)\Theta(p_0-|p|)\;,\label{eq:f0}
\end{equation}
where $\Theta$ is the Heaviside unit one-step function.
(This is the same as Eq.(\ref{eq:Water-Bag}).)

The parameters $\theta_0$ and $p_0$ of Eq.(\ref{eq:f0}) satisfy the relations $\sin\theta_0/\theta_0=M_0$, $p_0=\sqrt{6{\hat{E}}-3(1-M_0^2)}$, and $\hat{\eta}=1/(4\theta_0p_0)$ for initial magnetization $M_0$ and energy per particle ${\hat{E}}$ (see Section 4.2.1).
Using these relations, when we fix $\hat{\eta}$, we can deduce $\hat{E}$ from $M_0$.

In order to take advantage of the Vlasov incompressibility, that is, the dynamical conservation of $\hat{\eta}$, we classify simulation data by the common value of $\hat{\eta}$.
In the following part of this review, we consider $\hat{\eta}=0.15$ as the cocrete value.

\begin{figure}[htbp]
\begin{center}
\includegraphics[width=0.5\hsize,bb=0 0 260 179]{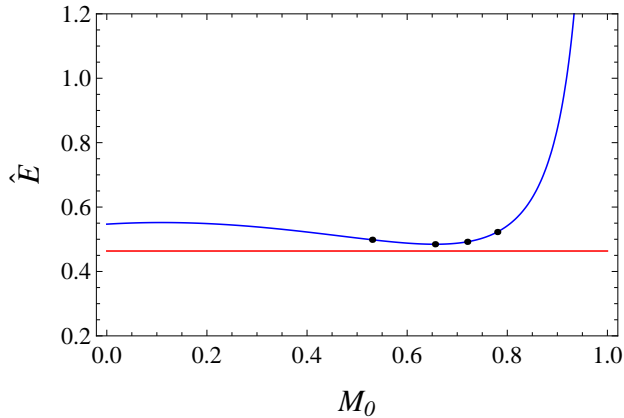}
\caption{The $\hat{\eta}=0.15$ configuration of four data $M_0=0.53,M_{{\rm{min}}},0.72,0.78$ on the $(M_0,{\hat{E}})$ plane to be used in Figs. 12 and 13. Here, $M_{{\rm{min}}}\sim 0.6556$ is the magnetization of the initial water-bag distribution Eq.(\ref{eq:f0}) at the minimum ${\hat{E}}$.
The blue and red curves represent the initial water-bag states and the energy per particle of the Vlasov stationary water-bag state $f_{\varepsilon_F}$ for ${\eta}=1500$, respectively.}
\end{center}
\end{figure}

Figure 9 shows that $M_0=M_{{\rm{min}}}$ gives the global minimum of the function $E(M_0)$ and in the neighborhood of this point, the function is convex.
This also holds for other values of $\hat{\eta}$.
Thus, it is natural to express some character of $f_0$ in terms of its total energy $E$.

Accordingly, we introduce the residual total energy ${{E}}_{{\rm{res}}}$ (refer to Fig. 9) which is equal to the total energy $E$ of the system minus the total energy ${E}_{\varepsilon_F}$ of the Vlasov stationary water-bag state $f_{\varepsilon_F}(\varepsilon)=\eta\Theta(\varepsilon_F-\varepsilon)$ for ${\eta}=1500$ (i.e., ${E}_{{\rm{res}}}={E}-{E}_{\varepsilon_F}$).
(In this context, the Vlasov stationary water-bag state $f_{\varepsilon_F}$ depends on three parameters, that is, the Fermi energy $\varepsilon_F$, the magnetization and the total energy $E_{\varepsilon_F}$.
These are determined by the two conservation laws and the self-consistency condition: see Section 4.2.4.)

The purpose of the introduction of $f_{\varepsilon_F}$ lies in its role in $E_{{\rm{res}}}$.
To show this, we note that $f_{\varepsilon_F}$ has a total energy lower than that of any $f_0$ with the common value of $\eta$ and cannot be accessible by Vlasov dynamics starting from $f_0$ due to energy conservation.

\begin{figure}[htbp]
\begin{center}
\includegraphics[width=0.5\hsize,bb=0 0 260 290]{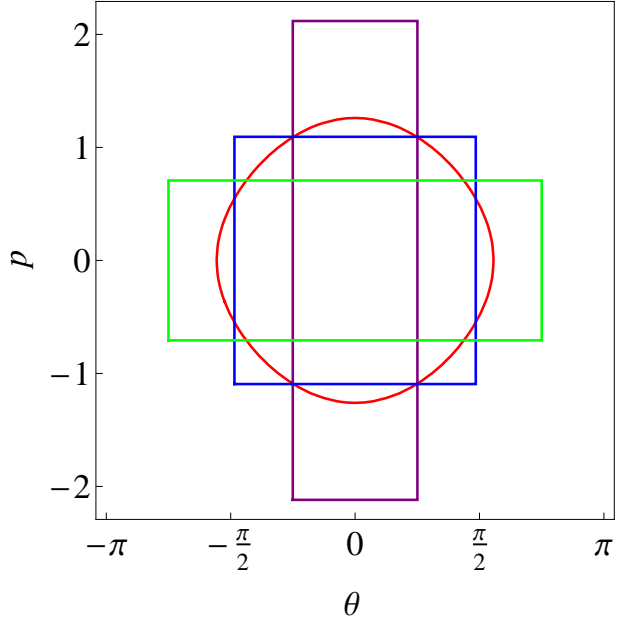}
\caption{Examples of shapes of water-bag distributions with $\hat{\eta}=0.15$: green ($M_0=0.2$), blue ($M_0=M_{{\rm{min}}}$), purple ($M_0=0.9$) and red (the Vlasov stationary one).}
\end{center}
\end{figure}

To clarify the meaning of $E_{{\rm{res}}}$, we consider the dynamics of the system on the phase space by referring to Fig. 10.

When the dynamics start from $f_0$, its center $f_{\varepsilon_F}$ is Vlasov stationary, and there is a total energy gap ${E}_{{\rm{res}}}>0$ between them.
So, by using $E_{{\rm{res}}}$ the system creates the halo of high-energy particles in the outer site, then, the inner part approaches the Vlasov stationary water-bag state due to energy conservation.
Thus, ${{E}}_{{\rm{res}}}$ measures the degree of the creation of the high-energy tail of the halo, which causes the system to deviate from the Lynden-Bell equilibrium.
That is, we argue that $E_{{\rm{res}}}$ is an {\it{a priori}} measure of the deviation of the system from the Lynden-Bell equilibrium.

For $M_0\le M_{{\rm{min}}}$, the Vlasov stationary water-bag distributions that inscribe and circumscribe the initial distribution Eq.(\ref{eq:f0}) (refer to Fig. 1) are close to each other.
So, in these cases, the validity of this argument weakens.

In Fig. 11, we illustrate this argument by the almost monotone correspondence between the residual total energy and the residue of the Lynden-Bell entropy of the Lynden-Bell equilibrium against that of the system.

\begin{figure}[htbp]
\begin{center}
\includegraphics[width=0.5\hsize,bb=0 0 260 179]{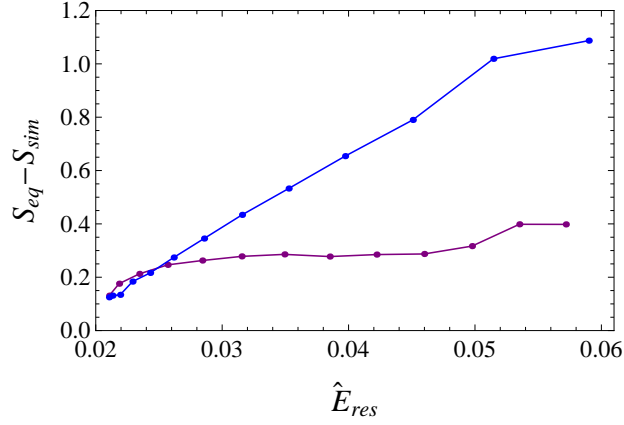}
\caption{This figure shows the residue of the Lynden-Bell entropy of the Lynden-Bell equilibrium $S_{{\rm{eq}}}$ for given ${{E}}$ and ${\eta}=1500$ against that of the simulation result $S_{{\rm{sim}}}$ as a function of the residual energy per particle ${\hat{E}}_{{\rm{res}}}$. 
Blue and purple plotted dots represent, respectively, the cases of $M_0=0.5$--$0.78$ (with $0.01$ increments) and $M_0=0.41$--$0.65$ (with $0.02$ increments) using $10$ run averages.}
\end{center}
\end{figure}

As already confirmed, the minimization condition on the residual total energy $({\pa {{E}}}_{{\rm{res}}}/{\pa M_0})_{{\eta}}=0$ matches the GVC for the HMF model.
So, this argument has an advantage over the GVC formulation.

As illustrated in Fig. 12, as the residual total energy ${{E}}_{{\rm{res}}}$ increases, the high-energy tail of the simulation resultant $f(\varepsilon)$ grows.
This high-energy tail causes the simulation resultant $f(\varepsilon)$ to deviate from the Lynden-Bell equilibrium.
\begin{figure}[htbp]
\begin{center}
\includegraphics[width=0.33\hsize,bb=0 0 260 179]{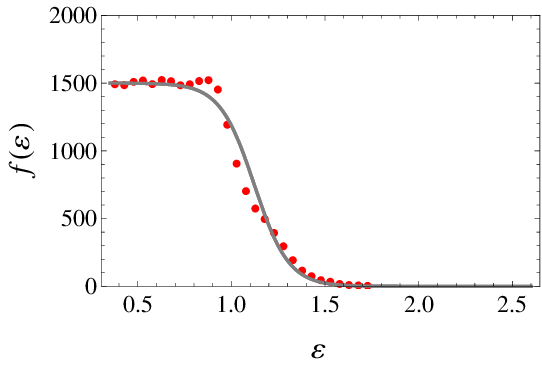}
\includegraphics[width=0.33\hsize]{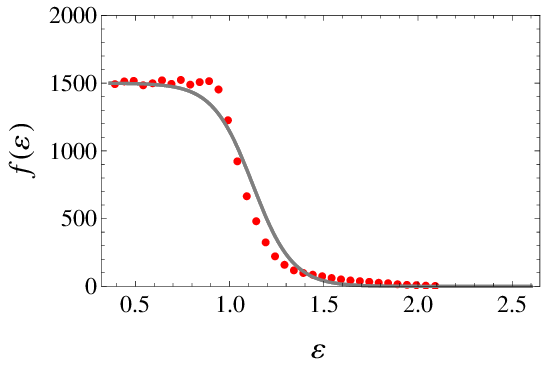}\includegraphics[width=0.33\hsize]{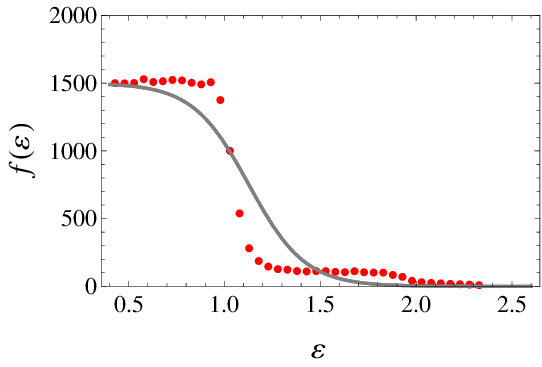}\\
\includegraphics[width=0.33\hsize]{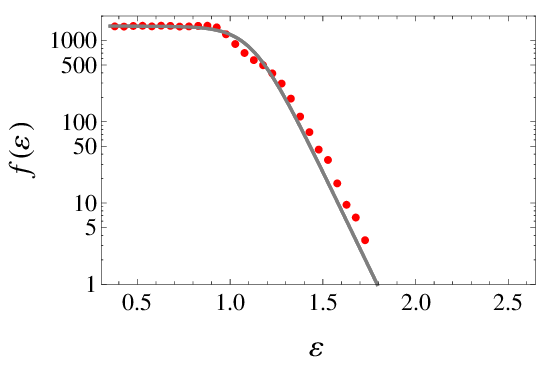}\includegraphics[width=0.33\hsize]{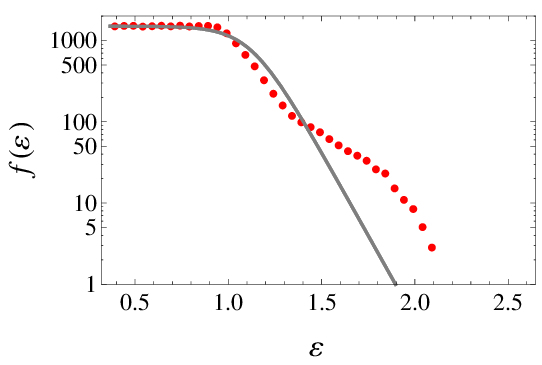}\includegraphics[width=0.33\hsize]{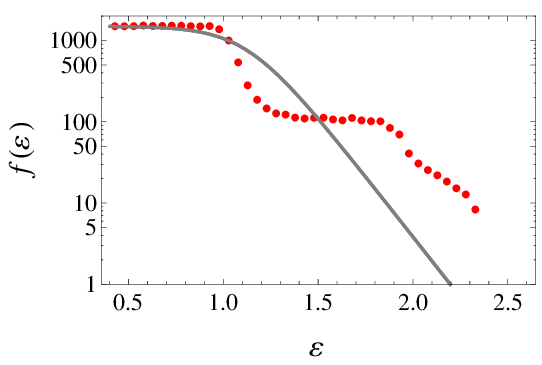}
\caption{These figures show the deviation of the simulation resultant $f(\varepsilon)$ (red dots) averaged over $20$ runs from the single Lynden-Bell equilibrium (gray curve) for $M_0=M_{{\rm{min}}},0.72,0.78$ (from left to right).}
\end{center}
\end{figure}
\begin{figure}[htbp]
\begin{center}
\includegraphics[width=0.33\hsize,bb=0 0 260 179]{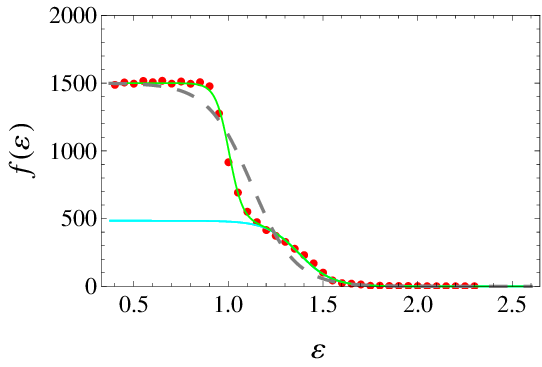}\includegraphics[width=0.33\hsize,bb=0 0 260 179]{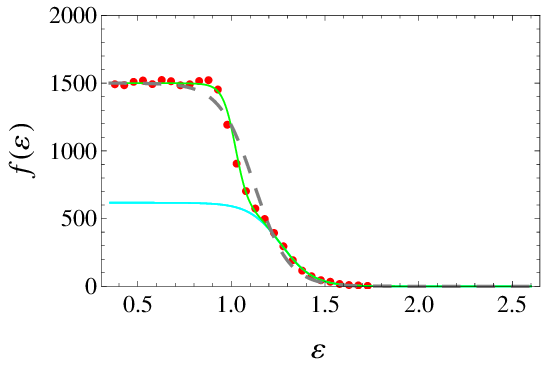}\includegraphics[width=0.33\hsize,bb=0 0 260 179]{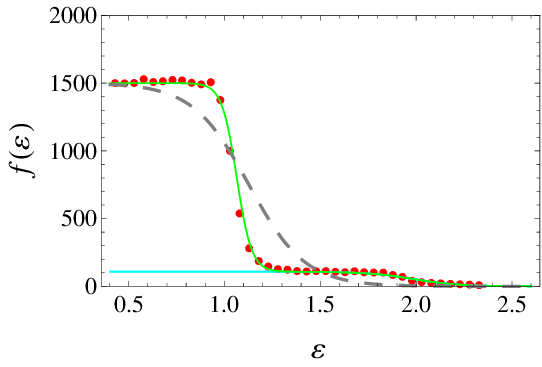}\\
\includegraphics[width=0.33\hsize,bb=0 0 260 179]{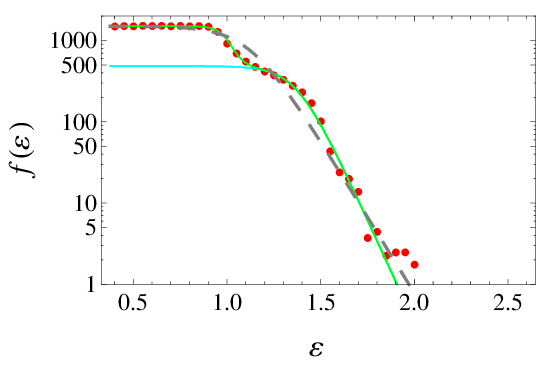}\includegraphics[width=0.33\hsize]{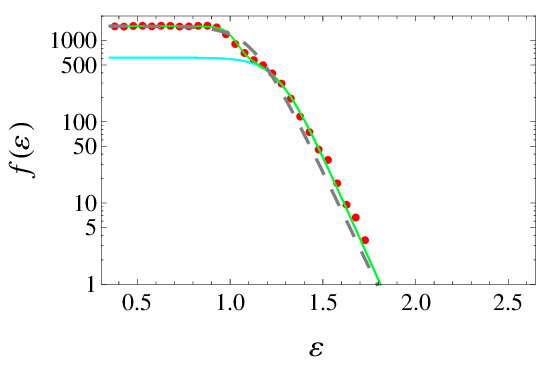}\includegraphics[width=0.33\hsize]{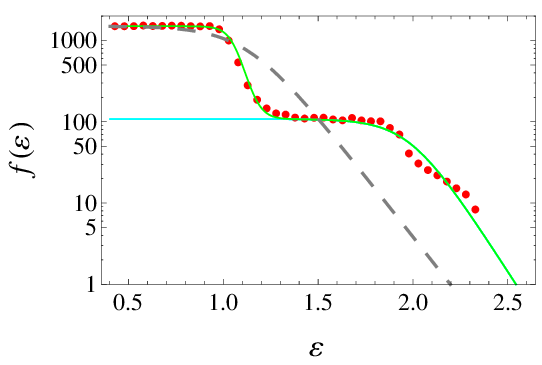}
\caption{These figures show the $M_0=0.53,M_{{\rm{min}}},0.78$ (from left to right) double Lynden-Bell theoretical semi-predictions of simulation resultant $f(\varepsilon)$ (red dots) averaged over $20$ runs. Green and cyan curves represent the full and halo parts of the double Lynden-Bell distribution, respectively. The former is constrained to satisfy the three conservation laws and the self-consistency condition and by adjusting values of the Lynden-Bell entropy, stationary magnetization $M_s$ and ${\eta}_c$ by hand. Gray dashed curves represent the Lynden-Bell equilibrium for given ${{E}}$ and ${\eta}=1500$.}
\end{center}
\end{figure}

In the double Lynden-Bell scenario, we argue that this deviation part is fitted by the halo part of the distribution, $f_h$.

As an illustration of this argument, in Fig. 13 we show the {\it{theoretical semi-predictions}} using the double Lynden-Bell distributions for the three initial magnetizations $M_0=0.53,M_{{\rm{min}}}$ and $0.78$.

A double Lynden-Bell distribution has seven degrees of freedom.
In Fig. 13, by adjusting three parameters by hand, we solve the four conditions, that is, three conservation laws for mass, energy and phase-space density, and the self-consistency condition, and derive the double Lynden-Bell distributions.
So, Fig. 13 is not just a fitting but is also a theoretical result.

The three parameters adjusted by hand to produce Fig. 13 include the Lynden-Bell entropy.
By setting the Lynden-Bell entropy to be lower than that of the Lynden-Bell equilibrium, we accurately reproduce the $N$-body simulation results.
These accurate reproductions corroborate the double Lynden-Bell scenario.

\subsubsection{Vlasov stationary water-bag states}

In this subsection, to supplement the contents of Section 4.2.3, for the Vlasov stationary water-bag distribution $f_{\varepsilon_F}(\varepsilon)$ with given phase space density $\eta$, we give the set of equations that determine its Fermi energy $\varepsilon_F$, magnetization $M$ and energy per particle $\hat{E}$.

First, by introducing the elliptic integrals
\begin{eqnarray}
E(x,k)&=&\int_0^x\sqrt{1-k^2\sin^2\theta}d\theta\;,\\
F(x,k)&=&\int_0^x\frac{1}{\sqrt{1-k^2\sin^2\theta}}d\theta\;,
\end{eqnarray}
we define the following special functions
(the complete elliptic integral is equal to the real part of the incomplete elliptic integral):
\begin{eqnarray}
F_0(\varepsilon,M)&=&\int_0^{\theta^0(\varepsilon)}\sqrt{2(\varepsilon+M\cos\theta)}d\theta\\
&=&{\rm{Re}}\biggl(2\sqrt{2(\varepsilon+M)}E\biggl(\frac{1}{2}{{\arccos}}\biggl(-\frac{\varepsilon}{M}\biggr),\sqrt{\frac{2M}{\varepsilon+M}}\biggr)\biggr)\;,\\
G_0(\varepsilon,M)&=&\int_0^{\theta^0(\varepsilon)}\sqrt{2(\varepsilon+M\cos \theta)}\cos\theta d\theta\\
&=&{\rm{Re}}\biggl(\frac{1}{3M}2\sqrt{2}\biggl(\varepsilon\sqrt{\varepsilon+M} E\biggl(\frac{1}{2}{{\arccos}}\biggl(-\frac{\varepsilon}{M}\biggr),\sqrt{\frac{2M}{\varepsilon+M}}\biggr)\nonumber\\
&&-\frac{1}{\sqrt{\varepsilon+M}}(\varepsilon^2-M^2)F\biggl(\frac{1}{2}{{\arccos}}\biggl(-\frac{\varepsilon}{M}\biggr),\sqrt{\frac{2M}{\varepsilon+M}}\biggr)\biggr)\biggr)\;,\\
H_0(\varepsilon,M)&=&\int_0^{\theta^0(\varepsilon)}(\sqrt{2(\varepsilon+M\cos \theta)})^3d\theta \\
&=&{\rm{Re}}\biggl(\frac{2}{3}\sqrt{2}\biggl(8\varepsilon\sqrt{\varepsilon+M}E\biggl(\frac{1}{2}{{\arccos}}\biggl(-\frac{\varepsilon}{M}\biggr),\sqrt{\frac{2M}{\varepsilon+M}}\biggr)\nonumber\\
&&-2\sqrt{\frac{1}{\varepsilon+M}}(\varepsilon^2-M^2)F\biggl(\frac{1}{2}{{\arccos}}\biggl(-\frac{\varepsilon}{M}\biggr),\sqrt{\frac{2M}{\varepsilon+M}}\biggr)\biggr)\biggr)\;,
\end{eqnarray}
where we set
\begin{equation}
\theta^0(\varepsilon)=\left\{\begin{array}{cc}{{\arccos}}(-\frac{\varepsilon}{M})&-1<-\frac{\varepsilon}{M}\\
\pi&{\rm{Otherwise}}\end{array}\right.
\end{equation}

For the simplicity of the equations, we define 
\begin{eqnarray}
F(\varepsilon,M)&=&4F_0(\varepsilon,M)\;,\\
G(\varepsilon,M)&=&4G_0(\varepsilon,M)\;,\\
H(\varepsilon,M)&=&\frac{2}{3}H_0(\varepsilon,M)\;.
\end{eqnarray}

Using these integral quantities, the conditions which the Vlasov stationary water-bag distribution $f_{\varepsilon_F}(\varepsilon)$ satisfies are
\begin{eqnarray}
\hat{\eta}F(\varepsilon_F-1,M)&=&1\ \ ({\rm{normalization}})\;,\\
\hat{\eta}G(\varepsilon_F-1,M)&=&M\ \ ({\rm{magnetization}})\;,\\
\hat{\eta}H(\varepsilon_F-1,M)+\frac{1-M^2}{2}&=&\hat{E}\ \ ({\rm{energy\ per\ particle}})\;.
\end{eqnarray}
We solve these numerically and determine $\varepsilon_F$, $M$ and $\hat{E}$.

We will explain how we derive these conditions at the next complicated level: see Section 4.2.5.
\subsubsection{Double Lynden-Bell existence region}
The existence region of the double Lynden-Bell distributions in core's $({\hat{N}}_c,{\hat{E}}_c,\hat{\eta}_c)$ space, where $M_s$ is given by hand, is not dense.
More precisely, on the $({\hat{N}}_c,{\hat{E}}_c)$ and $({\hat{N}}_c,\hat{\eta}_c)$ planes for fixed $\hat{\eta}_c$ and $\hat{E}_c$, respectively, the off-shell existence region for the double Lynden-Bell distribution, which we will call the {\it{double Lynden-Bell region}}, is restricted to a thin, spindle-shaped region.\cite{KS}

The double Lynden-Bell region has the following two main structures.

First, on the $({\hat{N}}_c,{\hat{E}}_c)$ and $({\hat{N}}_c,\hat{\eta}_c)$ planes for fixed $\hat{\eta}_c$ and $\hat{E}_c$, respectively, it has two edges where the energy-distribution is the superposition of two Vlasov stationary water-bag distributions.
The boundaries connecting these edges represent states in which a part of the components has a Vlasov stationary water-bag distribution.
That is, at the edges and boundaries of the double Lynden-Bell region, the temperature of the corresponding component becomes zero (i.e., $\beta_a\to \infty$), so $\hat{f}_a(\varepsilon)$ reduces to $\hat{\eta}_a\Theta(\mu_a-\varepsilon)$.\cite{KS}

Second, on the $({\hat{N}}_c,{\hat{E}}_c)$ plane, the center of this spindle-shaped double Lynden-Bell region is the off-shell maximization point of the double Lynden-Bell entropy Eq.(\ref{eq:S}) for fixed $\hat{\eta}_c$, namely $\pa S/\pa \hat{N}_c=\pa S/\pa \hat{E}_c=0$ holds.
Using the results of the derivative of the double Lynden-Bell entropy by the macroscopic variables (see next subsection), it can be shown numerically that the corresponding energy-distribution satisfies ${\beta_c}/{\hat{\eta}_c}\sim {\beta_h}/{\hat{\eta}_h}$ and $\mu_c\sim \mu_h$ and thus is almost the one of a {{single}} Lynden-Bell.\cite{KS}

In the following, we consider the two-step energy water-bag distribution
\begin{eqnarray}
\hat{f}(\varepsilon)&=&\hat{\eta}_c\Theta(\varepsilon_c-\varepsilon)+\hat{\eta}_h\Theta(\varepsilon_h-\varepsilon)\;,\label{eq:fs}\\
\varepsilon(\theta,p,M)&=&\frac{p^2}{2}+1-M\cos\theta\;,\\
\Theta(\varepsilon)&=&\left\{\begin{array}{cc}0 &\varepsilon<0\\ 1&\varepsilon\ge0\end{array}\right.
\end{eqnarray}
that corresponds to the two edges of the double Lynden-Bell region.
By introducing the variable $\chi$, we rewrite this as
\begin{eqnarray}
\hat{f}(\varepsilon)=(1-\chi)\hat{\eta}\Theta(\varepsilon_c-\varepsilon)+\chi\hat{\eta}\Theta(\varepsilon_h-\varepsilon)\;.
\end{eqnarray}
We denote the Fermi energies of the halo and core by $\varepsilon_h$ and $\varepsilon_c$ and their diluted phase-space densities by $\eta_h$ and $\eta_c$, respectively.

For the moment, we consider the off-shell case with respect to the magnetization $M$ and determine the edge distributions by using the conservation laws of mass and energy.
Namely, by giving $M$ and $\chi$, $\varepsilon_c$ and $\varepsilon_h$ are determined numerically by the two constraints
\begin{eqnarray}
\int_0^{2\pi} d\theta \int_{-\infty}^\infty dp \hat{f}(\theta,p)&=&1\;,\\
\int_0^{2\pi} d\theta \int_{-\infty}^\infty dp \hat{f}(\theta,p)\biggl(\frac{p^2}{2}+\frac{1-M\cos \theta}{2}\biggr)&=&\hat{E}\;.
\end{eqnarray}
We give the energy per particle $\hat{E}$ a concrete number. 
These two constraints can be written as
\begin{eqnarray}
\int d\theta \int dp &=&1\;,\\
\int d\theta\int dp \biggl(\frac{p^2}{2}+\frac{1-M\cos \theta}{2}\biggr)&=&\hat{E}\;.
\end{eqnarray}
Here, the $\theta$--$p$ integrals are explicitly
\begin{eqnarray}
\int d\theta \int dp&=&4\hat{\eta}(1-\chi)\int_0^{\theta_c} d\theta \int_{0}^{p_c(\theta)}dp+4\hat{\eta}\chi\int_0^{\theta_h} d\theta\int_{0}^{p_h(\theta)}dp\;,\\
p_c(\theta)&=&\sqrt{2(\varepsilon_c-1+M\cos\theta)}\;,\\
p_h(\theta)&=&\sqrt{2(\varepsilon_h-1+M\cos \theta)}
\end{eqnarray}
and we perform them within the integral domain where the integrands are real-valued.
(This integral domain is simplified by using the even function property of the integrand.)

Without having to consider the meanings of the step functions, these integrals can be derived by the following way:
\begin{eqnarray}
\int d\theta \int dp\Theta(\varepsilon_c-\varepsilon)&=&2\int d\theta \int d\varepsilon \frac{\pa (\theta,p)}{\pa (\theta,\varepsilon)}\Theta (\varepsilon_c-\varepsilon)\\
&=&2\int d\theta \int_{\varepsilon_{min}}^{\varepsilon_c}d\varepsilon \frac{\pa p}{\pa \varepsilon}\\
&=&2\int d\theta p_c(\theta)
\end{eqnarray}
etc. The factor of $2$ arises because the transformation of the variable is two to one.

The classification of the integral domain is done by excluding the domain where the quantities in the square roots in $p_c(\theta)$ and $p_h(\theta)$ are negative-valued
\begin{equation}
\theta>\theta_a^0={{\arccos}}\biggl(\frac{1-\varepsilon_a}{M}\biggr)>0\;,\ \ a=c,h\;.
\end{equation}

From the calculations done above, the total mass conservation and total energy conservation are
\begin{eqnarray}
1&=&\hat{\eta}_h F(\varepsilon_h-1,M)+\hat{\eta}_cF(\varepsilon_c-1,M)\;,\\
\hat{E}&=&\hat{\eta}_h H(\varepsilon_h-1,M)+\hat{\eta}_cH(\varepsilon_c-1,M) +\frac{1-M(\hat{\eta}_h G(\varepsilon_h-1,M)+\hat{\eta}_cG(\varepsilon_c-1,M))}{2}\;,\end{eqnarray}
where the self-consistency condition on the magnetization is unlocked.

By solving these, we obtain the Fermi energies $\varepsilon_c$ and $\varepsilon_h$.

At the edges of these existence domains, the number of particles and the energy of the core are
\begin{eqnarray}
\hat{N}(M,\eta_c)&=&\hat{\eta}_c F(\varepsilon_c-1,M)\;,\label{eq:bc1}\\
\hat{E}(M,\eta_c)&=&\hat{\eta}_c H(\varepsilon_c-1,M)+\frac{1}{2}\hat{\eta}_c F(\varepsilon_c-1,M)- M\hat{\eta}_c \frac{G(\varepsilon_c-1,M)}{2}\;.\label{eq:bc2}
\end{eqnarray}

To give the boundary curves $\hat{E}_\pm(\hat{N})$ of the existence domain, first by giving $\hat{N}$, we solve Eq.(\ref{eq:bc1}) for two possible values of $\varepsilon_c$ (denoted by $\varepsilon_c^+$ and $\varepsilon_c^-$ in decreasing order) and then, substitute $\varepsilon_c^\pm$ into Eq.(\ref{eq:bc2}) to produce $\hat{E}_\pm$ for $\hat{N}$.

\subsubsection{On-shell entropy maximization}

In this subsection, we examine whether or not the resultant QSSs from the simulation complete the relaxation between the core and halo Lynden-Bell distributions, which is a weaker criterion than Lynden-Bell relaxation.
The relaxation criterion to be considered can be expressed as the {\it{on-shell}} maximization of the double Lynden-Bell entropy in Eq.(\ref{eq:S}) (here we note that $S=S^{(c)}+S^{(h)}$):
\begin{equation}
\biggl[\frac{\pa S}{\pa X_c}-\lambda\frac{\pa(M-M_s)}{\pa {X}_c}\biggr]\biggl|_{M=M_s}=0\;,\ \ X=N,E,\eta\label{eq:OSEM1}
\end{equation}
for Lagrange multiplier $\lambda$, where the stationary magnetization $M_s$ is fixed by hand.
Eq.(\ref{eq:OSEM1}) leads to
\begin{equation}
\frac{\pa S}{\pa N_c}\biggl/\frac{\pa M}{\pa N_c}=\frac{\pa S}{\pa E_c}\biggl/\frac{\pa M}{\pa E_c}=\frac{\pa S}{\pa \eta_c}\biggl/\frac{\pa M}{\pa \eta_c}\label{eq:OSEM}
\end{equation}
at $M=M_s$.
To clarify the on-shell maximization of the double Lynden-Bell entropy Eq.(\ref{eq:OSEM}), we need to calculate the {{off-shell}} derivatives of the Lynden-Bell entropy Eq.(\ref{eq:LB}) and the magnetization by the macro-variables ${N}_a$, ${{E}}_a$ and $\eta_a$.

In the following, first, we perform the calculations with respect to the Lynden-Bell entropy.
We focus on the Lynden-Bell component $f=f_c,f_h$ and use $S$ to denote its Lynden-Bell entropy {\it{from here till Eq.(\ref{eq:finish})}}.
We define the derivatives $\pa S/\pa N$, $\pa S/\pa E$ and $\pa S/\pa \eta$ as follows.

From the definition, $N$ and $E$ are functions of the independent variables $\beta$, $\mu$ and $\eta$ that determine the Lynden-Bell distribution $f$ (the magnetization is fixed).
By reversing these relations, the Lagrange multipliers $\beta$ and $\mu$ can be regarded as functions of the independent variables $N$, $E$ and $\eta$:
\begin{eqnarray}
\left\{\begin{array}{ccc}\beta&=&\beta(N_0,E_0,\eta)\\ 
\mu&=&\mu(N_0,E_0,\eta)\end{array}\right.
\end{eqnarray}
which are equivalent to
\begin{eqnarray}
\left\{\begin{array}{ccc}N(\beta,\mu,\eta)&=&N_0\\
E(\beta,\mu,\eta)&=&E_0\end{array}\right.
\end{eqnarray}
In the right hand sides of the above equations, $N_0$ and $E_0$ are just numbers.

 In the representation using $\beta$ and $\mu$, the Lynden-Bell entropy is defined by
\begin{eqnarray}
S^{(a)}(\beta,\mu)&=&(-1)\int_0^{2\pi}d\theta\int_{-\infty}^\infty dp\ s(\theta,p;\beta,\mu)\\
s&=&\varrho\ln \varrho+(1-\varrho)\ln(1-\varrho)\;,\ \ \varrho=\frac{f}{\eta}\;,\end{eqnarray}
where $f$ is each Lynden-Bell component.
In the representation using $N$, $E$ and $\eta$, the Lynden-Bell entropy is defined by
\begin{equation}
S^{(b)}(N,E,\eta)=S^{(a)}(\beta(N,E,\eta),\mu(N,E,\eta))\;.
\end{equation}
In the following, we calculate
\begin{equation}
\frac{\pa S^{(b)}}{\pa N}\biggl|_{E,\eta:{\rm{fixed}}}\;,\ \ \frac{\pa S^{(b)}}{\pa E}\biggl|_{N,\eta:{\rm{fixed}}}\;,\ \ \frac{\pa S^{(b)}}{\pa \eta}\biggl|_{N,E:{\rm{fixed}}}\;,
\end{equation}
where we omit the right upper index of $S$ that specifies the set of variables and let
\begin{equation}
\frac{\pa}{\pa \beta}=\frac{\pa}{\pa \beta}\biggl|_{\mu,\eta:{\rm{fixed}}}\;,\ \ \frac{\pa }{\pa \mu}=\frac{\pa }{\pa \mu}\biggl|_{\beta,\eta:{\rm{fixed}}}\;,\ \ \frac{\pa}{\pa N}=\frac{\pa}{\pa N}\biggl|_{{E},\eta:{\rm{fixed}}}\;,\ \ \frac{\pa}{\pa {E}}=\frac{\pa}{\pa {E}}\biggl|_{N,\eta:{\rm{fixed}}}\;.
\end{equation}

First of all, we calculate the derivatives of $S$ by the Lagrange multipliers $\beta$ and $\mu$.
The derivatives of the integrand $s(\theta,p;\beta,\mu)$ without the sign of $S$ are
\begin{eqnarray}
\frac{\pa s}{\pa \lambda}&=&\frac{\pa }{\pa \lambda}(\varrho \ln \varrho+(1-\varrho)\ln(1-\varrho))\\
&=&\frac{\pa \varrho}{\pa \lambda}(\ln \varrho+1)-\frac{\pa \varrho}{\pa \lambda}(\ln(1-\varrho)+1)\\
&=&\frac{\pa \varrho}{\pa \lambda}\biggl(\ln \frac{\varrho}{1-\varrho}\biggr)\\
&=&\frac{1}{\eta}\frac{\pa f}{\pa \lambda}\ln(\exp(-\beta(\varepsilon-\mu))\\
&=&\frac{(-\beta(\varepsilon-\mu))}{\eta}\frac{\pa f}{\pa \lambda}\;,\ \ \lambda=\beta,\mu\;.
\end{eqnarray}
Due to the off-shell assumption with respect to $M$,
\begin{eqnarray}
\varepsilon\frac{\pa f}{\pa \lambda}&=&\frac{\pa (\varepsilon f)}{\pa \lambda}-\frac{\pa \varepsilon}{\pa \lambda}f\\
&=&\frac{\pa (\varepsilon f)}{\pa \lambda}\;,\ \ \lambda=\beta,\mu
\end{eqnarray}
holds. Using this,
\begin{eqnarray}
\int_0^{2\pi} d\theta \int_{-\infty}^\infty dp \frac{\pa (\cdots)}{\pa \lambda}&=&\frac{\pa }{\pa \lambda}\int_0^{2\pi} d\theta \int_{-\infty}^\infty dp(\cdots)\;,\ \ \lambda=\beta,\mu
\end{eqnarray}
and
\begin{eqnarray}
\int_0^{2\pi}d\theta \int_{-\infty}^\infty dp (\varepsilon f)&=&K+2V \\
&=&E+V\;,
\end{eqnarray}
where we define
\begin{eqnarray}
K&=&\int_0^{2\pi}d\theta\int_{-\infty}^\infty dp\frac{p^2}{2}f(\theta,p)\;,\\
V&=&\int_0^{2\pi}d\theta\int_{-\infty}^\infty dp\frac{\Phi(\theta)}{2}f(\theta,p)\\
&=&N\biggl(\frac{1-M_fM}{2}\biggr)\;,\\
M_f&=&\frac{1}{N}\int_0^{2\pi}d\theta \int_{-\infty}^\infty dp\cos\theta f(\theta,p)\;, \label{eq:Mag1}
\end{eqnarray}
we obtain
\begin{eqnarray}
\frac{\pa S}{\pa \lambda}&=&(-1)\biggl(\frac{-\beta}{\eta}\frac{\pa (E+V)}{\pa \lambda}+\frac{\beta \mu}{\eta}\frac{\pa N}{\pa \lambda}\biggr)\\
&=&\frac{\beta}{\eta}\biggl(\frac{\pa (E+V)}{\pa \lambda}-\mu\frac{\pa N}{\pa \lambda}\biggr)\;,\ \ \lambda=\beta,\mu\;.\label{eq:Sc}
\end{eqnarray}
In the following, we use the formula
\begin{eqnarray}
\frac{\pa V}{\pa \lambda}&=&\frac{V}{N}\frac{\pa N}{\pa \lambda}-\frac{N}{2}\frac{\pa M_f}{\pa \lambda}M\;,\ \ \lambda=\beta,\mu\;.\label{eq:Vc}
\end{eqnarray}
Using $\pa S/\pa \eta=0\ (\beta,\mu:{\rm{fixed}})$ or $\pa \eta/\pa N=\pa \eta/\pa E=0$, we obtain
\begin{eqnarray}
\frac{\pa S}{\pa X}&=&\frac{\pa S}{\pa \beta}\frac{\pa \beta}{\pa X}+\frac{\pa S}{\pa \mu}\frac{\pa \mu}{\pa X}\;,\ \ X=N,E\;.
\end{eqnarray}
We define two kinds of two-component vectors for macro-variable $X$ and Lagrange multiplier $\lambda$ by
\begin{equation}
|X)_\lambda=\frac{\pa \lambda}{\pa X}\;,\ \ \ _\lambda(X|=\frac{\pa X}{\pa \lambda}\;.
\end{equation}
Then,
\begin{eqnarray}
(X|Y)&=&\frac{\pa X}{\pa Y}\\
&=&\delta_{XY}\;,\ \ X,Y=E,N
\end{eqnarray}
holds.

We now calculate the derivative of the entropy with respect to the macro-variables $N$ and $E$:
\begin{eqnarray}
\frac{\pa S}{\pa N}&=&\frac{\pa S}{\pa \beta}\frac{\pa \beta}{\pa N}+\frac{\pa S}{\pa \mu}\frac{\pa \mu}{\pa N}\\
&=&\frac{\beta}{\eta}\biggl\{\biggl(\frac{\pa E}{\pa \beta}+\frac{\pa V}{\pa \beta}-\mu\frac{\pa N}{\pa \beta}\biggr)\frac{\pa \beta}{\pa N}+\biggl(\frac{\pa E}{\pa\mu }+\frac{\pa V}{\pa \mu}-\mu\frac{\pa N}{\pa \mu}\biggr)\frac{\pa \mu}{\pa N}\biggr\}\\
&=&\frac{\beta}{\eta}\biggl\{\biggl({\frac{V}{N}\frac{\pa N}{\pa \beta}}-\frac{N}{2}\frac{\pa M_f}{\pa \beta}M-\mu\frac{\pa N}{\pa \beta}\biggr)\frac{\pa \beta}{\pa N}\nonumber\\&&+\biggl(\frac{V}{N}\frac{\pa N}{\pa \mu}-\frac{N}{2}\frac{\pa M_f}{\pa \mu}M-\mu\frac{\pa N}{\pa \mu}\biggr)\frac{\pa \mu}{\pa N}+\biggl(\frac{\pa E}{\pa \beta}\frac{\pa \beta}{\pa N}+\frac{\pa E}{\pa \mu}\frac{\pa \mu}{\pa N}\biggr)\biggr\}\\
&=&\frac{\beta}{\eta}\biggl\{(N|N)\biggl(\frac{V}{N}-\mu\biggr)
+\frac{N}{2}\biggl(-\frac{\pa M_f}{\pa \beta}M\biggr)\frac{\pa \beta}{\pa N}+\frac{N}{2}\biggl(-\frac{\pa M_f}{\pa \mu}M\biggr)\frac{\pa \mu}{\pa N}+(E|N)
\biggr\}\\
&=&\frac{\beta}{\eta}\biggl(\frac{V}{N}-\mu-\frac{N}{2}\frac{\pa M_f}{\pa N}M\biggr)
\end{eqnarray}
and
\begin{eqnarray}
\frac{\pa S}{\pa E}&=&\frac{\pa S}{\pa \beta}\frac{\pa \beta}{\pa E}+\frac{\pa S}{\pa \mu}\frac{\pa \mu}{\pa E}\\
&=&\frac{\beta}{\eta}\biggl\{\biggl(\frac{\pa E}{\pa \beta}+\frac{\pa V}{\pa \beta}-\mu\frac{\pa N}{\pa \beta}\biggr)\frac{\pa \beta}{\pa E}+\biggl(\frac{\pa E}{\pa\mu }+\frac{\pa V}{\pa \mu}-\mu\frac{\pa N}{\pa \mu}\biggr)\frac{\pa \mu}{\pa E}\biggr\}\\
&=&\frac{\beta}{\eta}\biggl\{\biggl(\frac{V}{N}\frac{\pa N}{\pa \beta}-\frac{N}{2}\frac{\pa M_f}{\pa \beta}M-\mu\frac{\pa N}{\pa \beta}\biggr)\frac{\pa \beta}{\pa E}\nonumber\\&&
+\biggl(\frac{V}{N}\frac{\pa N}{\pa \mu}-\frac{N}{2}\frac{\pa M_f}{\pa \mu}M-\mu\frac{\pa N}{\pa \mu}\biggr)\frac{\pa \mu}{\pa E}+\biggl(\frac{\pa E}{\pa \beta}\frac{\pa \beta}{\pa E}+\frac{\pa E}{\pa \mu}\frac{\pa \mu}{\pa E}\biggr)\biggr\}\\
&=&\frac{\beta}{\eta}\biggl\{(N|E)\biggl(\frac{V}{N}-\mu\biggr)
+\frac{N}{2}\biggl(-\frac{\pa M_f}{\pa \beta}M\biggr)\frac{\pa \beta}{\pa E}+\frac{N}{2}\biggl(-\frac{\pa M_f}{\pa \mu}M\biggr)\frac{\pa \mu}{\pa E}+(E|E)\biggr\}\\
&=&\frac{\beta}{\eta}\biggl(1-\frac{N}{2}\frac{\pa M_f}{\pa E}M\biggr)\;.
\end{eqnarray}

Next, we calculate $\pa S/\pa \eta$ ($N,{E}$:fixed).
We note the relationships
\begin{eqnarray}
\{N,{E},\eta\}&\Rightarrow& \{\beta ,\mu\}\;,\\
\{\beta,\mu,\eta\}&\Rightarrow& \{N,{E}\}\;,\\
\{\beta,\mu\}&\Rightarrow& S\;,
\end{eqnarray}
and that in general
\begin{equation}
\biggl(\frac{\pa S}{\pa \eta}\biggr)_{N,E}\neq 0\;.
\end{equation}
Due to the chain rule
\begin{eqnarray}
0&=&\biggl(\frac{\pa S}{\pa \eta}\biggr)_{\beta,\mu}\\
&=&\biggl(\frac{\pa S}{\pa N}\biggr)_{E,\eta}\biggl(\frac{\pa N}{\pa \eta}\biggr)_{\beta,\mu}+\biggl(\frac{\pa S}{\pa {E}}\biggr)_{N,\eta}\biggl(\frac{\pa {E}}{\pa \eta}\biggr)_{\beta,\mu}+\biggl(\frac{\pa S}{\pa \eta}\biggr)_{N,E}\biggl(\frac{\pa\eta}{\pa \eta}\biggr)_{\beta,\mu}\;,
\end{eqnarray}
we obtain the formula
\begin{equation}
\biggl(\frac{\pa S}{\pa \eta}\biggr)_{N,E}=(-1)\biggl(\biggl(\frac{\pa S}{\pa N}\biggr)_{E,\eta}\biggl(\frac{\pa N}{\pa \eta}\biggr)_{\beta,\mu}+\biggl(\frac{\pa S}{\pa {E}}\biggr)_{N,\eta}\biggl(\frac{\pa {E}}{\pa \eta}\biggr)_{\beta,\mu}\biggr)\;.
\end{equation}
Here,
\begin{equation}
\biggl(\frac{\pa X}{\pa \eta}\biggr)_{\beta,\mu}=\frac{X}{\eta}\;,\ \ X=N,E\;.
\end{equation}

To summarize, we have the following results for the derivatives of entropy:
\begin{eqnarray}
\frac{\pa S}{\pa N}&=&\frac{\beta}{\eta} \biggl(-\mu+\frac{V}{ N}-\frac{N}{2}\frac{\pa M_f}{\pa N}M\biggr)\;,\\
\frac{\pa S}{\pa E}&=&\frac{\beta}{\eta}\biggl(1-\frac{N}{2}\frac{\pa M_f}{\pa E}M\biggr)\;,\\
\biggl(\frac{\pa S}{\pa \eta}\biggr)_{N,E}&=&-\biggl(\frac{\pa S}{\pa N}\frac{ N}{\eta}+\frac{\pa S}{\pa {E}}\frac{{E}}{\eta}\biggr)\;.
\end{eqnarray}

Next, we calculate the derivatives of the magnetization by the macro variables $N$ and $E$.
Since
\begin{eqnarray}
\frac{\pa NM_f}{\pa N}&=&M_f+N\frac{\pa M_f}{\pa N}\;,\\
\frac{\pa NM_f}{\pa E}&=&N\frac{\pa M_f}{\pa E}\;,\\
\biggl(\frac{\pa NM_f}{\pa \eta}\biggr)_{N,E}&=&-\biggl(\frac{\pa NM_f}{\pa N}\frac{N}{\eta}+\frac{\pa NM_f}{\pa E}\frac{E}{\eta}-\frac{NM_f}{\eta}\biggr)
\end{eqnarray}
and $M=\hat{N}_cM_c+\hat{N}_hM_h$ hold, it is sufficient to calculate the derivatives of $M_f$ (Eq.(\ref{eq:Mag1})) by $N$ and $E$.
Due to
\begin{eqnarray}
\frac{\pa f}{\pa N}&=&\frac{1}{\det}\biggl(\frac{\pa f}{\pa \beta}\frac{\pa {E}}{\pa \mu}-\frac{\pa f}{\pa \mu}\frac{\pa {E}}{\pa \beta}\biggr)\;,\label{eq:fN}\\
\frac{\pa f}{\pa {E}}&=&\frac{1}{\det}\biggl(-\frac{\pa f}{\pa \beta}\frac{\pa N}{\pa \mu}+\frac{\pa f}{\pa \mu}\frac{\pa N}{\pa \beta}\biggr)\;,\label{eq:fE}
\end{eqnarray}
where we set
\begin{equation}
\det=\frac{\pa {E}}{\pa \mu}\frac{\pa N}{\pa \beta}-\frac{\pa {E}}{\pa \beta}\frac{\pa N}{\pa \mu}\;,
\end{equation}
we obtain
\begin{eqnarray}
\frac{\pa M_f}{\pa N}&=&\frac{\pa}{\pa N}\biggl(\frac{1}{N}\int_0^{2\pi}d\theta \int_{-\infty}^\infty dp\cos\theta f(\theta,p)\biggr) \\
&=&-\frac{1}{N^2}\int_0^{2\pi} d\theta \int_{-\infty}^{\infty} dp \cos \theta f+\frac{1}{N}\int_0^{2\pi} d\theta \int_{-\infty}^{\infty} dp \cos \theta 
\frac{\pa f}{\pa N}
\\
&=&-\frac{1}{N^2}\int_0^{2\pi} d\theta \int_{-\infty}^{\infty} dp \cos \theta f+\frac{1}{N}\int_0^{2\pi} d\theta \int_{-\infty}^{\infty} dp \cos \theta
\frac{1}{\det}\biggl(\frac{\pa f}{\pa \beta}\frac{\pa E}{\pa \mu}-\frac{\pa f}{\pa \mu}\frac{\pa U}{\pa \beta}\biggr)
\end{eqnarray}
and
\begin{eqnarray}
\frac{\pa M_f}{\pa E}&=&\frac{\pa}{\pa E}\biggl(\frac{1}{N}\int_0^{2\pi}d\theta \int_{-\infty}^\infty dp\cos\theta f(\theta,p)\biggr) \\
&=&\frac{1}{N}\int_0^{2\pi} d\theta \int_{-\infty}^{\infty} dp \cos \theta
\frac{\pa f}{\pa E}
\\
&=&\frac{1}{N}\int_0^{2\pi} d\theta \int_{-\infty}^{\infty} dp \cos \theta
\frac{1}{\det}\biggl(-\frac{\pa f}{\pa \beta}\frac{\pa N}{\pa \mu}+\frac{\pa f}{\pa \mu}\frac{\pa N}{\pa \beta}\biggr)
\;.\label{eq:finish}
\end{eqnarray}
By these formulae, we have clarified Eq.(\ref{eq:OSEM}).

The on-shell entropy maximization criterion Eq.(\ref{eq:OSEM}) can be expressed geometrically as the tangency of contour surfaces of $S$ and $M$ in $({\hat{N}}_c,{\hat{E}}_c,\hat{\eta}_c)$ space.
For $\hat{\eta}=0.15$, the case $M_0=0.72$ fulfills this criterion for $M_s=0.6345$, which is within the oscillation range of the stationary magnetization (see Figs. 14 and 15).\cite{KS}
\begin{figure}[htbp]
\begin{center}
\includegraphics[width=0.4\hsize]{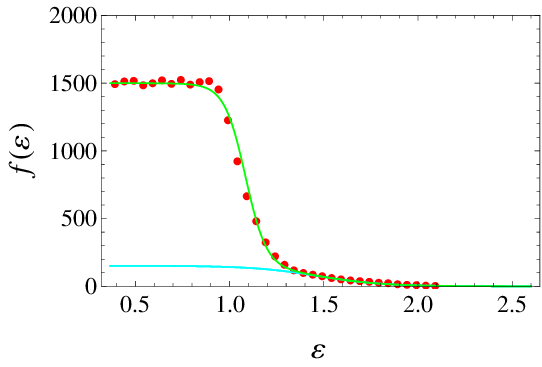}\ \ \ \ \includegraphics[width=0.4\hsize]{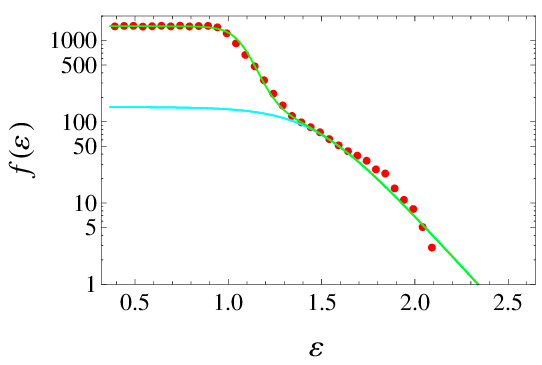}
\caption{These figures show the $M_0=0.72$ theoretical result for $f(\varepsilon)$ obtained by solving the on-shell entropy maximization condition for $M_s=0.6345$ (the full and halo part are drawn as green and cyan curves, respectively), and the simulation resultant $f(\varepsilon)$ averaged over $20$ runs (red dots).}
\end{center}
\end{figure}

\begin{figure}[htbp]
\begin{center}
\includegraphics[width=0.39626\hsize]{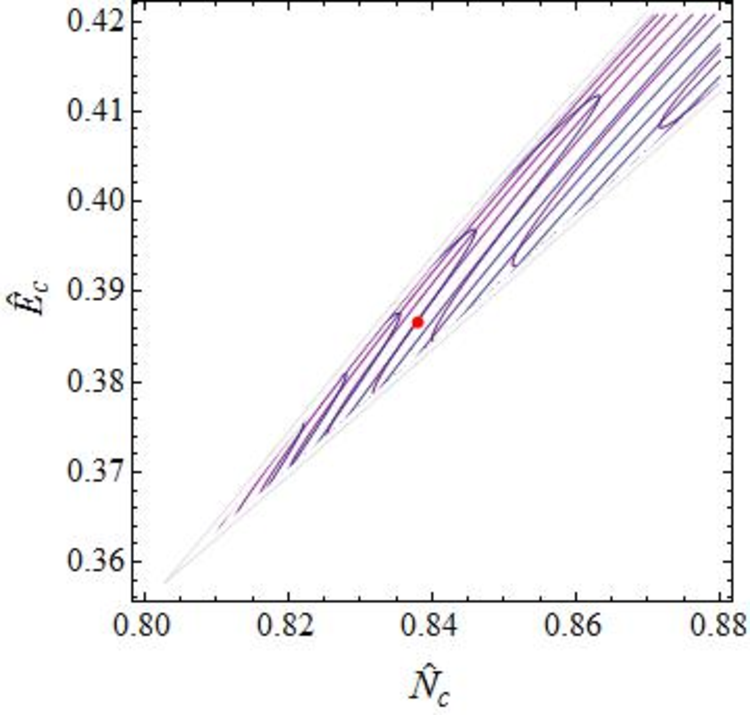}\ \ \ \ \includegraphics[width=0.40374\hsize]{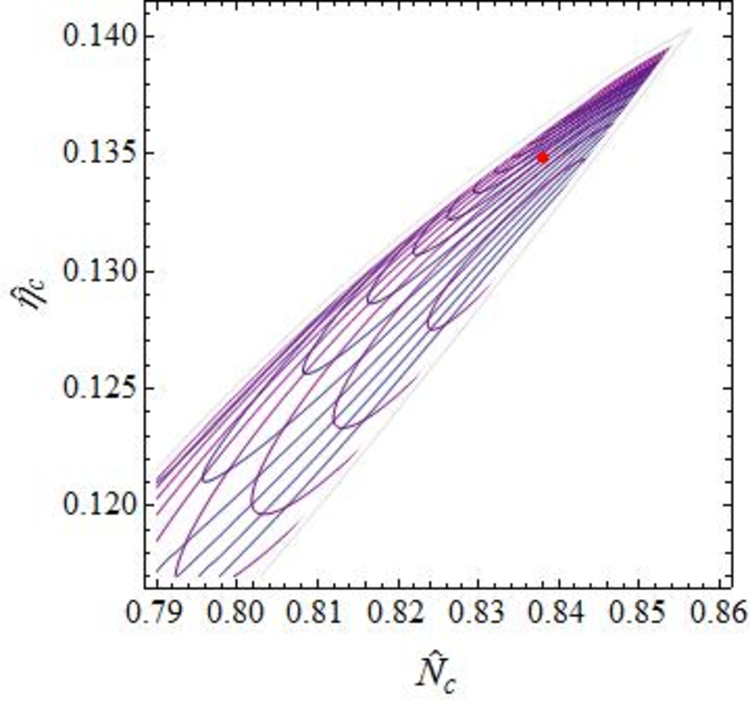}
\caption{These figures show $M_0=0.72$ contour curves of $S$ (purple) and $M$ (cyan) on the $({\hat{N}}_c,{\hat{E}}_c)$ and $({\hat{N}}_c,\hat{\eta}_c)$ planes for $M_s=0.6345$ at the simulation resultants $\hat{\eta}_c$ and ${\hat{E}}_c$, respectively. The red dot is the simulation resultant point.}
\end{center}
\end{figure}

However, in other values of $M_0$, the simulation results do not fulfill the on-shell entropy maximization criterion and are regarded as cases of incomplete relaxation of Eq.(\ref{eq:S}).
This can be seen from the fact that the structure of contours of $M$ on $({\hat{N}}_c,{\hat{E}}_c)$ plane, that is, slices of contour surfaces of $M$ at the simulation resultant $\hat{\eta}_c$, changes from the {{convex curves}} for $M_0=0.72$ to almost {{straight lines}} for other values of $M_0$, whose family has rightward or leftward monotonous growth directions of values on this plane (both cases can be realized), while the contour of $S$ is always convex.
In these cases, the tangent point between contours of $S$ and $M$ on this plane recedes, and the simulation results do not fulfil the on-shell entropy maximization criterion and are regarded in the case of incomplete relaxation of Eq.(\ref{eq:S}).\cite{KS}

\section{Conclusion}
In this section, we briefly summarize the results appearing in this review and discuss the open issues for the double Lynden-Bell scenario.

Based on preliminary results given in Sections 2 and 3, in Section 4, we studied QSSs with the core-halo structure in the HMF model.
In the latter part of Section 4, based on the author's original idea of describing the QSS form as a superposition of the core and the halo, we have systematically studied the core-halo structure of the QSSs starting from unsteady ($M_0\neq 0$ or Vlasov unstable; $\hat{E}\le 7/12$\cite{HMF}) initial rectangular water-bag distributions with $\hat{\eta}=0.15$ by means of $N$-body simulation. We have also corroborated the double Lynden-Bell scenario, in which the QSS distribution functions result in the superposition of two independent Lynden-Bell distribution functions, at least at low energies per particle.

{{Here, we briefly review the mechanism of the double Lynden-Bell scenario.
When we admit the bifurcation of a QSS distribution into a superposition, the Lynden-Bell distributions of the core and the halo arise from the two facts observed in the molecular dynamics simulation: in the phase mixing process the dynamical relaxation Eq.(\ref{eq:dr}) between the core and the halo plays the role of the incompressibility constraint on the phase-space elements of the core and the halo; and the halo's phase mixing progresses significantly due to its high-energy extension (see Figs. 7 and 8).}}

{{In comparison with the previous research, the author believes that the double Lynden-Bell scenario substantially improves our understanding of core-halo QSSs.
The grounds for this assessment are as follows.
First of all, from the standpoint of the double Lynden-Bell scenario, although the Pakter-Levin ansatz in Eq.(\ref{eq:fsPL}) captures the essence of the core-halo QSS distribution function, its core and halo have no relationship with the Lynden-Bell statistics, and it can be applied to low-temperature cases of our core and halo only.
Moreover, in the previous research, the reason for the degeneration of the halo in Eq.(\ref{eq:fsPL}) was unclear.
From the new viewpoint, this degeneration is an obvious consequence of the Lynden-Bell distribution of the halo.
Secondly, while in the previous research the existence of ergodicity breaking had been stressed for core-halo QSSs, ergodicity is independently maintained for the halo and the core in the double Lynden-Bell sense.
This is a significant conceptual advance arising from the double Lynden-Bell scenario.}}

In the double Lynden-Bell scenario, we have also examined the completeness of the collisionless relaxation by considering two entropies.
By using the Lynden-Bell entropy, we found that the systems being considered do not reach equilibrium and for higher total energy the degree of incompleteness of the relaxation\cite{Chavanis1,CSR,Chavanis2} increases.
By using the double Lynden-Bell entropy, for $\hat{\eta}=0.15$, in the case of $M_0=0.72$, the system completes the relaxation (i.e., can be determined by statistical mechanical methods); however, for other values of $M_0$ this does not happen.

Next, we suggest some issues of the present double Lynden-Bell scenario.

{{The main issue is that the seven parameters in the double Lynden-Bell distribution cannot be fully determined at present.
Here, note that the zero-temperature double Lynden-Bell distribution and the Pakter-Levin ansatz coincide.
Since the Pakter-Levin ansatz has no fitting parameter, the parameters $\beta_1$ and $\beta_2$ that describe the resolution of degeneracy should be regarded as the extra parameters in the double Lynden-Bell distribution from the aspect of the theoretical prediction.
Regarding this main issue, however, the author presumes that the present result  (see Sec. 4.2.6) may be the best one obtainable by a purely statistical mechanical approach, due to the following two facts.
Firstly, to determine the stationary magnetization of a system, we need to rely on the kinetic approach.
Secondly, as just mentioned, the results show that the on-shell double Lynden-Bell entropy maximization holds only in special cases, and other cases are regarded as incomplete relaxation.
So, we need to rely on kinetic theory beyond the statistical approach.
However, we note that using these kinetic approaches seem to be very difficult with the techniques that have been invented so far.}}

Besides this main issue, the double Lynden-Bell scenario has four other significant open issues.
First, we need to apply the double Lynden-Bell scenario to unsteady systems at higher energies per particle and understand the limits of its application.
Second, since the on-shell entropy maximization works in only special cases, an {\it{a priori}} measure of the deviation from complete relaxation between the core and the halo Lynden-Bell distributions needs to be found.
Related to this issue, we have shown that an {\it{a priori}} measure of the deviation from complete single Lynden-Bell relaxation is given by the residual energy of the system.
Third, the long-term evolution of the system after the double Lynden-Bell QSS until the Boltzmann-Gibbs equilibrium is reached needs to be studied.
Finally, to understand more deeply the reason why the system bifurcates into a superposition of the core and the halo is a fundamental issue.

\bigskip

{\it{Acknowledgements.---}}
{{The author wishes to thank Professor Masa-aki Sakagami for his collaboration in our original work and Professor Takayuki Tatekawa for providing the Fortran code for the HMF simulation.
}}

\begin{appendix}
\section{Elliptic Integrals}
In this section, we calculate the elliptic integrals appearing in Sections 4.2.4 and 4.2.5.

\subsection{Calculation of $\int_0^x d\theta\sqrt{a+b\cos \theta}$}

We define the elliptic integrals:
\begin{equation}
E(x,k)=\int_0^x\sqrt{1-k^2\sin^2\theta}d\theta\;,\ \ F(x,k)=\int_0^x\frac{1}{\sqrt{1-k^2\sin^2\theta}}d\theta\;.
\end{equation}

Then,
\begin{eqnarray}
\int^x_0 d\theta\sqrt{a+b\cos \theta}&=&\int_0^{\frac{x}{2}} d(2\theta) \sqrt{a+b\cos 2\theta}\\
&=&2\int_0^{\frac{x}{2}} d\theta \sqrt{a+b(\cos^2\theta-\sin^2\theta)}\\
&=&2\int_0^{\frac{x}{2}} d\theta \sqrt{(a+b)-2b\sin^2\theta}\\
&=&2\int_0^{\frac{x}{2}} d\theta\sqrt{a+b}\sqrt{1-\frac{2b}{a+b}\sin^2\theta}\\
&=&2\sqrt{a+b}E\biggl(\frac{x}{2},\sqrt{\frac{2b}{a+b}}\biggr)\;.
\end{eqnarray}

Furthermore,
\begin{eqnarray}
\int_0^xd\theta \frac{1}{\sqrt{a+b\cos \theta}}&=&\frac{2}{\sqrt{a+b}}\int_0^{\frac{x}{2}}\frac{d\theta}{\sqrt{1-\frac{2b}{a+b}\sin^2\theta}}\\
&=&\frac{2}{\sqrt{a+b}}F\biggl(\frac{x}{2},\sqrt{\frac{2b}{a+b}}\biggr)\;.
\end{eqnarray}

\subsection{Calculation of $\int_0^x d\theta \sqrt{a+b\cos\theta}\cos \theta$}

To calculate this integral, we rewrite $\sqrt{a+b\cos \theta}\cos \theta$ as the sum of terms which can be integrated elliptically.

First, we consider the following identities
\begin{eqnarray}
\frac{b+a\cos \theta}{\sqrt{a+b\cos \theta}}&=&\frac{b\sin^2\theta +(a+b\cos\theta)\cos \theta}{\sqrt{a+b\cos \theta}}\nonumber\\
&=&\frac{b\sin^2\theta}{\sqrt{a+b\cos\theta}}+\sqrt{a+b\cos\theta}\cos \theta\;,\label{eq:1}\\
\frac{b+a\cos \theta}{\sqrt{a+b\cos \theta}}&=&\alpha \sqrt{a+b\cos \theta}+\frac{\beta}{\sqrt{a+b\cos \theta}}\ \ \biggl(\alpha=\frac{a}{b}\;,\ \ \beta =\frac{b^2-a^2}{b}\biggr)\nonumber\\
&=&\frac{a}{b}\sqrt{a+b\cos\theta}-\frac{a^2-b^2}{b}\frac{1}{\sqrt{a+b\cos\theta}}\;.\label{eq:2}
\end{eqnarray}

Then, as the expressions on the right-hand sides of Eq.(\ref{eq:1}) and Eq.(\ref{eq:2}) are equal, we obtain
\begin{eqnarray}
\sqrt{a+b\cos\theta}\cos \theta=\frac{a}{b}\sqrt{a+b\cos \theta}-\frac{a^2-b^2}{b}\frac{1}{\sqrt{a+b\cos \theta}}-\frac{b\sin^2\theta}{\sqrt{a+b\cos\theta}}\;.
\end{eqnarray}

Since
\begin{equation}
-\frac{b\sin^2\theta}{\sqrt{a+b\cos \theta}}=2(\sqrt{a+b\cos \theta})^\prime \sin \theta\;,
\end{equation}
the integral to be calculated is
\begin{eqnarray}
I&\equiv&\int_0^x d\theta \sqrt{a+b\cos\theta}\cos \theta\\
&=&\frac{a}{b}\int_0^x d\theta \sqrt{a+b\cos \theta}-\frac{a^2-b^2}{b}\int_0^x d\theta \frac{1}{\sqrt{a+b\cos \theta}}+2\int_0^x d\theta(\sqrt{a+b\cos \theta})^\prime \sin \theta\;.
\end{eqnarray}
From this, and by noting
\begin{equation}
2\int_0^x d\theta (\sqrt{a+b\cos \theta })^\prime \sin \theta=2\sqrt{a+b\cos x}\sin x 
-2\int_0^x d\theta \sqrt{a+b\cos \theta}\cos \theta
\;,
\end{equation}
we obtain
\begin{eqnarray}
3I&=&\frac{a}{b}\int_0^x d\theta \sqrt{a+b\cos \theta }-\frac{a^2-b^2}{b}\int_0^x d\theta \frac{1}{\sqrt{a+b\cos \theta}}+2\sqrt{a+b\cos x }\sin x\;.
\end{eqnarray}

Namely,
\begin{eqnarray}
I&=&\frac{a}{3b}\int_0^x d\theta \sqrt{a+b\cos \theta}-\frac{a^2-b^2}{3b}\int_0^x d\theta \frac{1}{\sqrt{a+b\cos \theta}}+\frac{2}{3}\sqrt{a+b\cos x}\sin x\\
&=&\frac{a}{3b}2\sqrt{a+b}E\biggl(\frac{x}{2},\sqrt{\frac{2b}{a+b}}\biggr) -\frac{a^2-b^2}{b\sqrt{a+b}}\frac{2}{3}F\biggl(\frac{x}{2},\sqrt{\frac{2b}{a+b}}\biggr)+\frac{2}{3}\sqrt{a+b\cos x}\sin x\;.
\end{eqnarray}

The next integral is also calculated using the above results
\begin{eqnarray}
\int_0^x d\theta (\sqrt{a+b\cos \theta})^3&=&a\int_0^x d\theta \sqrt{a+b\cos \theta}+b\int_0^x d\theta \sqrt{a+b\cos \theta}\cos \theta \\
&=&2a\sqrt{a+b}E\biggl(\frac{x}{2},\sqrt{\frac{2b}{a+b}}\biggr)+b\biggl(\frac{a}{3b}2\sqrt{a+b}E\biggl(\frac{x}{2},\sqrt{\frac{2b}{a+b}}\biggr)\nonumber\\&& -\frac{a^2-b^2}{b\sqrt{a+b}}\frac{2}{3}F\biggl(\frac{x}{2},\sqrt{\frac{2b}{a+b}}\biggr)+\frac{2}{3}\sqrt{a+b\cos x}\sin x\biggr)\\
&=&\frac{8}{3}a\sqrt{a+b}E\biggl(\frac{x}{2},\sqrt{\frac{2b}{a+b}}\biggr)-\frac{a^2-b^2}{\sqrt{a+b}}\frac{2}{3}F\biggl(\frac{x}{2},\sqrt{\frac{2b}{a+b}}\biggr)\nonumber\\
&&+\frac{2b}{3}\sqrt{a+b\cos x}\sin x\;.
\end{eqnarray}

When $a>b$, $\cos^{-1}(-a/b)=\pi+\cos^{-1}(a/b)$. So, by setting $\cos^{-1}(a/b)$ to $ic$, we obtain
\begin{eqnarray}
&&{\rm{Re}}\biggl(\int_0^{\cos^{-1}-\frac{a}{b}}\sqrt{a+b\cos \theta}d\theta\biggl)\nonumber\\&&={\rm{Re}}\biggl(\int_0^{\pi}\sqrt{a+b\cos \theta}d\theta+\int_{\pi}^{\pi+\cos^{-1}(a/b)}\sqrt{a+b\cos \theta}d\theta\biggr)\\
&&={\rm{Re}}\biggl(\int_0^{\pi}\sqrt{a+b\cos \theta}d\theta+\int_{0}^{ic}\sqrt{a-b\cos \theta}d\theta\biggr)\\
&&={\rm{Re}}\biggl(\int_0^{\pi}\sqrt{a+b\cos \theta}d\theta+i\int_{0}^{c}\sqrt{a-b\cosh \theta^\prime}d\theta^\prime\biggr)\ \ (i\theta^\prime=\theta)\\
&&=\int_0^\pi \sqrt{a+b\cos \theta}d\theta\;.
\end{eqnarray}

For $a>b$, we also have
\begin{eqnarray}
{\rm{Re}}\biggl(\int_0^{\cos^{-1}-\frac{a}{b}}\frac{1}{\sqrt{a+b\cos \theta}}d\theta\biggr)&=&\int_0^\pi \frac{1}{\sqrt{a+b\cos \theta}}d\theta\;.
\end{eqnarray}
\section{Macroscopic Quantities of the HMF Lynden-Bell Distribution}

\subsection{Results}
In this section, we calculate the macroscopic quantities of the Lynden-Bell distribution\cite{LC} in the HMF model.

We consider the Lynden-Bell distribution
\begin{equation}
f(\varepsilon)=\frac{\eta}{\exp(\beta(\varepsilon-\mu))+1}\;.
\end{equation}

Here, the one-particle energy function is
\begin{equation}
\varepsilon=\frac{p^2}{2}+\Phi(\theta)\;,\ \ \Phi(\theta)=1-M\cos \theta\;.
\end{equation}

We introduce the complete Fermi-Dirac function\cite{FD}
\begin{equation}
F_n(x)=\int_0^\infty \frac{t^n}{\exp(t-x)+1}dt\;.
\end{equation}
By using it, we calculate the macroscopic quantities of the Lynden-Bell distribution, that is, the number of particles $N$ and the total kinetic energy $K$.

For the complete Fermi-Dirac function, the relation
\begin{equation}
\frac{d}{dx}F_n(x)=nF_{n-1}(x)
\end{equation}
can be shown directly by differentiation:
\begin{eqnarray}
\frac{d}{dx}F_n(x)&=&\int_0^\infty t^n\frac{\pa}{\pa x}\biggl(\frac{1}{\exp(t-x)+1}\biggr){dt}\\
&=&-\int_0^\infty t^n\frac{\pa}{\pa t}\biggl(\frac{1}{\exp(t-x)+1}\biggr)dt\\
&=&-
\biggl(t^n\frac{1}{\exp(t-x)+1}\biggr)\biggl|_0^\infty
+\int_0^\infty \frac{nt^{n-1}}{\exp(t-x)+1}dt\\
&=&nF_{n-1}(x)\;.
\end{eqnarray}

The change of the measure from the momentum integral $\int d p(\cdots)$ to the one-particle energy integral $\int d\varepsilon (\cdots)$ is
\begin{eqnarray}
dp&=&\frac{dp}{d\varepsilon}d\varepsilon\\
&=&\frac{1}{\frac{d\varepsilon}{dp}}d\varepsilon\\
&=&\frac{1}{p}d\varepsilon\\
&=&\frac{1}{\sqrt{2(\varepsilon-\Phi)}}d\varepsilon\;.
\end{eqnarray}

First, we calculate the number density of the particle $\varrho(\theta)$ and the kinetic energy density ${\cal{K}}(\theta)$:
\begin{eqnarray}
\varrho(\Phi(\theta))&=&\int_{-\infty}^{\infty}f(\varepsilon)
dp
\\
&=&2\int_{\Phi}^\infty f(\varepsilon)
\frac{d\varepsilon}{\sqrt{2(\varepsilon-\Phi)}}
\\
&=&\sqrt{2}\int_0^\infty f(\varepsilon+\Phi)\frac{d\varepsilon}{\sqrt{\varepsilon}}\\
&=&\sqrt{2}\eta \int_0^\infty \frac{1}{\exp(\beta((\varepsilon+\Phi)-\mu))+1}\frac{d\varepsilon }{\sqrt{\varepsilon}}\\
&=&\sqrt{\frac{2}{\beta}}\eta \int_0^\infty\frac{1}{\exp((\beta \varepsilon)-(\beta(\mu-\Phi)))+1}\frac{d(\beta \varepsilon)}{\sqrt{\beta \varepsilon}}\\
&=&\sqrt{\frac{2}{\beta}}\eta F_{-1/2}(\beta(\mu-\Phi))\end{eqnarray}
and
\begin{eqnarray}
{\cal{K}}(\Phi(\theta))&=&\int_{-\infty}^\infty f(\varepsilon)\frac{p^2}{2}
dp
\\
&=&2\int_{\Phi}^\infty f(\varepsilon)(\varepsilon-\Phi)
\frac{d\varepsilon}{\sqrt{2(\varepsilon-\Phi)}}
\\
&=&\sqrt{2}\int_{\Phi}^\infty f(\varepsilon)\sqrt{(\varepsilon-\Phi)}{d\varepsilon}\\
&=&\sqrt{2}\int_0^\infty f(\varepsilon+\Phi)\sqrt{\varepsilon}d\varepsilon \\
&=&\sqrt{2}\eta \int_0^\infty \frac{1}{\exp(\beta((\varepsilon+\Phi)-\mu))+1}\sqrt{\varepsilon}d\varepsilon \\
&=&\frac{1}{\beta}\sqrt{\frac{2}{\beta}}\eta \int_0^\infty\frac{1}{\exp((\beta \varepsilon)-(\beta(\mu-\Phi)))+1}\sqrt{\beta \varepsilon}d(\beta \varepsilon)\\
&=&\frac{1}{\beta}\sqrt{\frac{2}{\beta}}\eta F_{1/2}(\beta(\mu-\Phi))\;.
\end{eqnarray}

Here, the potential energy function is
\begin{equation}
\Phi(\theta)=1-M\cos \theta\;.
\end{equation}
So, the change of the integral measure from the position integral $\int d\theta (\cdots)$ to the potential energy integral $\int d\Phi(\cdots)$ is
\begin{eqnarray}
\frac{d\theta}{d\Phi}&=&\frac{1}{\frac{d\Phi}{d\theta}}\\
&=&\frac{1}{M\sin \theta}\\
&=&\left\{\begin{array}{c}\frac{1}{\sqrt{M^2-(\Phi-1)^2}}\ \ (0\le \theta\le \pi)\\
-\frac{1}{\sqrt{M^2-(\Phi-1)^2}}\ \ (\pi \le \theta \le 2\pi)
\end{array}\right.
\end{eqnarray}

In the position integral $\int d\theta(\cdots)$, the complete Fermi-Dirac function includes a cosine of the variable, but by changing to a potential energy integral $\int d\Phi (\cdots)$ we can remove the cosine.

Here we note that the correspondence is two to one when we change the integral variable from $\theta$ to $\Phi$.
So, the number of particles $N$ and the kinetic energy $K$ are 
\begin{eqnarray}
N&=&\int_0^{2\pi}\varrho d\theta\\
&=&\int_0^\pi \varrho d\theta +\int_\pi^{2\pi}\varrho d\theta \\
&=&\eta\sqrt{\frac{2}{\beta}}\biggl(\int_{1-M}^{1+M}\frac{F_{-1/2}(\beta (\mu-\Phi))}{\sqrt{M^2-(\Phi-1)^2}}d\Phi-\int_{1+M}^{1-M}\frac{F_{-1/2}(\beta (\mu-\Phi))}{\sqrt{M^2-(\Phi-1)^2}}d\Phi\biggr)\\
&=&2\eta\sqrt{\frac{2}{\beta}}\int_{1-M}^{1+M}\frac{F_{-1/2}(\beta (\mu-\Phi))}{\sqrt{M^2-(\Phi-1)^2}}d\Phi\\
&=&2\eta\sqrt{\frac{2}{\beta}}\int_{-M}^{M}\frac{F_{-1/2}(\beta (\mu-(\Phi+1)))}{\sqrt{M^2-\Phi^2}}d\Phi\\
&=&2\eta\sqrt{\frac{2}{\beta}}\int_{-1}^{1}\frac{F_{-1/2}(\beta (\mu-(M\Phi+1)))}{\sqrt{1-\Phi^2}}d\Phi\end{eqnarray}
and
\begin{eqnarray}
K&=&\int_0^{2\pi}{\cal{K}}d\theta \\
&=&\int_0^\pi {\cal{K}}d\theta +\int_\pi^{2\pi}{\cal{K}}d\theta \\
&=&\frac{\eta}{\beta}\sqrt{\frac{2}{\beta}}\biggl(\int_{1-M}^{1+M}\frac{F_{1/2}(\beta(\mu-\Phi))}{\sqrt{M^2-(\Phi-1)^2}}d\Phi-\int_{1+M}^{1-M}\frac{F_{1/2}(\beta(\mu-\Phi))}{\sqrt{M^2-(\Phi-1)^2}}d\Phi\biggr)\\
&=&2\frac{\eta}{\beta}\sqrt{\frac{2}{\beta}}\int_{1-M}^{1+M}\frac{F_{1/2}(\beta(\mu-\Phi))}{\sqrt{M^2-(\Phi-1)^2}}d\Phi\\
&=&2\frac{\eta}{\beta}\sqrt{\frac{2}{\beta}}\int_{-M}^{M}\frac{F_{1/2}(\beta(\mu-(\Phi+1)))}{\sqrt{M^2-\Phi^2}}d\Phi\\
&=&2\frac{\eta}{\beta}\sqrt{\frac{2}{\beta}}\int_{-1}^{1}\frac{F_{1/2}(\beta(\mu-(M\Phi+1)))}{\sqrt{1-\Phi^2}}d\Phi\;.
\end{eqnarray}

The self-consistency condition is due to
\begin{eqnarray}
\int_0^{2\pi}d\theta \int_{-\infty}^\infty dp (\cos\theta f)&=&\int_0^{2\pi}d\theta (\cos \theta \varrho)\\
&=&2\eta \sqrt{\frac{2}{\beta}}\int_{-M}^M\frac{(-\Phi/M) F_{-1/2}(\beta(\mu-(\Phi+1)))}{\sqrt{M^2-\Phi^2}}d\Phi\\
&=&-2\eta \sqrt{\frac{2}{\beta}}\int_{-1}^1\frac{\Phi F_{-1/2}(\beta(\mu-(M\Phi+1)))}{\sqrt{1-\Phi^2}}d\Phi\;,
\end{eqnarray}
and is given by
\begin{eqnarray}
M_f&=&\frac{1}{N}\int_0^{2\pi}d\theta \int_{-\infty}^\infty dp (\cos\theta f) \label{eq:Mag}\\
&=&-\frac{\int_{-1}^1\frac{\Phi F_{-1/2}(\beta(\mu-(M\Phi+1)))}{\sqrt{1-\Phi^2}}d\Phi}{\int_{-1}^1\frac{F_{-1/2}(\beta(\mu-(M\Phi+1)))}{\sqrt{1-\Phi^2}}d\Phi}\;.
\end{eqnarray}

The potential energy per particle $V$ is, using Eq.(\ref{eq:Mag}),
\begin{eqnarray}
V&=&\int_0^{2\pi}d\theta \int_{-\infty}^\infty dp\biggl(\frac{1-M\cos \theta}{2}\biggr) f(\theta,p)\\
&=&N\bigg(\frac{1-M_fM}{2}\biggr)\;.
\end{eqnarray}

\subsection{Degeneration limit}

The degeneration limit of the Fermi-Dirac function is\cite{FD}
\begin{eqnarray}
F_k(\eta)&\to&\frac{1}{k+1}\eta^{k+1}\ \ (\eta\gg 0)\;,\\
F_{-1/2}(\eta)&\to&2\eta^{1/2}\ \ (\eta\gg 0)\;,\label{eq:dg1}\\
F_{1/2}(\eta)&\to &\frac{2\eta^{3/2}}{3}\ \ (\eta\gg 0)\;.\label{eq:dg2}
\end{eqnarray}

Thus, in the degeneration limit, we obtain
\begin{eqnarray}
\varrho(\theta)&=&\int_{-\infty}^\infty (f) dp\\
&=&\sqrt{\frac{2}{\beta}}\eta 
F_{-1/2}(\beta (\mu-1+M\cos \theta))
\\
&\to&2\sqrt{\frac{2}{\beta}}\eta (\beta(\mu-1+M\cos \theta))^{1/2}\\
&=&2\sqrt{2}\eta(\mu-1+M\cos \theta)^{1/2}\label{eq:rhod}
\end{eqnarray}
and
\begin{eqnarray}
{\cal{K}}(\theta)&=&\int_{-\infty}^\infty \frac{p^2}{2}f dp\\
&=&\frac{1}{\beta}\sqrt{\frac{2}{\beta}}\eta 
F_{1/2}(\beta(\mu-1+M\cos \theta))
\\
&\to&\frac{1}{\beta}\sqrt{\frac{2}{\beta}}\eta \frac{2}{3}(\beta(\mu-1+M\cos \theta))^{3/2}\\
&=&\frac{2}{3}\sqrt{2}\eta (\mu-1+M\cos \theta)^{3/2}\;.\label{eq:Kd}
\end{eqnarray}

On the other hand, for the Vlasov stationary water-bag distribution $f$, we obtain
\begin{eqnarray}
\varrho(\theta)&=&\int_{-\infty}^\infty (f)dp
\\
&=&2\eta p\\
&=&2\eta \sqrt{2(\varepsilon_F-1+M\cos\theta)}\;,\label{eq:dga}\\
{\cal{K}}(\theta)&=&\int_{-\infty}^\infty\frac{p^2}{2}f dp\\
&=&2\eta\frac{1}{6}p^3\\
&=&\eta\frac{1}{3}(\sqrt{2(\varepsilon_F-1+M\cos \theta)})^3\;,\label{eq:dgb}
\end{eqnarray}
where $\varepsilon_F$ denotes the Fermi energy.

These results agree with Eqs.(\ref{eq:rhod}) and (\ref{eq:Kd}).
\end{appendix}

\end{document}